\documentclass[a4paper,fleqn,usenatbib]{mnras}

\usepackage[T1]{fontenc}
\usepackage{ae,aecompl}

\usepackage{graphicx}	
\usepackage{amsmath}	
\usepackage{xcolor}
\usepackage{threeparttable}
\usepackage{longtable}
\usepackage{supertabular}

\usepackage{ulem} 
\usepackage{newtxtext,newtxmath}

\def\amin{\hbox{$^{\prime}$}}

\def\NgalCubeOne{25}
\def\NgalCubeTwo{21}
\def\NgalCubeThree{11}
\def\NgalCubeFour{22}
\def\NgalallCube{79}
\def\NgalallGMOS{35}

\def\NgalGMOSandMUSE{12}
\def\NgalallGMOSnewClus{8}
\def\NgalallGMOScommon{10}
\def\NgalallMUSEGMOS{87} 

\def\redseqgals{210}

\newcommand{\NgalMUSEspec}{117}
\newcommand{\NgalALLspec}{139}
\newcommand{\tscin}{0.96$^{+0.31}_{-0.18}$}
\newcommand{\tscout}{2.60$^{+1.07}_{-0.53}$}
\newcommand{\massN}{2.44 $\pm$ 1.41}
\newcommand{\massS}{3.16 $\pm$ 1.88}

\newcommand{\massSNnE}{1.3}
\newcommand{\massSPT}{7.63 h$_{70}^{-1}$ $\pm$ 1.36}
\newcommand{\redseqgalsspec}{64}

\title[Clash of Titans: SPT-CL J0307-6225]{Clash of Titans: a MUSE dynamical study of the extreme cluster merger SPT-CL J0307-6225}

\author[Hern\'andez-Lang et al.]{Hern\'andez-Lang, D.,$^{1,2,3}$\thanks{E-mail: daniel.hernandez@physik.lmu.de}
Zenteno, A.,$^{2}$
Diaz-Ocampo, A.,$^{2,3}$
Cuevas, H.,$^{3}$
Clancy, J.,$^{4}$\newauthor
Prado P., H.,$^{3}$
Ald\'as, F.,$^{3}$
Pallero, D.,$^{3,5}$
Monteiro-Oliveira, R.,$^{6,7}$
G\'omez, F. A.,$^{3,8}$\newauthor
Ramirez, Amelia,$^{3}$
Wynter, J.,$^{4}$
Carrasco, E. R.,$^{9}$ 
Hau, G. K. T., $^{10}$
Stalder, B., $^{11,12}$ \newauthor 
McDonald, M., $^{13}$ 
Bayliss, M., $^{14}$ 
Floyd, B.,$^{15}$
Garmire, G.,$^{16}$
Katzenberger, A.,$^{17,18,19}$\newauthor
Kim, K. J.,$^{14}$
Klein, M.,$^{1}$
Mahler, G.,$^{20}$
Nilo Castellon, J. L.,$^{3, 21}$
Saro, A.,$^{22,23,24,25}$ \newauthor 
and Somboonpanyakul, T. $^{11}$ \\
\\
\emph{\normalsize Affiliations at the end of the paper.}
}

\date{Accepted XXX. Received YYY; in original form ZZZ}

\pubyear{2022}

\begin{document}
\label{firstpage}
\pagerange{\pageref{firstpage}--\pageref{lastpage}}
\maketitle

\begin{abstract}
We present MUSE spectroscopy, Megacam imaging, and {\it Chandra} X--ray emission for SPT-CL J0307-6225, a z=0.58 major merging galaxy cluster with a large BCG-SZ centroid separation and a highly disturbed X--ray morphology. The galaxy density distribution shows two main overdensities with separations of 0.144 and 0.017 arcmin to their respective BCGs. We characterize the central regions of the two colliding structures, namely 0307-6225N and 0307-6225S, finding  velocity derived masses of $M_{200,N}=$ \massN{} $\times10^{14}$ M$_\odot$ and $M_{200,S}=$ \massS{} $\times10^{14}$ M$_\odot$, with a line-of-sight velocity difference of $|\Delta v| = 342$ km s$^{-1}$. The total dynamically derived mass is consistent with the SZ derived mass of \massSPT{} $\times10^{14}$ M$_\odot$. We model the merger using the Monte Carlo Merger Analysis Code, estimating a merging angle of 36$^{+14}_{-12}$ degrees with respect to the plane of the sky. Comparing with simulations of a merging system with a mass ratio of 1:3, we find that the best scenario is that of an ongoing merger that began \tscin{} Gyr ago. We also characterize the galaxy population using H$\delta$ and [OII] $\lambda 3727$ \AA\ lines. We find that most of the emission-line galaxies belong to 0307-6225S, close to the X--ray peak position, with a third of them corresponding to red-cluster sequence galaxies, and the rest to blue galaxies with velocities consistent with recent periods of accretion. Moreover, we suggest that 0307-6225S suffered a previous merger, evidenced through the two equally bright BCGs at the center with a velocity difference of $\sim$674 km s$^{-1}$.
\end{abstract}




\section{Introduction}
Galaxy clusters are located at  the peaks  of the  (dark) matter  density field  and, as they evolve, they accrete galaxies, galaxy groups, and other clusters from the cosmic web.  Some of those merging events are among the most energetic and violent events in the Universe, releasing energies up to 10$^{64}$ ergs \citep{Sarazin2002, Sarazin2004}, providing extreme conditions to study a range of phenomena, from particle physics \citep[e.g.][]{markevitch04,harvey15,kim2017} to cosmology \citep[e.g.][]{clowe06,thompson15}, including galaxy evolution \citep[e.g.][]{ribeiro13,zenteno2020}.

The cluster assembly process affects galaxies via several physical processes, including harassment, galaxy-galaxy encounters \citep[e.g.,][]{ToomreToomre72},  tidal truncation, starvation, and ram pressure stripping \citep{GunnGott72}, which act upon the galaxies at different cluster centric distances \citep[e.g.,][]{treu03}. Such events not just change the galaxies  in  terms  of  stellar populations  and  morphologies \citep[e.g., ][]{Kapferer2009, mcpartland16, poggianti16, Kelkar20},  but   also  by  destroying  them,  as indicated by a  Halo Occupation Number index lower than 1 \citep[e.g.,][]{lin04a,zenteno11,zenteno16,hennig17}. 

In such extreme environments, galaxies are exposed to conditions that may quench \citep[e.g.][]{Poggianti2004, Pallero2021} or trigger star formation \citep[e.g.][]{Ferrari2003, Owers2012}. For example, \citet{Kalita2019} found evidence of a Jellyfish galaxy in the dissociative merging galaxy cluster A1758N ($z\sim0.3$), concluding that it suffered from ram-pressure striping due to the merging event. \citet{pranger14} studied the galaxy population of the post-merger system Abell 2384 (z$\sim$0.094), finding that the population of spiral galaxies at the center of the cluster does not show star formation activity, and proposing that this could be a consequence of ram-pressure stripping of spiral galaxies from the field falling into the cluster. \citet{ma10} discovered a fraction of lenticular post-starburst galaxies in the region in-between two colliding structures, in the merging galaxy cluster MACS J0025.4-1225 (z$\sim$0.59), finding that the starburst episode occurred during the first passage ($\sim$0.5-1 Gyr ago), while the morphology was already affected, being transformed into lenticular galaxies because of either ram-pressure events or tidal forces towards the central region.

On the other hand, \citet{Yoon2020} found evidence of increase in the star formation activity of galaxies in merging galaxy clusters, alleging that it could be due to an increment of barred galaxies in this systems \citep{Yoon2019}. \citet{Stroe2014} found an increase of H$\alpha$ emission in star-forming galaxies in the merging cluster ``Sausage''(CIZA J2242.8+5301) and, by comparing the galaxy population with the more evolved merger cluster ``Toothbrush'' (1RXS J0603.3+4213), concluded that merger shocks could enhance the star formation activity of galaxies, causing them to exhaust their gas reservoirs faster \citep{Stroe2015}. Furthermore,  \cite{stroe17} using a sample of 19 clusters, at $0.15 < z < 0.31$, found excess of H$\alpha$ emission in merging clusters with respect to relaxed cluster, specially closer to the cluster's core. Such results were further confirmed with an spectroscopic examination of 800 H$\alpha$-selected cluster galaxies \citep{stroe21}. 

To understand how the merger process impacts cluster galaxies,  it is crucial to assemble large samples of merging clusters and determine their corresponding merger phase: pre, ongoing or post. The SZ-selected samples are ideal among the available cluster samples, as they are composed of the most massive clusters in the Universe and are bound to be the source of the most extreme events. The South Pole Telescope \citep[SPT,][]{carlstrom11} has completed a thermal SZ survey, finding 677 cluster candidates \citep{bleem15b}, providing a well understood sample to study the impact of cluster mergers on their galaxy population.  There is rich  available information on those clusters,  including the gas centroids (via SZ and/or X--ray), optical imaging, near-infrared imaging, cluster  masses,  photometric redshifts, etc.  Furthermore, as the SPT cluster selection is nearly independent of redshift, a merging cluster sample will also allow evolutionary studies to high redshifts.

Using SPT-SZ selected clusters and optical imaging, \citet[][]{song12b} reported the brightest cluster galaxy (BCG) positions on 158 SPT cluster candidates and, by using the separation between the cluster BCG and the SZ centroid as a dynamical state proxy, found that SPT-CL J0307-6225 is the most disturbed galaxy cluster of the sample, i.e., with the highest separation. Recently, \cite{zenteno2020} employed optical data from the first three years of the Dark Energy Survey  \citep[DES, ][]{abbott18,morganson18,des16} to use the BCG in 288 SPT SZ-selected clusters \citep{bleem15b} to classify their dynamical state. They identified the 43 most extreme systems, all with a separation greater than 0.4 $r_{200}$, including once again SPT-CL J0307-6225. 

SPT-CL J0307-6225 is a merger candidate at $z=0.5801$ \citep{bayliss16}, with a mass estimate from SPT data of $M_{500} = 5.06\pm0.90 \times 10^{14} h_{70}^{-1}$ M$_\odot$ \citep{bleem15b}. SPT-CL J0307-6225 has (1) $gri$ optical data observed with the Megacam instrument on the Magellan Clay telescope \citep{chiu16b}, (2) X--ray data obtained with the {\it Chandra} telescope \citep[][]{mcdonald13}, and (3) spectroscopic information taken with the Gemini Multi-Object Spectrograph \citep[GMOS;][]{bayliss16}. \cite{Dietrich2019} used the Megacam data to measure the weak lensing mass density and, although the cluster was observed under the best seeing conditions in the sample (0.55-0.65 arcsec), the resulting WL mass distribution is of low significance, with the recovered center located away from the gas distribution or the galaxies (see their Fig.~B.4). 

In the absence of precise WL measurements, the galaxy-gas offset can be used to constrain self-interacting dark matter models as shown by \cite{Wittman2018}. The separation between the X--ray centroid of SPT-CL J0307-6225, estimated using \textit{Chandra} data \citep{mcdonald13}, and the BCG \citep{zenteno2020} is 1.98 arcmin ($\sim$790 kpc). This would be the largest gas-galaxy offset within the \cite{Wittman2018} sample of merging galaxy clusters, implying a high potential for SPT-CL J0307-6225 to constrain such models. Using GMOS spectroscopic data, \cite{bayliss16} studied the velocity distribution of the SPT-GMOS sample (62 galaxy clusters), finding SPT-CL J0307-6225 to be one of the 9 clusters with a non-Gaussian (i.e., disturbed) velocity distribution (2-$\sigma$ level). \cite{Nurgaliev2017} used the \textit{Chandra} data to make an estimate of the X--ray asymmetry for this system, finding it to be the second\footnote{In \citet{zenteno2020}, the most asymmetric system, SPT-CL J2332-5053, was said to be a cluster in pre-merger state with a close companion, which would then contaminate the estimated asymmetry index. Excluding SPT-CL J2332-5053 would make SPT-CL J0307-6225 the most asymmetric system in the sample.} most asymmetric system in the full SPT-Chandra sample (over 90 galaxy clusters), with an X--ray morphology as disturbed as El Gordo, a well-known major merger \citep[][]{williamson11,Menanteau2012}, making this cluster an interesting system to test the impact of a massive merging event in galaxy evolution, the goal of this paper.

We use VLT/MUSE integral field and Gemini/GMOS spectroscopy, X--ray data from {\it Chandra}, and Megacam optical imaging to characterize the SPT-CL J0307-6225 merger stage, and its impact on galaxy population. The paper is organized as follow: in $\S$\ref{sec:observations} we provide details of the observations and data reduction. In $\S$\ref{sec:analysis} we show the analysis for the spectroscopic and optical data, while in $\S$\ref{sec:results} we report our findings for both the merging scenario and the galaxy population. In $\S$\ref{sec:discussion} we propose an scenario for the merging event and connect it to the galaxy population. In  $\S$\ref{sec:conclusions} we give a summary of the results. Throughout the paper we assume a flat Universe, with a $\Lambda$CDM cosmology, $h=0.7$, $\Omega_m = 0.27$ \citep[][]{Komatsu2011}. Within this cosmology, 1 arcsec at the redshift of the cluster ($z\approx0.58$) corresponds to $\sim$6.66 kpc.

\begin{figure*}
    \centering
    \includegraphics[width=\linewidth]{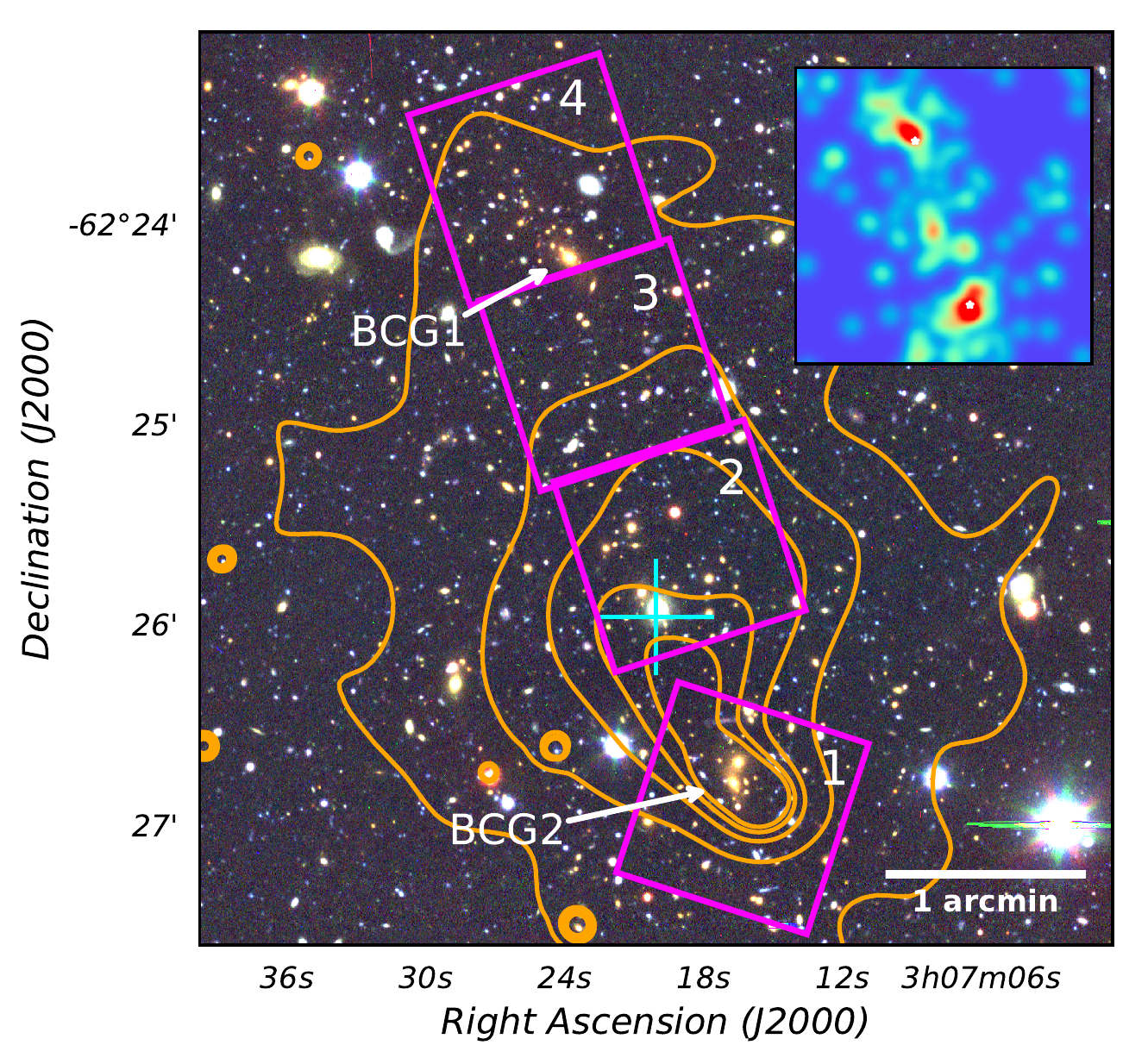}
    \vskip-0.13in
    \caption{Pseudo-color image, from $gri$ filters combination, of the central area of SPT-CL J0307-6225. Magenta squares show the MUSE footprints, where the numbers on the top-right corner of each square shows the cube's number. Orange contours where derived from archival \textit{Chandra} images. The cyan plus-sign marks the X--ray centroid \citep{mcdonald13}. The arrows show the positions of the two brightest galaxies of the cluster. The white bar on the bottom shows the scale of 1 arcmin. The inset shows the 2D galaxy number density (which matches the size of the main figure), where the two highest intensity areas correspond to the areas around the BCGs, which are shown as white stars.}
    \label{fig:rgb_image}
\end{figure*}

\section{Observations and Data Reduction}
\label{sec:observations}

\subsection{Optical Imaging}
\label{sec:imaging}
\cite{chiu16b} obtained optical images using Magellan Clay with Megacam during a single night on November 26, 2011 (UT). They reduced  and calibrated the data following \cite{High2012}. Megacam has a 24\amin x 24\amin field-of-view, which at redshift $\sim$0.58 correspond to $\sim$10 Mpc. Several dithered exposures were taken in $g$, $r$, and $i$ filters for a total time of 1200 s, 1800 s, and 2400 s  respectively. The median seeing of the images was approximately 0.79 arcsec or about 5 kpc, with a better seeing in r-band, averaging 0.60 arcsec. The 10$\sigma$ limit magnitudes in $gri$ are 24.24, 24.83, and 23.58, respectively \citep{chiu16b}. In Fig.~\ref{fig:rgb_image} we show the $gri$ pseudo-color image, centered on the SZ cluster position of SPT-CL J0307-6225, with the white bar on the bottom right showing the corresponding scale.

The catalogs for the photometric calibration were created following \citet{High2012} and \citet{Dietrich2019} including standard bias subtraction and bad-pixel masking, as well as flat fielding, illumination, and fringe (for i-band only) corrections. To calibrate the zeropoint of the data, the stellar locus regression technique was used \citep{High2009}, together with constraints by cross-matching with 2MASS catalogs \citep{Skrutskie2006}, giving uncertainties in absolute magnitude of 0.05 mag and in color of 0.03 mag \citep{Desai2012, song12b}.

For the creation of the galaxy photometric catalogs, we use a combination of Source Extractor \citep[\texttt{SExtractor};][]{bertin96} and the Point Spread Function Extractor \citep[\textsc{PSFex};][]{Bertin2011} softwares. \textsc{SExtractor} is run in dual mode, using the $i$-band image as the reference given the redshift of the cluster\footnote{ At $z\approx0.58$, the $i$-band is located redwards the 4000\AA\ break.}. We extract all detected sources with at least 6 pixels connected above the 4$\sigma$ threshold, using a 5 pix Gaussian kernel. Deblending is performed with 64 sub-thresholds and a minimum contrast of 0.0005. Galaxy magnitudes are \textsc{SExtractor}'s \texttt{MAG\_AUTO} estimation, whereas colors are derived from aperture magnitudes.

The star-galaxy separation in our sample is performed following \citet{Drlica-Wagner2018}, by using the \textsc{SExtractor} parameter \textsc{spread\_model}, and its corresponding error, \textsc{spreaderr\_model}, derived from the $i$-band image, for objects within $R_{200}$ from the SZ center \citep[$R_{200} = 3.84'$;][]{song12b, zenteno2020}. \citet{Drlica-Wagner2018} classified a source as a star if it satisfies

\begin{equation}\label{eq:phot_gals}
    |\textsc{spread\_model} + \left(\frac{5}{3}\right)\times\textsc{spreaderr\_model} | < 0.002
\end{equation}
\noindent
With this, we remove stars from our catalogue and, to improve upon this selection, we apply a magnitude cut, such that $i_{\rm{auto}}<18.5$ mag, which is $\sim0.5$ mag brighter than the BCG. On the faint end the cut is set at $i_{\rm{auto}} < m^*+3 = 23.39$, which is beyond the limit of our spectroscopic catalogue (see Appendix \ref{sec:completeness}). With this we obtain 639 photometric galaxies.

\subsection{Spectroscopic data}
\label{sec:reductions}
\subsubsection{MUSE data}
The Multi Unit Spectroscopic Explorer \citep[MUSE,][]{bacon12} observations were taken on August 22nd, 23rd and 24th, 2016  (program id: 097.A-0922(A), PI: Zenteno), and November 10 and December 20, 2017 (program id: 100.A-0645(A), PI: Zenteno). The observations consisted of four pointings, with a total exposure time of 1.25 hours per data cube, with an airmass = 1.4 (see Table~\ref{tab:muse_data}). MUSE in nominal mode covers the wavelength range 4800-9300 \AA, with resolution of 1700 < R < 3500, covering redshifted emission lines such as [OII] $\lambda$3727 \AA\ and [OIII] $\lambda$5007 \AA, as well as absorption lines such as the Hydrogen Balmer series H$\delta$, H$\gamma$ and H$\beta$. The positions of the pointings were selected to cover the two BCGs (labeled as BCG1 and BCG2 on Fig.~\ref{fig:rgb_image}) and the area between them. The MUSE footprints for the 4 observed data cubes are shown as magenta squares on Fig.~\ref{fig:rgb_image}, with the cubes enumerated in the top right corner of each square. We use these numbers to refer to the cubes throughout the paper.

\begin{table}
    \caption{Central coordinates and seeing conditions of the observed MUSE fields}
    \label{tab:muse_data}
    \centering
    \resizebox{\linewidth}{!}{%
    \begin{tabular}{c c c c c}
    \hline
    \hline
        CUBE & Program & \multicolumn{2}{c}{Coordinates} & Seeing \\
         & ID & R.A. (J2000) & Dec. (J2000) & (arcsec)\\
        \hline
        1 & 097.A-0922(A) & $03^{\rm h}\ 07^{\rm m}\ 16.34^{\rm s}$ & $-62^\circ\ 26^{\prime}\ 54.98^{\prime\prime}$ & 0.56 \\
        2 & 097.A-0922(A) & $03^{\rm h}\ 07^{\rm m}\ 19.052^{\rm s}$ & $-62^\circ\ 25^{\prime}\ 36.430^{\prime\prime}$ & 0.70 \\ 
        3 & 0100.A-0645(A) & $03^{\rm h}\ 07^{\rm m}\ 22.271^{\rm s}$ & $-62^\circ\ 24^{\prime}\ 42.140^{\prime\prime}$ & 0.68 \\
        4 & 0100.A-0645(A) & $03^{\rm h}\ 07^{\rm m}\ 25.302^{\rm s}$ & $-62^\circ\ 23^{\prime}\ 46.570^{\prime\prime}$ & 0.97 \\
        \hline
    \end{tabular}
    }
\end{table}

The data was taken in WFM-NOAO-N mode, with a position angle of 18 deg for three of the cubes and 72 deg for the one to the south, and using the dithering pattern recommended for best calibration: 4 exposures with offsets of 1 arcsec and 90 degrees rotations (MUSE User Manual ver. 1.3.0). The raw data were reduced through the MUSE pipeline \citep{weilbacher14, Weilbacher2016} provided by ESO.

We construct 1D spectra from the MUSE cube using the {\tt MUSELET} software \citep{bacon16}. {\tt MUSELET} finds source objects by constructing line-weighted (spectrally) 5x1.25 \AA\ wide  narrow band images and running \texttt{SExtractor} on them. In order to create well fitted masks to their respective sources, the parameter {\tt DETECT\_THRESH} is set to be 2.5. If the chosen value is below that, {\tt SExtractor} will detect noise and output wrong shapes in the segmentation map. We proceed to use the source file to extract the {\tt SExtractor} parameters {\tt A\_WORLD}, {\tt B\_WORLD} and {\tt THETA\_WORLD} to create an elliptical mask centered in each source. 

Finally, we use the {\tt MUSELET} routines {\tt mask\_ellipse} and {\tt sum}  to create the 1D weighted spectra of the sources. To make sure the objects fit into their apertures, the SExtractor parameter {\tt PHOT\_FLUXFRAC} is set at 0.9, which means that 90\% of the source's flux will be contained within the mask's radius. 

\subsubsection{GMOS data}
We complement MUSE redshifts with Gemini/GMOS data published by \citet{bayliss16}. The Bayliss galaxy redshift sample consists in \NgalallGMOS\ galaxies redshifts, with \NgalallGMOSnewClus\ not present in our MUSE data. The spectroscopic data from their sample can be found online at the \textsc{VizieR Catalogue Service} \citep{Vizier}, with the details on the data reduction described in \cite{bayliss16} and \cite{Bayliss2017}. For SPT-CL J0307-6225, they used 2 spectroscopic masks with an exposure time of 1 hour each. The target selection consisted mostly of galaxies from the red sequence (selected as an overdensity in the color-magnitude and color-color spaces) up to $m^* + 1$, prioritising BCG candidates.

\subsection{X--ray data} 
\label{subsec:xraydata}
SPT-CL J0307-6225 was observed by {\it Chandra} as part of a larger, multi-cycle effort to follow up the 100 most massive SPT-selected clusters spanning $0.3 < z < 1.8$ \citep{mcdonald13,McDonald2017}. In particular, this observation (12191) was obtained via the ACIS Guaranteed Time program (PI: Garmire). A total of 24.7\,ks was obtained with ACIS-I in VFAINT mode, centering the cluster $\sim$1.5$^{\prime}$ from the central chip gap. The data was reprocessed using \textsc{ciao} v4.10 and \textsc{caldb} v.4.8.0. For details of the observations and data processing, see \cite{mcdonald13}. The derived X--ray centroid is shown as a cyan plus-sign on Fig.~\ref{fig:rgb_image}.

An image in the 0.5--4.0\,keV bandpass was extracted and adaptively smoothed using \textsc{csmooth}\footnote{\url{https://cxc.harvard.edu/ciao/ahelp/csmooth.html}}. This smoothed image, shown as orange contours in Fig.~\ref{fig:rgb_image}, reveals a highly asymmetric X--ray morphology, with a bright, dense core offset from the large-scale centroid by $\sim$1$^{\prime}$ ($\sim$400\,kpc).

\begin{figure}
    \centering
    \includegraphics[width=\linewidth]{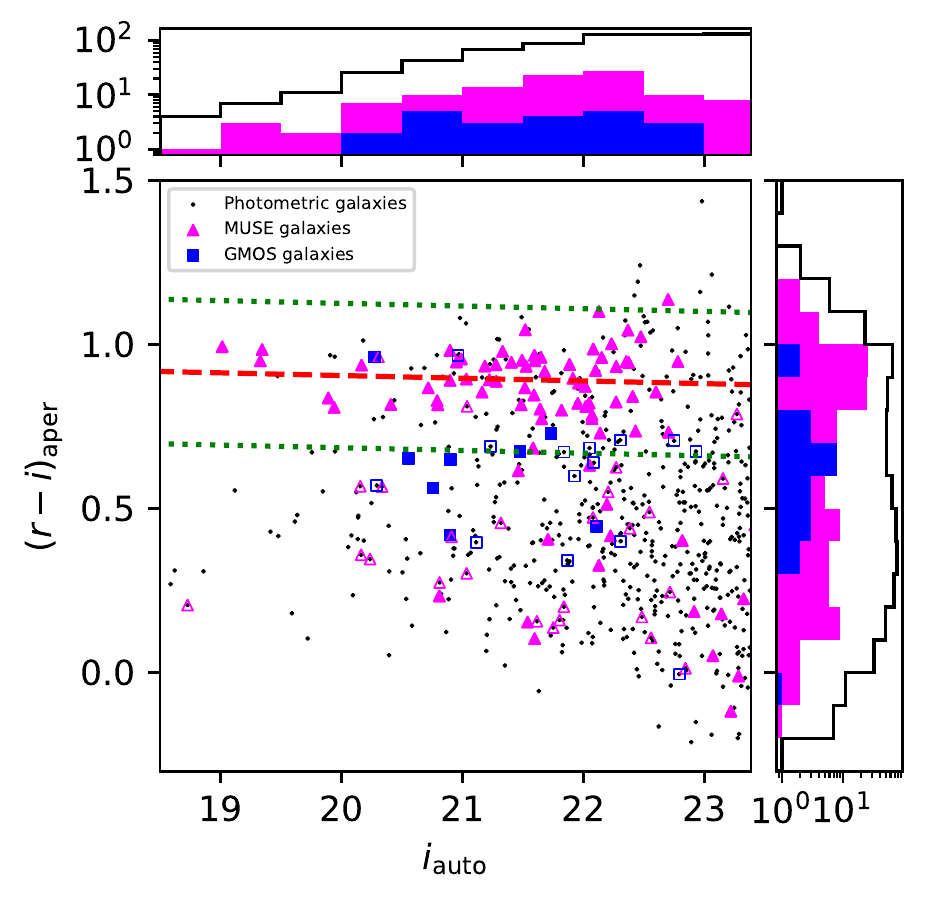} 
    \vskip-0.13in
    \caption{Color-magnitude diagram (CMD) of SPT-CL J0307-6225 from Megacam data within R$_{200}$. The $y$-axis shows the color index $r-i$ estimated from aperture magnitudes, with a fix aperture of $\sim$40 kpc ($\sim$6 arcsec) at the cluster redshift, while the $x$-axis shows \textsc{SExtractor}'s \texttt{MAG\_AUTO}. Magenta triangles and blue galaxies represent galaxies from our MUSE and GMOS data, respectively, filled for those that belong to the cluster, whereas black dots are galaxies from our photometric sample. The red cluster sequence (RCS) estimated for the cluster is shown as a red-dashed line, while the green dotted lines are the 0.22 mag width established for the RCS.}
    \label{fig:cmd}
\end{figure}

\section{Analysis}
\label{sec:analysis}
\subsection{Color-Magnitude Diagram and RCS selection}
\label{sec:photometric}

The color-magnitude diagram (CMD) for the cluster is shown in Fig.~\ref{fig:cmd}, where the magenta triangles and the blue squares are galaxies from the MUSE and GMOS spectroscopic samples, respectively, and the dots represent galaxies from our photometric sample (selected as described in \S~\ref{sec:imaging}). For the selection of the red cluster sequence (RCS) galaxies, which consist mostly of passive galaxies which are likely to be at the redshift of the cluster \citep{Gladders2000}, we examine the location of the galaxies from our spectroscopic sample in the CMD. We then select all galaxies with $r-i>0.65$ and perform a 3$\sigma$-clipping cut on the color index to remove outliers. We keep all the galaxies from our previous magnitude cut in $\S$~\ref{sec:imaging} ($i_{\rm auto}< 23.39$). Finally, we fit a linear regression to the remaining objects, which is shown with a red dashed line in Fig.~\ref{fig:cmd}. The green dotted lines denote the limits for the RCS, chosen to be $\pm$0.22 [mag] from the fit, which corresponds to the average scatter of the RCS at 3$\sigma$ \citep{lopez04}. This gives us a total of \redseqgals\ optically selected RCS galaxy candidates, with \redseqgalsspec\ of those being spectroscopically confirmed members. 

\subsection{Spectroscopic catalog}

\subsubsection{Galaxy redshifts}
\label{sec:redshifts}
\begin{figure}
    \centering
    \includegraphics[width=\linewidth]{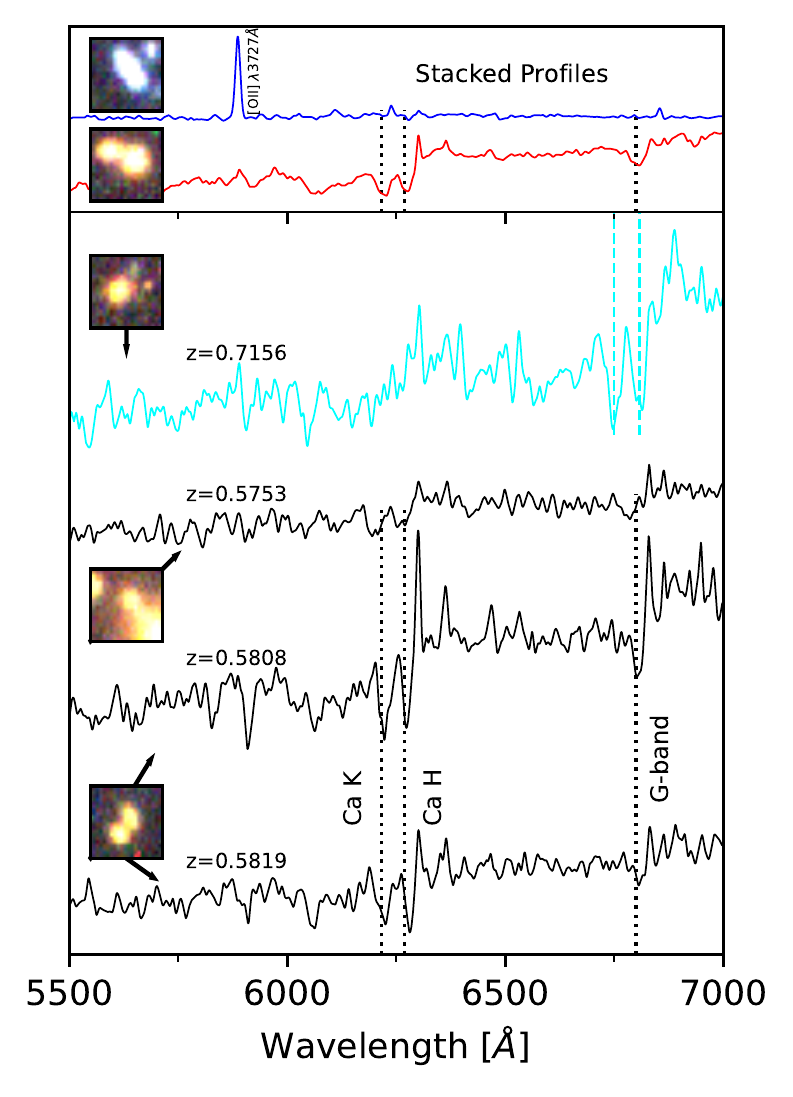}
    \vskip-0.13in
    \caption{\textit{top}: Stacked spectra of a couple of blue and red galaxies at the cluster's redshift, shown in blue and red, respectively. The cutout on the left shows an example of a galaxy from each profile. Black dotted lines mark the Calcium $H$ and $K$ lines, together with the G-band feature at 4304 \AA, redshifted to $z=0.58$.. \textit{bottom}: Spectrum of the sources with redshifts estimated manually (black) and that of a galaxy with similar characteristic to those of the cluster, but at $z=0.716$. A small cutout of $5\times5$ arcsec$^2$ is shown on the left for each galaxy, with a black arrow pointing at the respective spectra. The redshift found with \textsc{MARZ} of each source is written on top of each spectrum. Dotted lines are the same as in the upper panel, with the cyan dashed lines marking the Calcium $H$ and $K$ lines redshifted to $z=0.716$.}
    \label{fig:manual_redshift}
\end{figure}

To obtain the redshifts, we use an adapted version of \textsc{MARZ} \citep{Hinton2016} for MUSE spectra\footnotemark \footnotetext{\url{http://saimn.github.io/Marz/\#/overview} (Hinton, private communication)}. \textsc{MARZ} takes the 1D spectra of each object as an input, obtaining the spectral type (late-type galaxy, star, quasar, etc.) and the redshift that best fits as an output. The results are examined visually for each of the objects, calibrating them using the 4000\AA\ break and the Calcium $H$ and $K$ lines. Heliocentric correction was applied to all redshifts using the \textsc{rvcorrect} task from \textsc{iraf}. The upper panel of Fig.~\ref{fig:manual_redshift} shows the stacked spectra of a couple of blue and red galaxies.

There are three sources in the cube 4 region which appeared to be part of the cluster, but were not well fitted by \textsc{MARZ}. These sources are shown in the bottom panel of Fig.~\ref{fig:manual_redshift}, with their spectra shown in black and the cutouts of the galaxies in the left. The cyan spectra shows a galaxy with an estimated redshift higher than that of the cluster but with a $r-i$ color within our RCS selection. We manually estimate the redshifts of these 3 sources using \textsc{MARZ}.

In total we estimate spectroscopic redshifts for \NgalMUSEspec\ objects within the MUSE fields, with 4 of them classified as stars. In Table~\ref{tab:all_objs_properties} we show the redshifts and magnitudes for this objects. For details of the different columns please refer to Appendix~\ref{sec:appendix_catalog}.

In Table~\ref{tab:all_objs_properties} we show the properties of 22 objects from GMOS, excluding the 12 in common with MUSE and the potential cluster member from our measured redshifts. In Appendix \ref{sec:gmos_comparison} we give further details into the estimation of the GMOS spectra redshifts, the comparison to our estimates with MUSE and the exclusion of potential members. GMOS redshifts in Table~\ref{tab:all_objs_properties} correspond to the ones measured using \textsc{fxcor}. Our final spectroscopic catalog is composed of \NgalALLspec{} objects; 134 galaxies and 5 stars.

\subsubsection{Cluster redshift estimation}
\label{sec:cluster_redshifts}

The cluster's redshift is estimated following the biweight average estimator from \citet{Beers1990}, using the median redshift from all objects with measured redshift in our sample. This estimated redshift is then used instead of the median in their equation, in order to estimate a new redshift. This process is iterated 3 times. We select only spectroscopic sources with a peculiar velocity within $\pm$5000 km s$^{-1}$ from the cluster's estimated redshift, in order to exclude most of the foreground and background objects \citep[eg.][]{Bosch2013,pranger14}. We then estimate the velocity dispersion ($\sigma_v$)  using the biweight sample variance presented in \citet{Ruel2014}, so that

\begin{equation}\label{eq:vdisp}
    \sigma_{\rm bi}^2 = N \frac{\sum_{|u_i|<1} (1-u_i^2)^4(v_i-\bar{v})^2}{D(D-1)}
\end{equation}    
\begin{equation}
    D = \sum_{|u_i|<1} (1-u_i^2)(1-5u_i^2)
\end{equation}
\noindent
where the peculiar velocities of the galaxies, $v_i$, and the biweight weighting, $u_i$, are estimated as 
\begin{equation}
    v_i = \frac{c (z_i - z_{\rm cl})}{1+z_{\rm cl}}
\end{equation}
\begin{equation}
    u_i = \frac{v_i - \bar{v}}{9\rm{MAD}(v_i)}
\end{equation}
\noindent
with $c$ being the speed of light, $\rm{MAD}$ corresponds to the median absolute deviation and $z_i$, $z$ being the redshifts of the galaxies and the biweight estimation of the redshift of the sample, respectively. Then, the velocity dispersion is estimated as the square root of $\sigma_{\rm bi}^2$, with its uncertainty estimated as 0.92$\sigma_{\rm bi} \times \sqrt{ N_{\rm members} - 1}$. To obtain a final redshift for the cluster we use a 3$\sigma$-clipping iteration (with $\sigma=\sigma_v$), obtaining $z_{\rm cl}=0.5803\pm0.0006$, where the error is estimated as the standard error, i.e., the standard deviation over the square root of the number of cluster members.

\subsubsection{Cluster member selection}
\label{sec:cluster_member_selection}

Observationally,  galaxies belonging to a cluster are selected by imposing restrictions on their distance to the center of the cluster and their relative velocities to the BCG. In this section, we studied the appropriate cut in the Line of Sight (LoS) velocity for a theoretical cluster with the same mass and the same redshift than SPT-CLJ0307-6225 using the Illustris TNG300 simulations. Illustris TNG is a suite of  cosmological-magnetohydrodynamic simulation which aims to study the physical processes that drive galaxy formation \citep{nelson17, pillepich17, springel17, naiman18, marinacci18}.  We used the TNG300 because it is the simulation with the largest volume,  having a side length of $L\sim 250 h^{-1 }$ Mpc. This volume contains $2000^3$ Dark Matter (DM) particles and $2000^3$ baryonic particles. The relatively large size of the simulated box allow us to identify a significant number of massive structures. The mass resolution of TNG300  is $5.9\times 10^7 M_{\odot} $, and $1.1\times 10^7 M_{\odot}$ for the DM and baryonic matter respectively. Also, the adopted softening length is 1 h$^{-1}$ kpc for the DM particles and 0.25 h$^{-1}$ kpc for the baryonic particles \citep{marinacci18}. 

This simulation have a total of 1150 structures with masses  between $ 10^{14}{M_\odot}\leq M_{200}\leq 9 \times 10^{14}{M_\odot}$, in a redshift range $0.1\leq z\leq 1$. Here $M_{200}$ is the mass within a sphere having a mean mass density equal to 200 times the critical density of the Universe.  To ensure that our results are not affected by numerical resolution effects, we only selected subhalos with at least 1000 dark matter particles per galaxy  ($M_{\rm DM}\geq 5.9 \times 10^{10}{M_\odot}$) and at least 100 stellar particles ($M_{\rm stellar}\geq 1.1 \times 10^{9}{M_\odot}$). 

We used the criteria proposed by  \citet{zenteno2020} to divide the clusters according their virialization stage. We consider a  that a cluster is disturbed when the offset between the position of the BCG and the center of mass of the gas is greater than $0.4\times R_{200}$ (used as a proxy for the Sunayev-Zeldovich effect) otherwise, we consider them as relaxed. The final sample used in this work is composed by  the 150 relaxed clusters and 150 disturbed clusters.

To stack information from the selected clusters we normalize the velocity distributions  using the $\sigma_v - M_{200}$ scaling relation from \citet{Munari2013}. This scaling relation  was obtained from a radiative simulation which included both (a) star formation and supernova triggered feedback, and (b) active galactic nucleus feedback (which they call the AGN-set). The equation is described as follows:

\begin{equation}\label{eq:mass}
   \sigma_{\rm 1D} = A_{\rm 1D} \left[ \dfrac{h(z) M_{200}}{10^{15} M_\odot} \right] ^\alpha
\end{equation}
\noindent

where $\sigma_{\rm 1D}$ is the one-dimensional velocity dispersion and h(z) = H(z)/100 km s$^{-1}$ Mpc$^{-1}$. We choose the values of $A_{\rm 1D} = 1177 \pm 4.2$ and $\alpha=0.364 \pm 0.0021$, obtained using galaxies associated to subhaloes in the AGN-set simulation \citep{Munari2013}. 

\begin{figure}
\centering
\includegraphics[width=\linewidth]{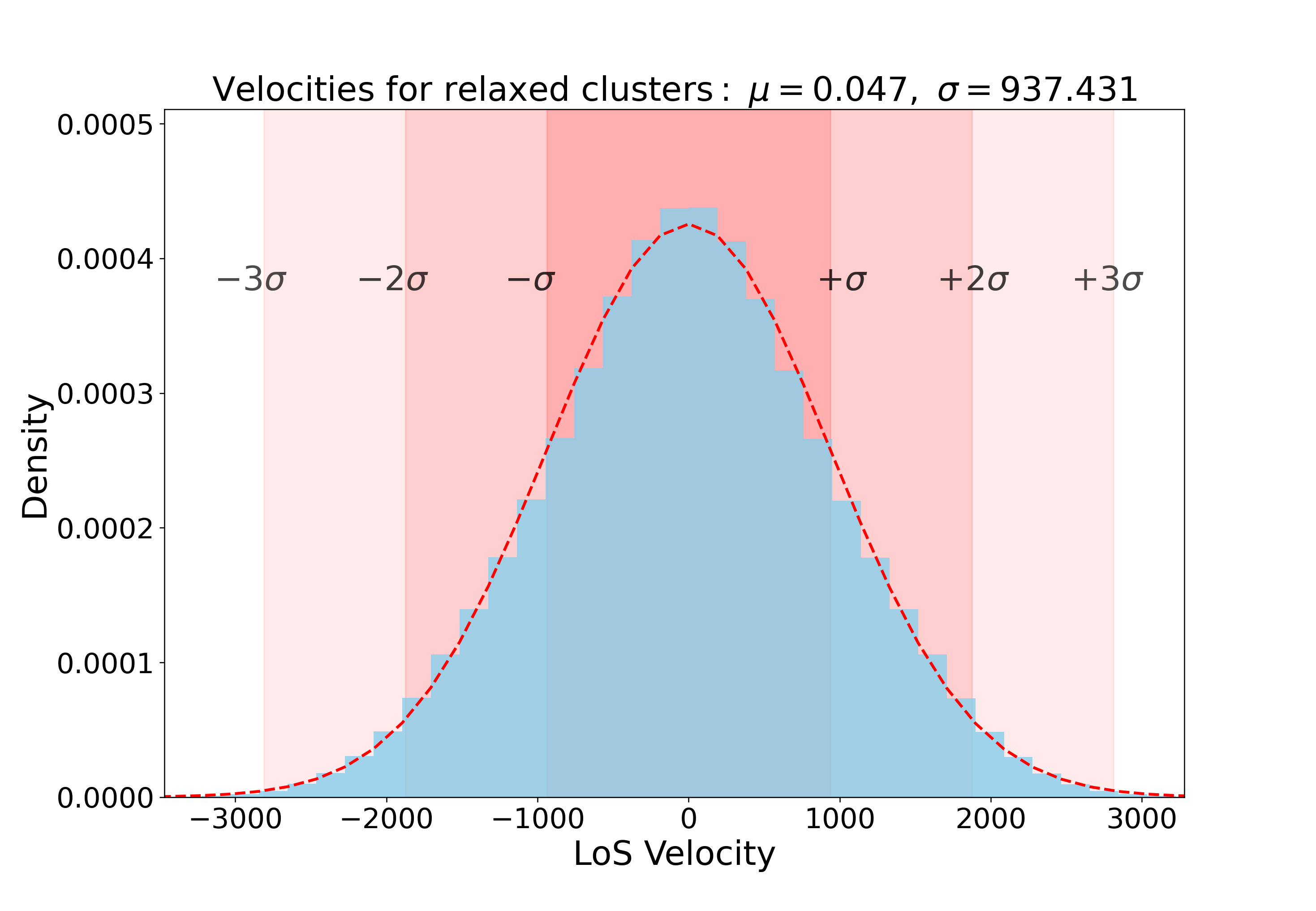}
\includegraphics[width=\linewidth]{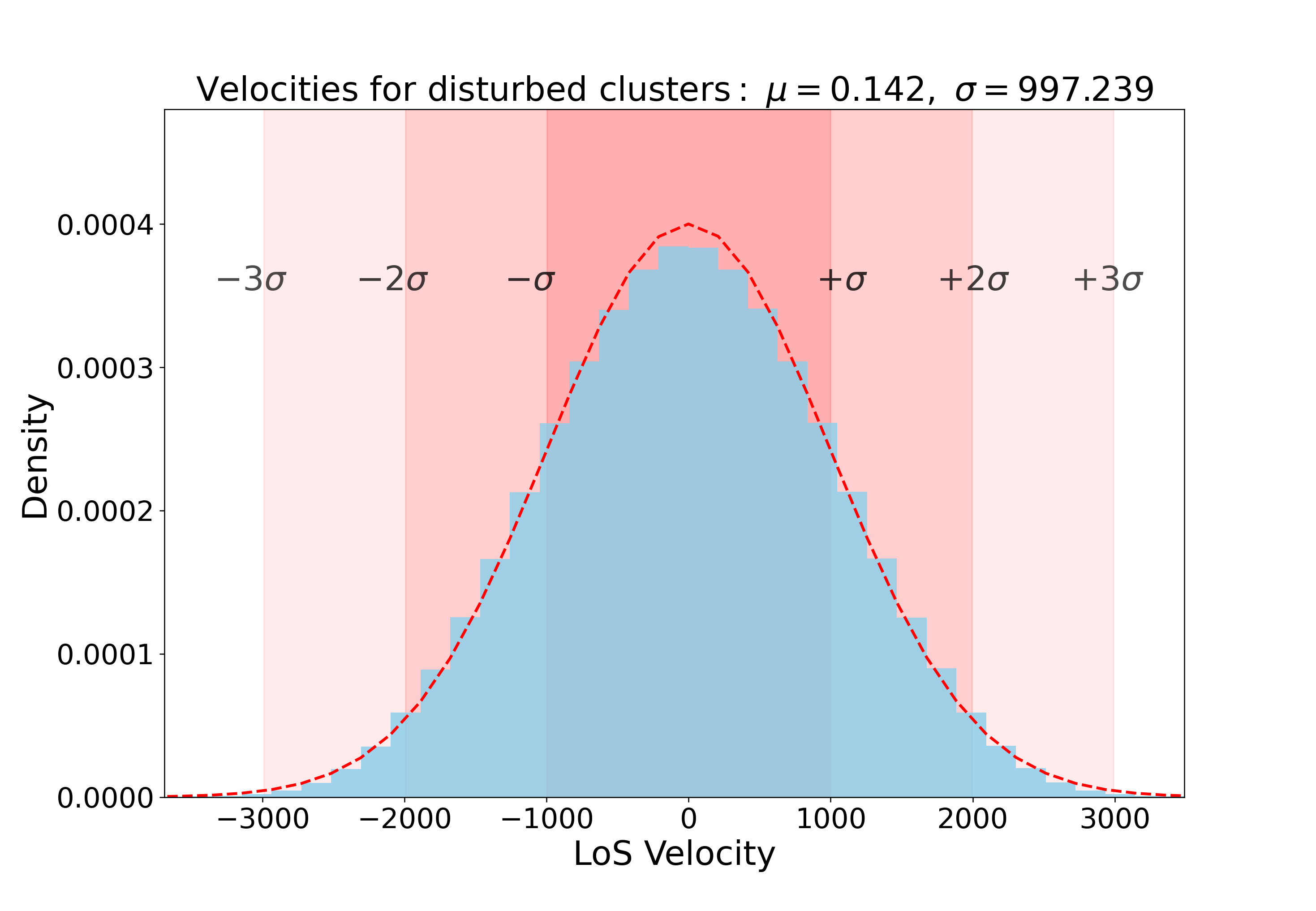}
\vskip-0.13in
\caption{Histogram for the LoS satellite velocities distribution for relaxed (top) and disturbed (bottom) clusters with masses $M_{200}=7.64 \times 10^{14}{M_\odot}$ at redshift $z=0.6$, in red the fitted normal distribution and in light red the confidence intervals. }
\label{fig:histogram_relaxed}
\end{figure}

To find the intrinsic Line of Sight (LoS) velocity distribution of a simulated cluster with mass $M_{200}=5 \times 10^{14}{M_\odot}$, at a given redshift of $z=0.6$, we followed the following procedure. We first fit the projected  1D velocity  distribution of the cluster galaxies relative to the BCG using a Gaussian distribution  with mean $\mu_0$ and dispersion $\sigma_0$. After, using the Equation \ref{eq:mass}, we compute the  value of the 1D velocity dispersion $\sigma_1$ that the cluster would have if it had a mass of $M_{200}=5 \times 10^{14}{M_\odot}$. Then,  we obtain the 1D velocities for each galaxy normalized by the mass and the redshift using the equation \ref{transformation}. Finally, we obtained the LoS velocities applying 200 different randomized rotations to each cluster, 
\begin{equation}
z=\sigma_1 \left( \frac{x-\mu_0}{\sigma_0} +\mu_0\right).
\label{transformation}
\end{equation}

Fig.~\ref{fig:histogram_relaxed} presents the histogram in the LoS velocity for the galaxies associated to the 150 relaxed (top) and disturbed (bottom) clusters stacked in different projections (blue histogram), the best fit normal distribution (red dashed line) and the confidence intervals shaded red areas. We conclude that for a relaxed cluster with mass of $M_{200}= 7.64\times 10^{14}$ the LoS velocity is distributed with a dispersion $\sigma_v= 940 km\;s^{-1}$. For disturbed clusters the velocities are normally distributed with a dispersion of $\sigma= 1000 km\;s^{-1}$. This means that $95 \%$ of the galaxies belonging to a disturbed cluster with $M_{200}=7.64\times 10^{14}$ would have LoS velocities lower than 2000 km s$^{-1}$, and $99\% $ of them have LoS velocities lower than 3000 km s$^{-1}$. In what follows we adopt a cut of 3,000 km s$^{-1}$. Our results shows that the distribution of LoS velocity is not significantly affected by the virialized status of the studied cluster.

Applying the $\pm$~3,000 km/s cut we obtain a total number of cluster redshifts of \NgalallMUSEGMOS{}, including  \NgalCubeOne{} members from cube 1, \NgalCubeTwo{} from cube 2, \NgalCubeThree{} from cube 3, \NgalCubeFour{} from cube 4 and 8 from the GMOS data. 

\begin{figure}
    \centering
    \includegraphics[width=\linewidth]{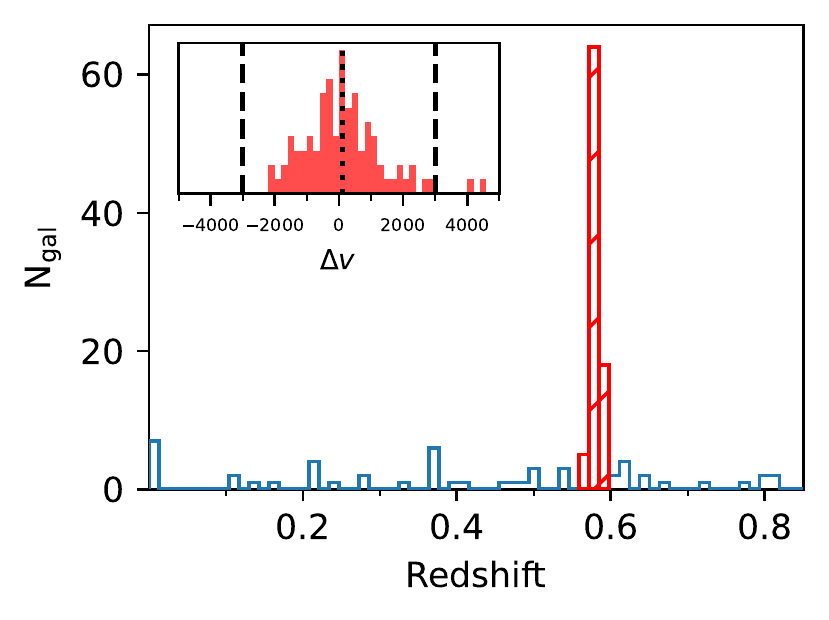}
    \vskip-0.13in
    \caption{Redshift distribution of spectroscopic sources with good measurement from \textsc{MARZ} and \textsc{fxcor}. Hashed red bars represent the region within a range of $\pm$3000 km s$^{-1}$ in peculiar velocity from the cluster's redshift. The histogram insert on the top left shows the distribution of galaxies within this velocity range, where the black dashed and dotted lines represent the cuts at $\pm$3000 km s$^{-1}$ and the velocity of the BCG, respectively.}
    \label{fig:redshift_distribution}
\end{figure}

\subsubsection{Summary of spectroscopic catalog}
\label{sec:spec_summary}
In total, we obtain 87 galaxies with spectroscopic redshifts for SPT-CL J0307-6225.  Out of those, 79 come from the 1D MUSE objects from $\S$\ref{sec:reductions} and 8 from the GMOS archival spectroscopic data \citep{bayliss16}. The final redshift, estimated as the biweight average estimator, is $z_{\rm cl} = 0.5803 \pm 0.0006$. The final  galaxy cluster redshift distributions is shown in Fig.~\ref{fig:redshift_distribution}. The inset shows the peculiar velocity of these selected galaxies, with the black dashed lines denoting the velocity cut and the black dotted line marking the velocity of the BCG. The velocity dispersion for the cluster, estimated following Eq.~\ref{eq:vdisp}, is  $\sigma_v = 1093\pm108$ km $s^{-1}$.

\begin{table}
    \caption{Galaxy population classification}
    \label{tab:gal_pop_clas}
    \centering
    \begin{tabular}{l l}
    \hline
    \hline
    Type & Criteria \\
    \hline
    PSB & Galaxies with EW(H$\delta$) $\geq 5$ \AA\ and EW(OII) $< 5$ \AA\\
    SSB & Galaxies with EW(H$\delta$) $< 0$ \AA\ and EW(OII) $\geq 5$ \AA\\
    EL & Galaxies with EW(OII) $\geq 5$ \AA\ (SF, SSB and A+em)\\
    NEL & Galaxies with EW(OII) $< 5$ \AA\ (Passive and PSB) \\\\
    Red & Galaxies belonging to (or redder than) the RCS \\
    Blue & Galaxies with colors lower than the RCS \\
    \hline
    \end{tabular}
\end{table}

\begin{table*}\caption{Fraction of galaxy types at different magnitude ranges. The second column is the total number of galaxies for a given magnitude range, while the third column is the median S/N of the galaxies.}
    \label{tab:mag_properties}
    \centering
    \begin{threeparttable}
    \begin{tabular}{rrrrrrrrrr}
    \hline
    \hline
     & & & \multicolumn{2}{|c}{Photometric} & \multicolumn{5}{c|}{Spectroscopic} \\
    \multicolumn{1}{c}{Mag} & \multicolumn{1}{c}{N$_{\rm Total}$} & \multicolumn{1}{c}{S/N} & \multicolumn{1}{c}{Red} & \multicolumn{1}{c}{Blue} & \multicolumn{1}{c}{NEL} & \multicolumn{1}{c}{EL} & \multicolumn{1}{c}{Low S/N} & \multicolumn{1}{c}{SSB} & \multicolumn{1}{c}{PSB} \\
    & & & \multicolumn{1}{r}{\%} & \multicolumn{1}{r}{\%} & \multicolumn{1}{r}{\%} & \multicolumn{1}{r}{\%} & \multicolumn{1}{r}{\%} & \multicolumn{1}{r}{\%} & \multicolumn{1}{r}{\%}\\
    \hline
    $i_{\rm{auto}} < m^*$\ \ \ \ \ \ \ & 6 & 12.0 & 100.00 & 0.00 & 100.00 & 0.00 & 0.00 & 0.00 & 0.00\\
    $m^* \leq i_{\rm{auto}} < m^*+1$ & 16 & 7.8 & 93.75 & 6.25 & 81.25 & 12.50  & 6.25 & 0.00 & 0.00 \\
    $m^*+1 \leq i_{\rm{auto}} < m^*+2$ & 43 & 4.0 & 81.40 & 18.60 & 76.75 & 18.60 & 4.65 & 4.65 & 4.65\\
    $i_{\rm{auto}} \geq m^* + 2$ & 14 & 2.3 & 50.00 & 50.00 & - & - & - & - & -\\
    \hline
    \end{tabular}
        \begin{tablenotes}
      \small
      \item {\bf Notes.} SSB are a subpopulation of the EL galaxies, whereas PSB are a subpopulation of NEL galaxies. The red and blue populations add up to 100\% for the photometric classification, while the NEL, EL and Low S/N populations add up to 100\% in the spectroscopic classification. We do not use spectral classification for galaxies with $i_{\rm{auto}} \geq m^* + 2$.  
    \end{tablenotes}
    \end{threeparttable}
\end{table*}

\subsubsection{Spectral classification}
\label{sec:spectral_class}

To understand if the merger is playing a role in the star formation activity of the galaxies, we make use of two measurements; the equivalent widths (EW) of the [OII] $\lambda$3727 \AA\ and H$\delta$ lines. [OII] $\lambda$3727 \AA\ traces recent star formation activity in timescales $\leq$10 Myr, while the Balmer line H$\delta$ has a scale between 50 Myr and 1 Gyr \citep{Paulino2019}. A strong H$\delta$ absorption line is interpreted as evidence of an explosive episode of star formation which ended between 0.5-1.5 Gyrs ago \citep{Dressler1983}. To measure the equivalent widths of [OII] $\lambda$3727 \AA, EW(OII), and H$\delta$, EW(H$\delta$), the flux spectra for each object is integrated following the ranges described by \citet{balogh99} using the \textsc{IRAF} task \textsc{sbands}. Also, we only make use of MUSE galaxies, excluding the 8 GMOS galaxies added, given that the MUSE selection is unbiased. We do not expect this to change our main results since these galaxies are not located along the merger axis. 

We use the same scheme defined by \citet{balogh99} to classify our galaxies into different categories; passive, star forming (SF), short-starburst (SSB), post-starburst \citep[PSB, K+A in][]{balogh99} and A+em (which could be dusty star-forming galaxies). For this classification we only take into account galaxies with $i_{\rm auto} < m^* +2$, meaning over 80\% completeness (Appendix \ref{sec:completeness}), and a signal-to-noise ratio, S/N $>3$ (62 galaxies), given that galaxies with low S/N can affect the measurements of lines in crowded sections, like in the region of the [OII] $\lambda$3727 \AA\ line \citep{Paccagnella2019}. The median signal-to-noise ratio (S/N) of our MUSE galaxies is shown in Table~\ref{tab:mag_properties} for different magnitude ranges. We estimate the S/N in the entire spectral range of our data by using the \textsc{der\_S/N} algorithm \citep{Stoehr2007}.

For simplicity, we use the following notation (and their combinations) to refer to the different galaxy populations throughout the text; EL for emission-line galaxies (EW(OII) $\geq 5$ \AA), including SSB, star-forming (SF) and A+em, and NEL for non emission-line galaxies (passive and PSB). We also use the RCS selection from $\S$~\ref{sec:photometric} to separate red and blue galaxies. We also analyze in particular SSB and PSB galaxies. Table~\ref{tab:gal_pop_clas} summarizes the different criteria of each population.

\begin{figure}
    \centering
    \includegraphics[width=\linewidth]{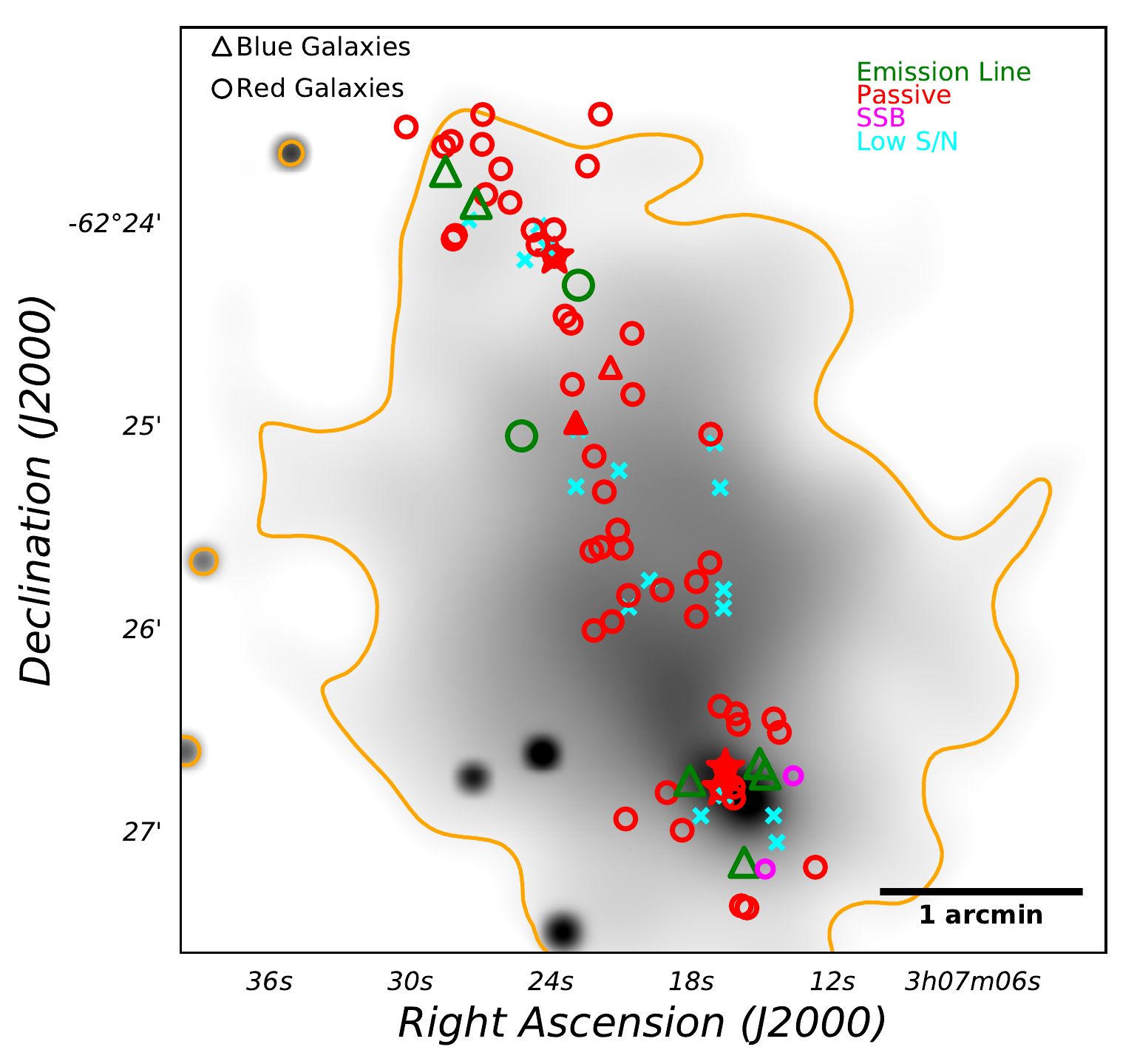}
    \vskip-0.13in
    \caption{ Sky positions of the MUSE cluster galaxies on top of the X--ray map, with the orange contour showing the outermost contour in Fig.~\ref{fig:rgb_image}. Circles are red galaxies and triangles are blue galaxies, color coded by their spectral type, with cyan crosses being galaxies with S/N < 3 or $i_{\rm{auto}} \geq m^*+2$ and the PSB shown as the red filled triangle.}
    \label{fig:sky_pos_xray}
\end{figure}

Table~\ref{tab:mag_properties} shows the fraction of galaxies for different magnitude ranges. The fractions are divided by the photometric classification (red or blue) and the spectroscopic classification (EL, NEL, SSB, PSB or Low S/N). Fig.~\ref{fig:sky_pos_xray} shows the sky positions of the galaxy population on top of the X--ray emission map. The results of this classification will be further discussed in $\S$\ref{sec:galpopulation}.

\subsection{Galaxies association}
\label{sec:subcluster}

One of the most common techniques to estimate the level of substructure in galaxy clusters is to analyze the galaxy velocity distribution on a 1D space, where it is assumed that for a relaxed cluster it should be close to a Gaussian shape \citep[e.g.][]{Menci1996, ribeiro13}. \citet{Hou2009} used Monte Carlo simulations to show that the Anderson-Darling (AD) test is among the most powerful to classify Gaussian (G) and non-Gaussian (NG) clusters.

\cite{Hou2009} use the $\alpha$ value (the significance value of the statistic) to assign the dynamical state of clusters (see Eq. 17 in their paper), where $\alpha<0.05$ indicates a NG distribution. \citet{Nurgaliev2017} uses the p-value of the statistic (p$_{\rm AD}$) and separates the clusters using p$_{\rm AD} < 0.05/n$ for NG clusters, where $n$ indicates the number of tests being conducted. We divide our data in 4 subsets for the application of the AD test; Cubes 2 and 3 for the middle overdensity, Cubes 1 and 4 to compare the two most overdense regions, all the data cubes and all the data cubes plus GMOS data. 

To test for 3D substructures (using the velocities and the on-sky positions), we use the Dressler-Shectman test \citep[DS-test,][]{dressler88}, which uses the information of the on-sky coordinates along with the velocity information, and can be used to trace perturbed structures \citep[e.g.][]{pranger14, Olave2018}. The DS-test uses the velocity information of the closest (projected) neighbors of each galaxy to estimate a $\Delta$ statistic, which is given by 
\begin{equation}
\Delta = \sum^{N_{\rm tot}}_i \delta_i,
\end{equation}
\noindent
where $N_{\rm tot}$ corresponds to the total number of members of the cluster and 
\begin{equation}
\delta^2 = \frac{N+1}{\sigma^2_{\rm cl}} \left[ (\bar{v}_{\rm loc}-\bar{v}_{\rm cl})^2 + (\sigma_{\rm loc} - \sigma_{\rm cl})^2  \right],
\end{equation}
\noindent
where $\delta$ is estimated for each galaxy. $N$ corresponds to the number of neighbors of the galaxy to use to estimate the statistic, estimated as $N = \sqrt[]{N_{\rm tot}}$ \citep{Pinkney1996}, $\sigma_{\rm cl}$ and $\sigma_{\rm loc}$ correspond to the velocity dispersion of the whole cluster and the velocity dispersion of the $N$ neighbors, respectively, and $\bar{v}_{\rm cl}$ and $\bar{v}_{\rm loc}$ correspond to the mean peculiar velocity of the cluster and the mean peculiar velocity of the $N$ neighbors, respectively. A value of $\Delta/N_{\rm tot} \leq 1$ implies that there are no substructures on the cluster. 

To calibrate our DS-test results, we perform $10^4$ Monte Carlo simulations by shuffling the velocities, i.e., randomly interchanging the velocities among the galaxies, while maintaining their sky coordinates (meaning that the neighbors are always the same). The p-value of the statistic (p$_{\Delta}$) is estimated by counting how many times the simulated $\Delta$ is higher than that of the original sample, and divide the result by the total number of simulations. Choosing p$_{\Delta} <  0.05$ ensures a low probability of false identification \citep{Hou2012} and is accepted for the distribution to be considered non-random. Both AD and DS test results are shown in Table~\ref{tab:tests}.

To test for 2D substructures (sky positions) we build surface density maps \citep[see, e.g., ][]{White2015, Monteiro2017, Monteiro2018, Monteiro2020, Yoon2019}. The galaxy surface density map at the top right of Fig.~\ref{fig:rgb_image} implies that there are at least two colliding-structures. To obtain the density map we use the RCS galaxy catalog and the \textsc{sklearn.neighbors.KernelDensity} python module, applying a gaussian kernel with a bandwidth of 50 kpc.

\section{Results}
\label{sec:results}

\subsection{Cluster substructures}
\label{sec:dynamics}

Table \ref{tab:tests} shows the results of both the AD-test and the DS-test applied to different subsets. The second column corresponds to the number of spectroscopic galaxies belonging to a given subsample. The subset which gives the smallest p-values for both the AD-test and the DS-test is the Cubes 1+4 subset, with these cubes located on top of the two density peaks, enclosing also the area next to the two brightest galaxies (see Fig.~\ref{fig:rgb_image}). We find that both the AD-test and the DS-test provide no evidence of substructure. Applying a 3$\sigma$-clipping iteration to the samples does not change the results. The results, along the X--ray morphology, show no evidence of substructure along the line of sight, and rather support a merger in the plane of the sky, thus we take a look into the spatial distribution of the galaxies.

\begin{table}\caption{Results for the substructure-identification tests applied to different subsamples.}\label{tab:tests}
\centering
\begin{tabular}{l r r r r r}
\hline
\hline
Subsample&N&\multicolumn{2}{c}{AD-test}&\multicolumn{2}{c}{DS-test}\\
&&$\alpha$&P-value&$\Delta/N_{\rm{tot}}$&P-value\\
\hline
Cubes 2+3 & 32 & 0.264 & 0.674 & 0.967 & 0.421 \\
Cubes 1+4 & 48 & 0.383 & 0.383 & 1.329 & 0.097 \\
All Cubes & \NgalallCube\ & 0.234 & 0.789 & 1.205 & 0.138 \\
MUSE+GMOS & \NgalallMUSEGMOS\ & 0.272 & 0.662 & 1.203 & 0.152 \\
\hline
\end{tabular}
\end{table}

\begin{figure}
    \centering
    \includegraphics[width=\linewidth]{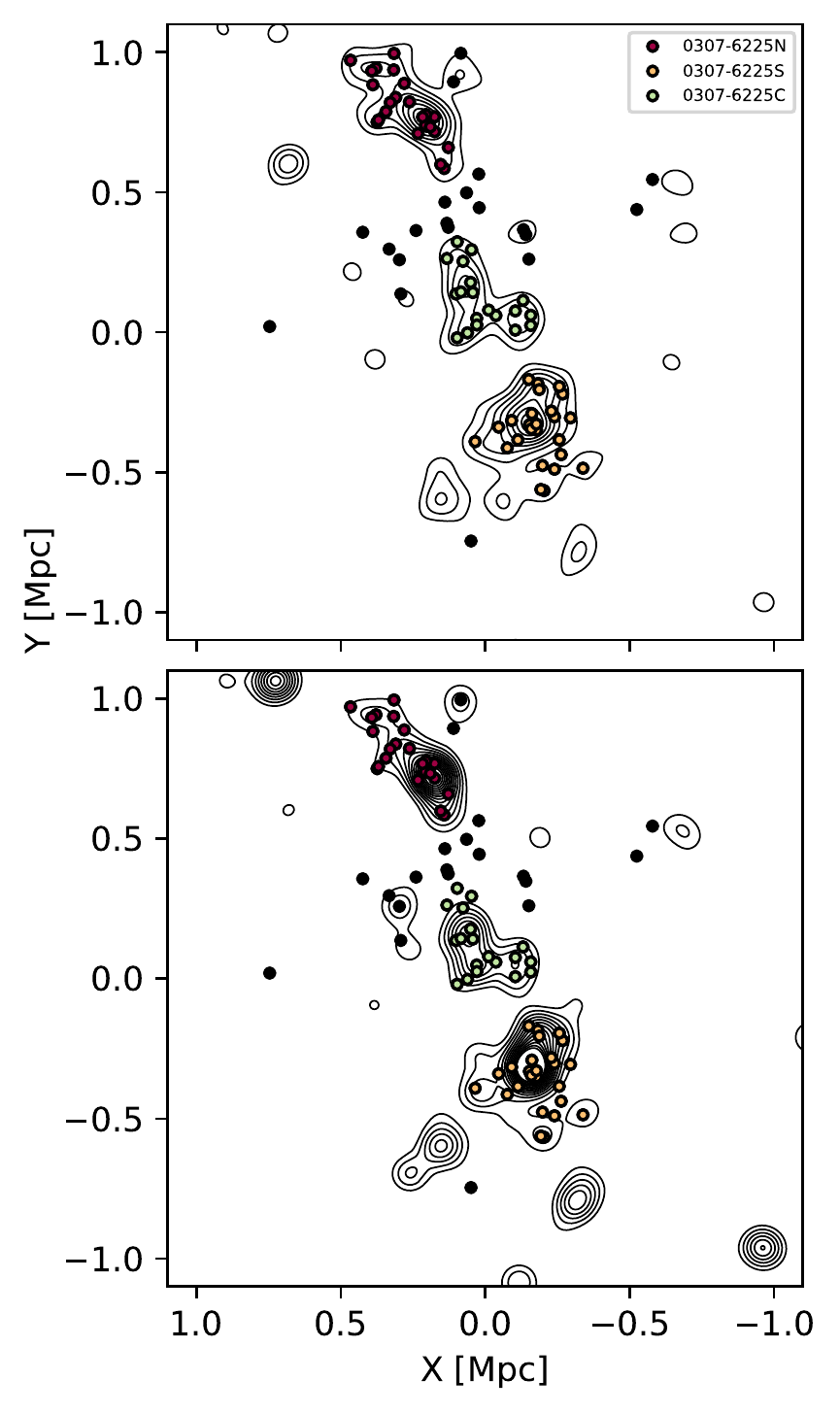}
    \vskip-0.13in
    \caption{Unweighted (top) and flux weighted (bottom) RCS galaxies (photometric and spectroscopic)  numerical density map is shown in black contours, where levels begin at 100 galaxies per Mpc$^{2}$ and the flux was estimated from the $i$ band. Galaxies not close to the density levels or classified as not being part of any structure by the \texttt{DBSCAN} algorithm are shown as black dots, while dots in different substructures according to the algorithm are shown with different colors according to the substructure; 0307-6225N (red), 0307-6225S (orange) and a in-between overdensity (green).}
    \label{fig:mpcdensitymap}
\end{figure}

Fig.~\ref{fig:mpcdensitymap} shows the contours of the unweighted and flux weighted density maps, top and bottom figures respectively, of the RCS galaxies. The contour levels begin at 100 gal Mpc$^{-2}$ and increase in intervals of 50 gal Mpc$^{-2}$. Dots correspond to galaxies from our spectroscopic samples. These figures, regardless of whether they are weighted or unweighted, show the core of the two main structures with corresponding BCGs, and a high density of galaxies in-between them. 

For the definition of the substructures we take into account only spectroscopic members within (or near) the limits of our density contours. To distinguish the galaxies with a higher probability of being part of each structure we use the Density-Based Spatial Clustering of Applications with Noise \citep[\texttt{DBSCAN},][]{Ester1996} algorithm. The advantage of using this algorithm is that the galaxies are not necessarily assigned to a given group, leaving some of them out. We use a \textsc{python}-based application of this algorithm, following the work of \citet[][substructure defined as at least three neighbouring galaxies within a separation of $\sim$140 kpc]{Olave2018}. 

Fig.~\ref{fig:mpcdensitymap} shows the results of the different found structures. Black dots represent galaxies that either were too far from our density contours or were discarded by the \texttt{DBSCAN} algorithm. We name the two most prominent structures, defined by DBSCAN, as 0307-6225N (red dots) and 0307-6225S (orange dots), comprised by 23 members and 25 members, respectively. The BCGs for 0307-6225S and 0307-6225N are marked in Table~\ref{tab:all_objs_properties} by the upper scripts $S_1$ and $N$, respectively. Both structures show a Gaussian velocity distribution when applying the AD test, and the distance between them is: $\sim$1.10 Mpc between their BCGs and $\sim$1.15 Mpc between the peaks of the density distribution. 

We also find a third substructure in-between the two colliding ones (green dots), which we name 0307-6225C, with 19 galaxies and no BCG-like galaxy. Fig.~\ref{fig:subs_veldist} shows the velocity distribution of the galaxies of each substructure, color coded  following Fig.~\ref{fig:mpcdensitymap}.
Table~\ref{tab:substructures} shows the sky coordinates of the substructures (estimated as the peak of the overdensity), along with their estimated redshifts, velocity dispersions and number of members.

\begin{figure}
    \centering
    \includegraphics[width=\linewidth]{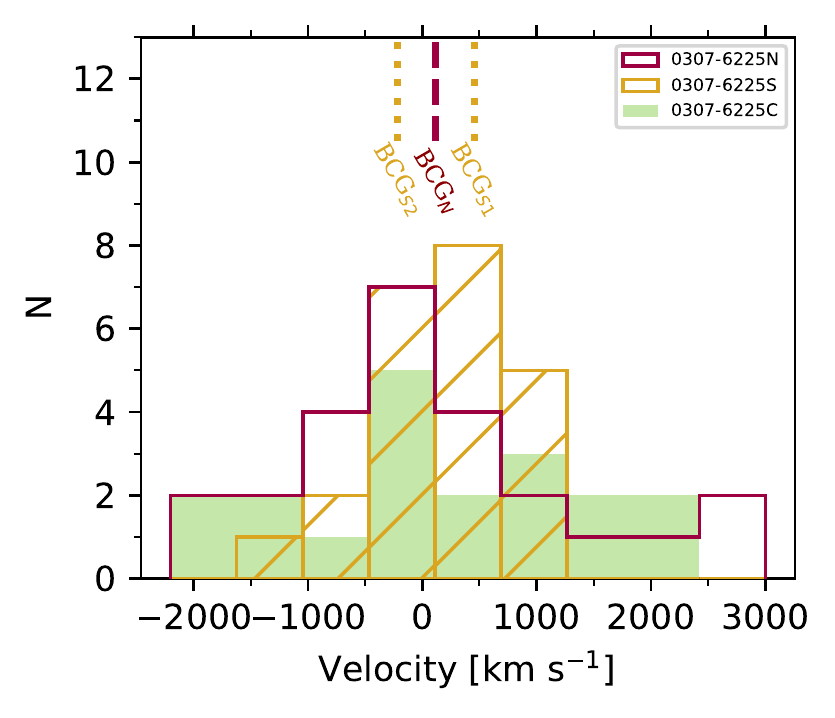}
    \vskip-0.13in
    \caption{Peculiar velocity distribution of the galaxies belonging to the three substructures; 0307-6225N, 0307-6225S and 0307-6225C. The velocity of the BCG of 0307-6225N is shown with a dashed line in the top axis, while the velocity of the two BCGs of 0307-6225S shown with dotted lines.}
    \label{fig:subs_veldist}
\end{figure}

\subsection{Cluster dynamical mass}\label{subsec:dmass}

We estimate the masses using \citet{Munari2013} scaling relations between the mass and the velocity dispersion of the cluster (see Eq.~ \ref{eq:mass}). The Gaussian velocity distribution together with the large separation between the center of both structures ($\sim$1.1 Mpc between the BCGs) and the fact that the velocity difference between them is $\Delta v_{N-S} = 342$ km s$^{-1}$ (at the cluster's frame of reference) strongly suggest a plane of the sky merger \citep[see, e.g.][]{Dawson2015, Mahler2020} and could therefore, imply that the overestimation of the masses using scaling relations is minimal \citep{Dawson2015}. We further explore this in $\S$\ref{sec:mass_overstimation_merging}. In order to minimize the possible overstimation of using scaling relations, we only use RCS spectroscopic galaxies to estimate $\sigma_v$, since in clusters with a high accretion rate, blue galaxies tend to raise the value of the velocity dispersion \citep{Zhang2012}. Note that, however, the number of members shown in Table~\ref{tab:substructures} also considers blue galaxies.

In Table~\ref{tab:substructures} we show the estimated masses of the substructures. The two prominent substructures, 0307-6225S and 0307-6225N, have similar masses with the most probable ratio of $M_{\rm S}/M_{\rm N}\approx$ \massSNnE{} with large uncertainties. Galaxies selected for the dynamical mass estimation are likely to belong to the core regions of the two clusters. Galaxies in these regions are expected to be virialized and should more closely follow the gravitational potential of the clusters during a collision, giving a better estimation of the masses when using the velocity dispersion.

\begin{table}
\caption{Substructure properties}
\label{tab:substructures}
\centering
\resizebox{\linewidth}{!}{%
\begin{tabular}{c c c c r c c}
\hline
\hline
Structure & R.A. & Dec. & $z$ & \multicolumn{1}{c}{$\sigma_{v}$} & M$_{200,\rm{dyn}}$ & N \\
0307-6225 & (J2000) & (J2000) & & \multicolumn{1}{c}{km s$^{-1}$} & $\times$10$^{14}$ M$_\odot$ & \\
\hline
S & 46.8195 & -62.4463 & 0.5792 $\pm$ 0.0002 & 756 $\pm$ 164 & 3.16 $\pm$ 1.88 & 25\\
N & 46.8526 & -62.4009 & 0.5810 $\pm$ 0.0002 & 688 $\pm$ 145 & 2.44 $\pm$ 1.41 & 23\\
C & 46.8396 & -62.4258 & 0.5803 $\pm$ 0.0004 & 1415 $\pm$ 336 & 17.67 $\pm$ 11.53 & 19 \\
\hline
\end{tabular}
}
\end{table}

\subsection{Cluster merger orbit}\label{sec:history}

With the masses estimated, the merging history can be recovered by using a two-body model \citep{Beers1990,Cortese2004,Gonzalez2018} or by using hydrodynamical simulations constrained with the observed properties of the merging system \citep[e.g.][]{Mastropietro2008, Machado+2015, Doubrawa2020,Moura21}, with the disadvantage being that the latter method is computationally expensive. To understand the merging event, we use the Monte Carlo Merger Analysis Code \citep[\texttt{MCMAC},][]{Dawson2013}, which is a good compromise between computational time and accuracy of the results, with a dynamical parameter estimation accuracy of about 10\% for two dissociative mergers; Bullet Cluster and Musket Ball Clusters. \texttt{MCMAC} analyzes the dynamics of the merger and outputs its kinematic parameters. The model assumes a two-body collision of two spherically symmetric halos with a NFW profile \citep[][]{NFW0, NFW1}, where the total energy is conserved and the impact parameters is assumed to be zero. The different parameters are estimated from the Monte Carlo analysis by randomly drawing from the probability density functions of the inputs.

The inputs required for each substructure are the redshift and the mass, with their respective errors, along with the distance between the structures with the errors on their positions. We use the values shown in Table \ref{tab:substructures} as our inputs, where the errors for the redshifts are estimated as the standard error, while the errors for the distance are given as the distances between the BCGs and the peak of the density distribution of each structure (0.144 arcmin and 0.017 arcmin for 0307-6225N and 0307-6225S, respectively). The results are obtained by sampling the possible results through 10$^5$ iterations, and are showed and described in Table \ref{tab:mcmac}, with the errors corresponding to the 1$\sigma$ level. 

\begin{table}
    \caption{Output from the \texttt{MCMAC} code, with the priors from Table \ref{tab:substructures}. Errors correspond to the $1\sigma$ level.}
    \label{tab:mcmac}
    \centering
    \begin{tabular}{l r c l}
        \hline
        \hline
        Param. & Median & Unit & Description\\
        \hline
        \vspace{3.5pt}
        $\alpha$ & 39$^{+13}_{-11}$ & deg & Merger axis angle\\
        \vspace{3.5pt}
        $d3D_{\rm obs}$ & 1.29$^{+0.32}_{-0.15}$ & Mpc & \footnotesize{3D distance of the halos at T$_{\rm{obs}}$.}\\
        \vspace{3.5pt}
        $d3D_{\rm{max}}$ & 1.72$^{+0.44}_{-0.22}$ & Mpc & \footnotesize{3D distance of the halos at apoapsis.}\\
        \vspace{3.5pt}
        $v3D_{\rm{col}}$ & 2300$^{+122}_{-96}$ & km/s & \footnotesize{3D velocity at collision time.}\\
        \vspace{3.5pt}
        $v3D_{\rm{obs}}$ & 547$^{+185}_{-103}$ & km/s & \footnotesize{3D velocity at T$_{\rm{obs}}$.}\\
        \vspace{3.5pt}
        $v_{\rm{rad}}$ & 339$^{+28}_{-28}$ & km/s & \footnotesize{Radial velocity of the halos at T$_{\rm{obs}}$.}\\
        \vspace{3.5pt}
        TSP0 & 0.96$^{+0.31}_{-0.18}$ & Gyr & \footnotesize{TSP for outgoing system.}\\
        \vspace{3.5pt}
        TSP1 & 2.60$^{+1.07}_{-0.53}$ & Gyr & \footnotesize{TSP for incoming system.}\\
        \hline
    \end{tabular}
\end{table}

\texttt{MCMAC} gives as outputs the merger axis angle $\alpha$, the estimated distances and velocities at different times and two possible current stages of the merger; outgoing after first pericentric passage and incoming after reaching apoapsis. The time since pericentric passage (TSP) for both possible scenarios are described as TSP0 for the outgoing scenario and TSP1 for the incoming one. This last two estimates are the ones that we will further discuss when recovering the merger orbit of the system.

To further constrain the stage of the merger we compare the observational features with simulations.  We use the Galaxy Cluster Merger Catalog \citep{ZuHone2018}\footnote{\url{http://gcmc.hub.yt/simulations.html}}, in particular, the ``A Parameter Space Exploration of Galaxy Cluster Mergers'' simulation \citep{ZuHone2011}, which consists of an adaptive mesh refinement grid-based hydrodynamical simulation of a binary collision between two galaxy clusters, with a box size of 14.26 Mpc. The binary merger initial configuration separates the two clusters by a distance on the order of the sum of their virial radii, with their gas profiles in hydrostatic equilibrium. With this simulation one can explore the properties of a collision of clusters with a mass ratio of 1:1, 1:3 and 1:10, where the mass of the primary cluster is $M_{200}=6\times10^{14}$ M$_\odot$, similar to the SZ derived mass of $M_{200}=7.63  \times h^{-1}_{70} 10^{14}$ M$_\odot$ for SPT-CL J0307-6225 \citep{bleem15b}, and with different impact parameters ($b = 0, 500, 1000$ kpc).

We use both a merger mass ratio of 1:3 and 1:1. Since we cannot constrain the impact parameter, we use all of them and study their differences, where, for example, the bigger the impact parameter, the longer it takes for the merging clusters to reach the apoapsis. We also note that for our analysis we use a projection on the $z$-axis, since evidence suggests a collision taking place on the plane of the sky. 

\subsubsection{Determining TSP0 and TSP1 from the simulations}

We use the dark matter distribution of both objects to determine the collision time, focusing on the distance between their density cusps at different snapshots. Also, to determine the snapshots for an outgoing and an incoming scenario, which would be the closest to what we see in our system, we look for the snapshot where the separation between the peaks is similar to the projected distance between our BCGs ($\sim$1.10 Mpc).

\begin{table}
\caption{Estimated collision times and times since collision (TSP0$_{\rm sim}$ and TSP1$_{\rm sim}$) for the simulations with different impact parameters $b$ and mass ratios.}
    \label{tab:simulation_prop}
    \centering
    \begin{threeparttable}
    \begin{tabular}{r c c c c}
         \hline 
         \hline
          $b$ & Mass ratio & Collision time & TSP0$_{\rm sim}$ & TSP1$_{\rm sim}$  \\
          kpc & & Gyr & Gyr & Gyr \\
         \hline
         0 & 1:3 & 1.22 $\pm$ 0.02 & 0.78 $\pm$ 0.20 & -\\
         500 & 1:3 & 1.24 $\pm$ 0.02 & 0.66 $\pm$ 0.20 & 0.96 $\pm$ 0.20 \\
         1000 & 1:3 & 1.34 $\pm$ 0.02 & 0.56 $\pm$ 0.20 & 1.46 $\pm$ 0.20 \\
         0 & 1:1 & 1.32 $\pm$ 0.02 & 0.68 $\pm$ 0.20 & -\\
         500 & 1:1 & 1.34 $\pm$ 0.02 & 0.46 $\pm$ 0.20 & - \\
         1000 & 1:1 & 1.40 $\pm$ 0.02 & 0.80 $\pm$ 0.20 & 1.00 $\pm$ 0.20 \\

        \hline
    \end{tabular}
    \begin{tablenotes}
      \small
      \item {\bf Notes.} No TSP1 value is provided when we cannot separate between the outgoing and incoming scenarios by requiring a distance of  $\sim$1.1 Mpc.
    \end{tablenotes}
    \end{threeparttable}
\end{table}

In Table \ref{tab:simulation_prop} we show the results for the different impact parameters, where the second column indicates the mass ratio. The third column shows the simulation time where the distance between the two halos is minimal (pericentric passage time). The errors are the temporal resolution of the simulation at the chosen snapshot. Following the previous nomenclature, the fourth column, TSP0$_{\rm sim}$, corresponds to the amount of time from the first pericentric passage (minimum approach), while the fifth column, TSP1$_{\rm sim}$, corresponds to the amount of time from the pericentric passage, to the first turn around, and heading towards the second passage. Times are either the snapshot time or an average between two snapshots if the estimated separations are nearly equally close to the $\sim$1.10 Mpc distance. 

For $b=0$ kpc, the maximum achieved distance between the two dark matter halos in the 1:3 mass ratio simulation was 1.05 Mpc, while for the 1:1 mass ratio it was 0.99 Mpc, meaning that we cannot separate between both scenarios when comparing the projected distance of 0307-6225N and 0307-6225S.

\begin{figure}
    \centering
    \includegraphics[width=\linewidth]{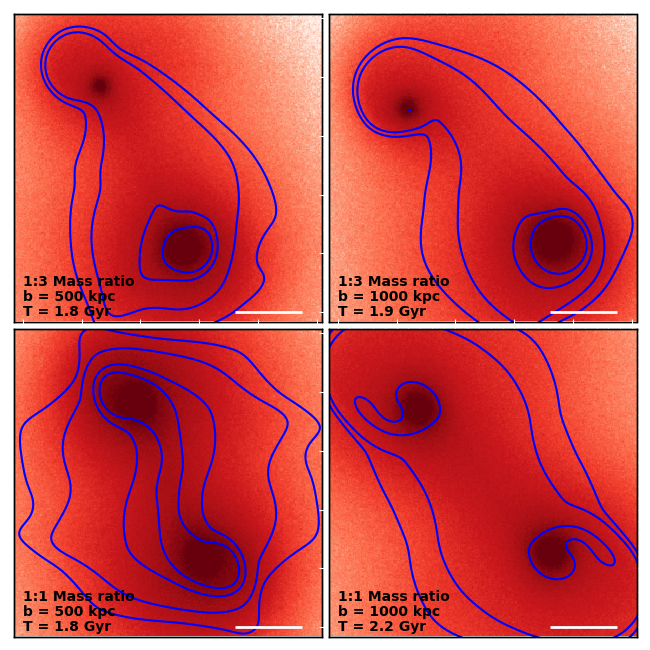}
    \vskip-0.13in
    \caption{Density and X--ray contours of the different simulations. The simulation times are shown on the bottom left corner, and correspond to (or are close to in case of averaging over two snapshots) the collision time plus the TSP0 time since collision (see Table \ref{tab:simulation_prop}). The projected total density of the simulations is shown in red in the background, with the contrast starting at $1\times10^{7}$ M$_\odot$ kpc$^{-2}$. Blue contours where derived from the projected X--ray emission, with the levels being $0.5, 1, 5, 10, 15\times10^{-8}$ photons/s/cm$^{2}$/arcsec$^{2}$. Simulations are divided according to their mass ratio (1:3 on top and 1:1 on the bottom) and according to the impact parameter (500 kpc on the left panels and 1000 kpc on the right panels). The used box size is the same to the one used in Fig.~\ref{fig:rgb_image}. The white bar also corresponds to the same length of 1 arcmin shown in Fig.~\ref{fig:rgb_image}.
    }
    \label{fig:sim_TSP0_xray}
\end{figure}

\begin{figure}
    \centering
    \includegraphics[width=\linewidth]{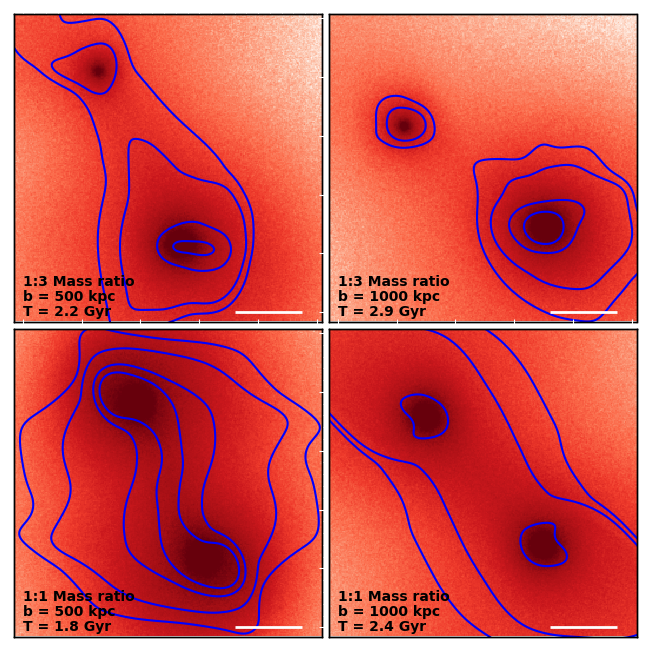}
    \vskip-0.13in
    \caption{Same as Fig.~\ref{fig:sim_TSP0_xray}, but derived from the simulations at the TSP1 times.}
    \label{fig:sim_TSP1_xray}
\end{figure}

\subsubsection{X--ray morphology}
The hydrodynamical simulations render a gas distribution that can be directly compared to the observations. Fig.~\ref{fig:sim_TSP0_xray} shows the snapshots of the outgoing scenario, while Fig.~\ref{fig:sim_TSP1_xray} shows the snapshots of the incoming scenario, where the X--ray projected emission is overplotted as blue contours on top of the projected total density, for the simulation snapshots close to the derived TSP (Table \ref{tab:simulation_prop}), with the simulation time shown on the bottom left of each panel. Note however that for the 1:1 mass ratio and $b=500$ kpc, the system has the $\sim$1.1 Mpc distance at turnaround, which means that we cannot differentiate between and outgoing and incoming scenario. We decide to keep the same snapshot in both Figures \ref{fig:sim_TSP0_xray} and \ref{fig:sim_TSP1_xray} just for comparison. The scenarios for 1:3 mass ratio closest resemble the gas distribution from our {\it Chandra} observations (orange contours on Fig.~\ref{fig:rgb_image}). We comeback to this in $\S$~\ref{sec:constraining_tsc_with_sims}.

\subsection{The impact of the merging event in the galaxy populations}
\label{sec:galpopulation}

\begin{figure}
    \centering
    \includegraphics[width=\linewidth]{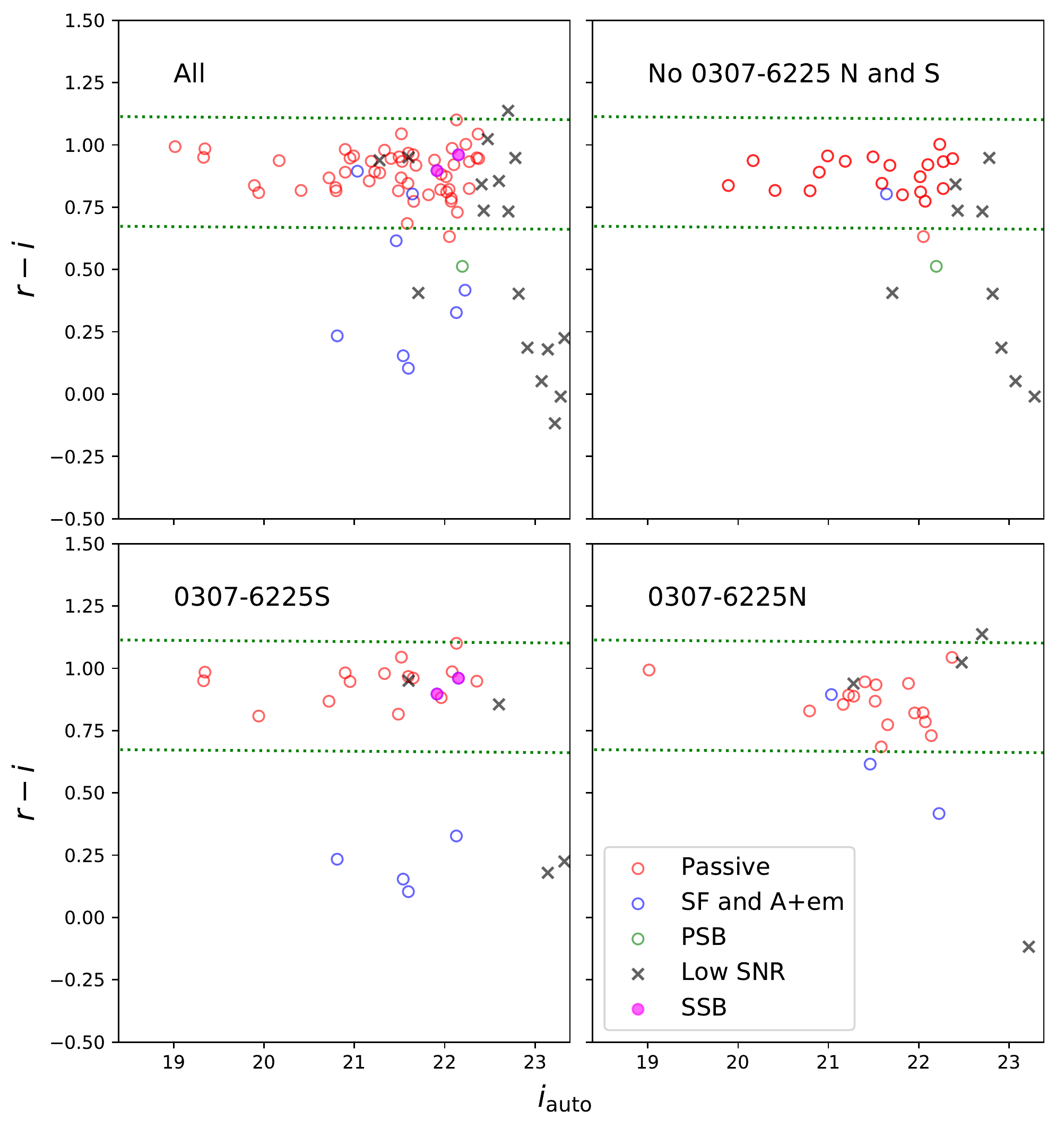}
    \vskip-0.13in
    \caption{CMD of the cluster for the different samples. Galaxies are color-coded depending on their spectral classification described in ~\S\ref{sec:galpopulation}. \textit{top left}: entire spectroscopic data sample. \textit{top right}: sample comprising galaxies not belonging to 0307-6225N and 0307-6225S, i.e., galaxies from 0307-6225C plus galaxies not belonging to any substructure according to \textsc{DBSCAN}. \textit{bottom}: 0307-6225S and 0307-6225N samples shown in left and right panels, respectively. The green dotted lines are the limits for the RCS zone. Black crosses are galaxies with S/N < 3 or $i_{\rm auto} \geq m^* + 2$. Filled colors are galaxies classified as SSB.}
    \label{fig:gal_properties}
\end{figure}

In Fig.~\ref{fig:gal_properties} we show the CMD for each subsample; all galaxies, galaxies belonging to 0307-6225N and 0307-6225S, and galaxies not belonging to either of them. Galaxies are color coded according to their spectral classification. Most of the star-forming galaxies are located within the two main structures (9 out of 10 SF+SSB galaxies), with some of them being classified as RCS galaxies (4; 2 SF and 2 SSB). Galaxies with S/N < 3 and/or $i_{\rm auto}$ > $m^* + 2$ are plotted as black crosses. 

\begin{figure}
    \centering
    \includegraphics[width=\linewidth]{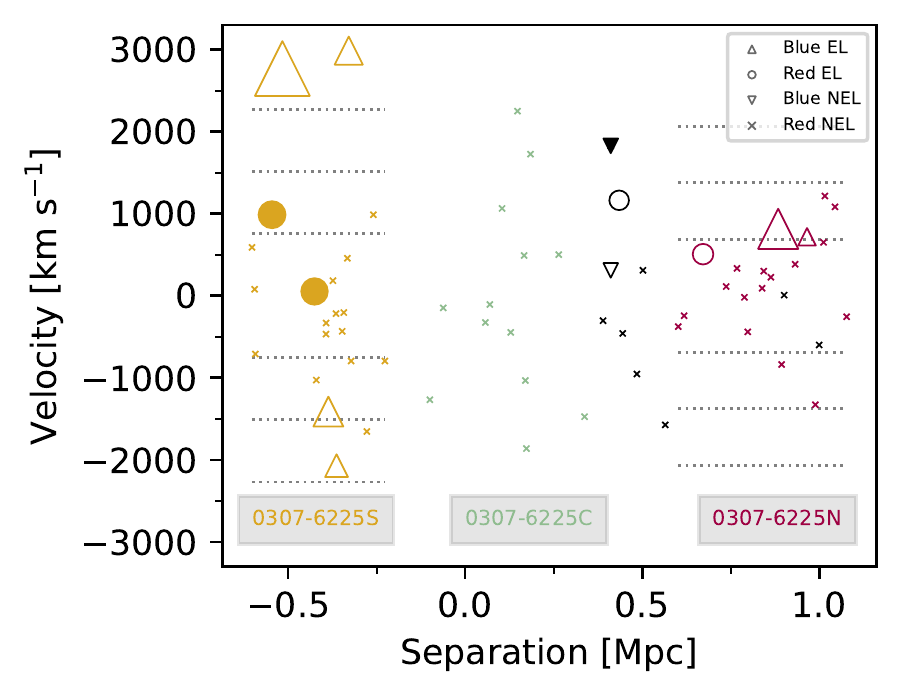}
    \vskip-0.13in
    \caption{Phase-space diagram of spectroscopic members with S/N $\geq3$ and $i_{\rm auto} < m^* + 2$. The separation is measured with respect to the SZ-center, negative for objects to the south of it. Galaxies are colored as dark red, dark orange, dark green and black if they were classified as belonging to the 0307-62255N, 0307-6225S, 0307-6225C or to neither of them, respectively. Crosses are galaxies classified as non-emission line galaxies. Emission line galaxies which belong to (or have redder colors than) the RCS are plotted as circles, triangles are galaxies with colors lower than the RCS, whereas inverted triangles are blue post-starburst (filled) or passive (unfilled) galaxies. The sizes of EL galaxies are correlated with their EW(OII) strength. Filled circles correspond to SSB galaxies. Black dotted lines mark $\pm$1$\sigma_v$, $\pm$2$\sigma_v$ and $\pm$3$\sigma_v$ for the two main substructures.}
    \label{fig:phase_spectro}
\end{figure}

Given that most of the SF galaxies seem to be located in the substructures, especially the red SF galaxies, it is plausible that they were part of the merging event, instead of being accreted after it. In Fig.~\ref{fig:phase_spectro} we show a phase-space diagram, with the X-axis being the separation from the SZ-center. Galaxies are color coded following the substructure to which they belong. In Fig.~\ref{fig:cutouts_spectra} we show small crops of 7$\times$7 arcsec$^2$ (47$\times$47 kpc$^2$ at the cluster's redshift) of the EL galaxies plus the two blue NEL galaxies, separating by different substructures and with the spectra of each galaxy shown to the right.

\subsection{The particular case of 0307-6225S}\label{sec:southernstr}

\begin{figure}
    \centering
    \includegraphics[width=\linewidth]{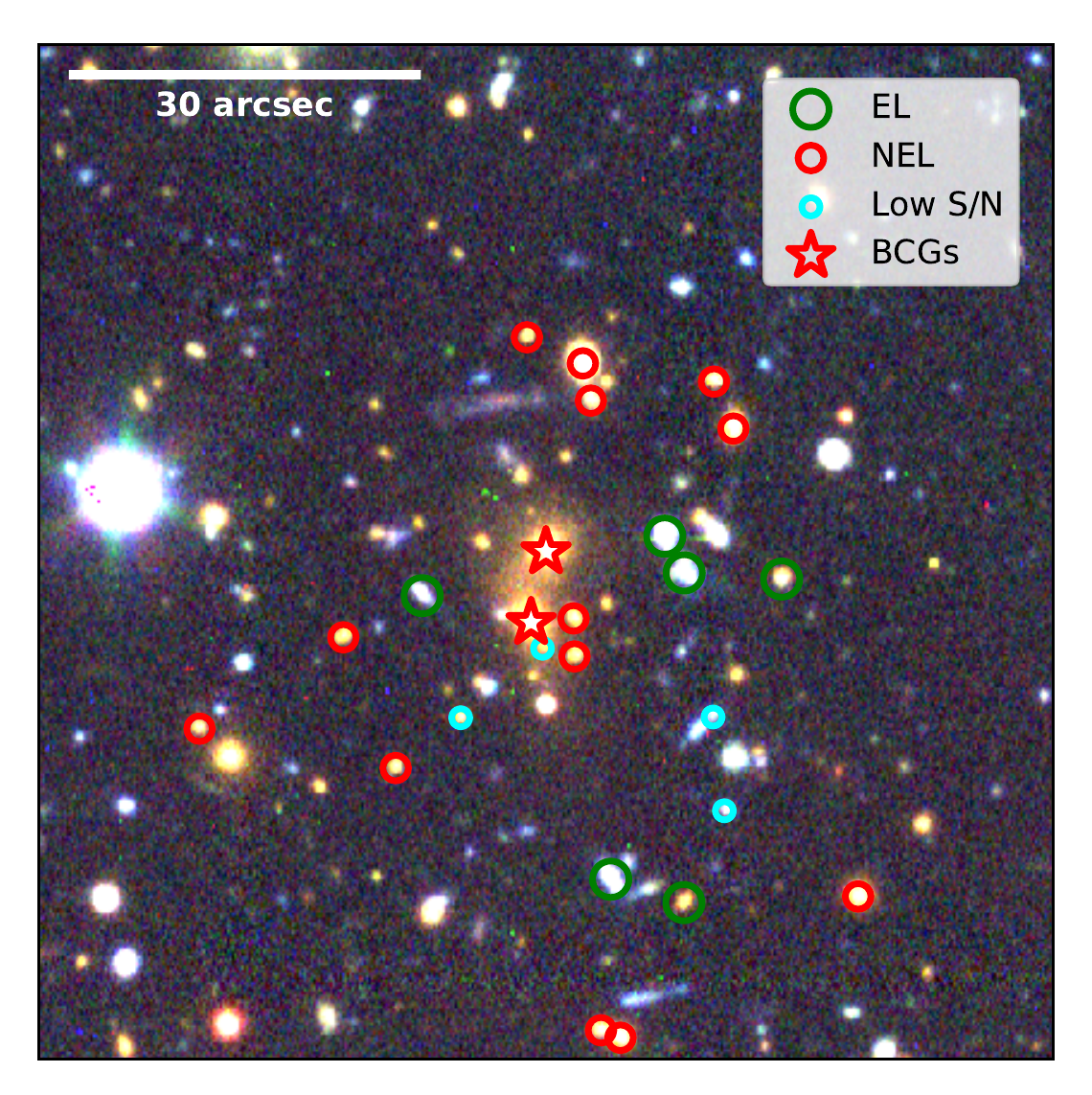}
    \vskip-0.13in
    \caption{Zoom from Fig.~\ref{fig:rgb_image} into 0307S, with the white bar on the top left showing the scale of the image. Spectroscopic members with S/N < 3 or $i_{\rm auto} \geq m^* + 2$ are shown as cyan circles, while red and green circles/stars represent passive and emission-line cluster galaxies, respectively, where emission-line refers SF or SSB galaxies. The 2 brightest galaxies are marked with stars. 
    }
    \label{fig:southernstr}
\end{figure}

Fig.~\ref{fig:gal_properties} shows that 0307-6225S has (1) the bluest members from our sample and (2) two very bright galaxies with nearly the same magnitudes (galaxies with ID 35 and 46 from the MUSE-1 field in Table~\ref{tab:all_objs_properties}, marked with an upper script $S_1$ and $S_2$, respectively). In Fig.~\ref{fig:southernstr} we provide a zoom from Fig.~\ref{fig:rgb_image},  to show in more detail the southern structure. Red circles mark spectroscopic members for this region with S/N $>3$ and $i_{\rm auto} < m^* + 2$. The two brightest galaxies are the two elliptical galaxies in the middle marked with red stars, with $\Delta m_i = 0.0152 \pm 0.0063$ and $\Delta  v = 600 km/s$.  The on-sky separation between the center of them ($\sim$41 kpc), suggests that these galaxies could be interacting with each other. 

\section{Discussion}
\label{sec:discussion}

\subsection{Merging history of 0307-6225S and 0307-6225N}\label{sec:disc_merger}

\subsubsection{Mass estimation of a merging cluster}\label{sec:mass_overstimation_merging}

Being able to recover the merging history of two observed galaxy clusters is not trivial. Most methods require a mass estimation of the colliding components, which is not always an easy task \citep[see merging effect on cluster mass in][]{takizawa10,nelson12,nelson14}. 

The velocity dispersion (along the line-of-sight) of the galaxies of a cluster can be used to infer its mass, using for example the virial theorem \citep[e.g.][]{Rines2013,White2015} or scaling relations \citep[e.g.][]{Evrard2008, Saro2013, Munari2013, Dawson2015, Monteiro-Oliveira21}. For the mass estimations of our structures we use the later one, although it is important to note that these measurements are also affected by the merging event, as colliding structures could show alterations in the velocities of their members. \citet{White2015} argues that the masses of merging systems estimated by using scaling relations can be overestimated by a factor of two. Evidence suggests that the merger between 0307-6225S and 0307-6225N is taking place close to the plane of the sky, with a low velocity difference between the two, similar to what \citet{Mahler2020} find for the dissociative merging galaxy cluster SPT-CLJ0356-5337. The velocity difference between the BCGs and the redshift of each substructure is $\leq$20 km s$^{-1}$ for both substructures, which might indicate that the two merging substructures were not too dynamically perturbed by the merger.

It is worth noting that recently \cite{Ferragamo2020} suggested correction factors on both $\sigma_v$ and the estimated mass to account for cases with a low number of galaxies. They also apply other correction factors to turn $\sigma_v$ into an unbiased estimator by taking into account, for example, interlopers and the radius in which the sources are enclosed. However, applying these changes does not change our results drastically, with the new derived masses being within the errors of the previously derived ones.

\begin{figure}
    \centering
    \includegraphics[width=\linewidth]{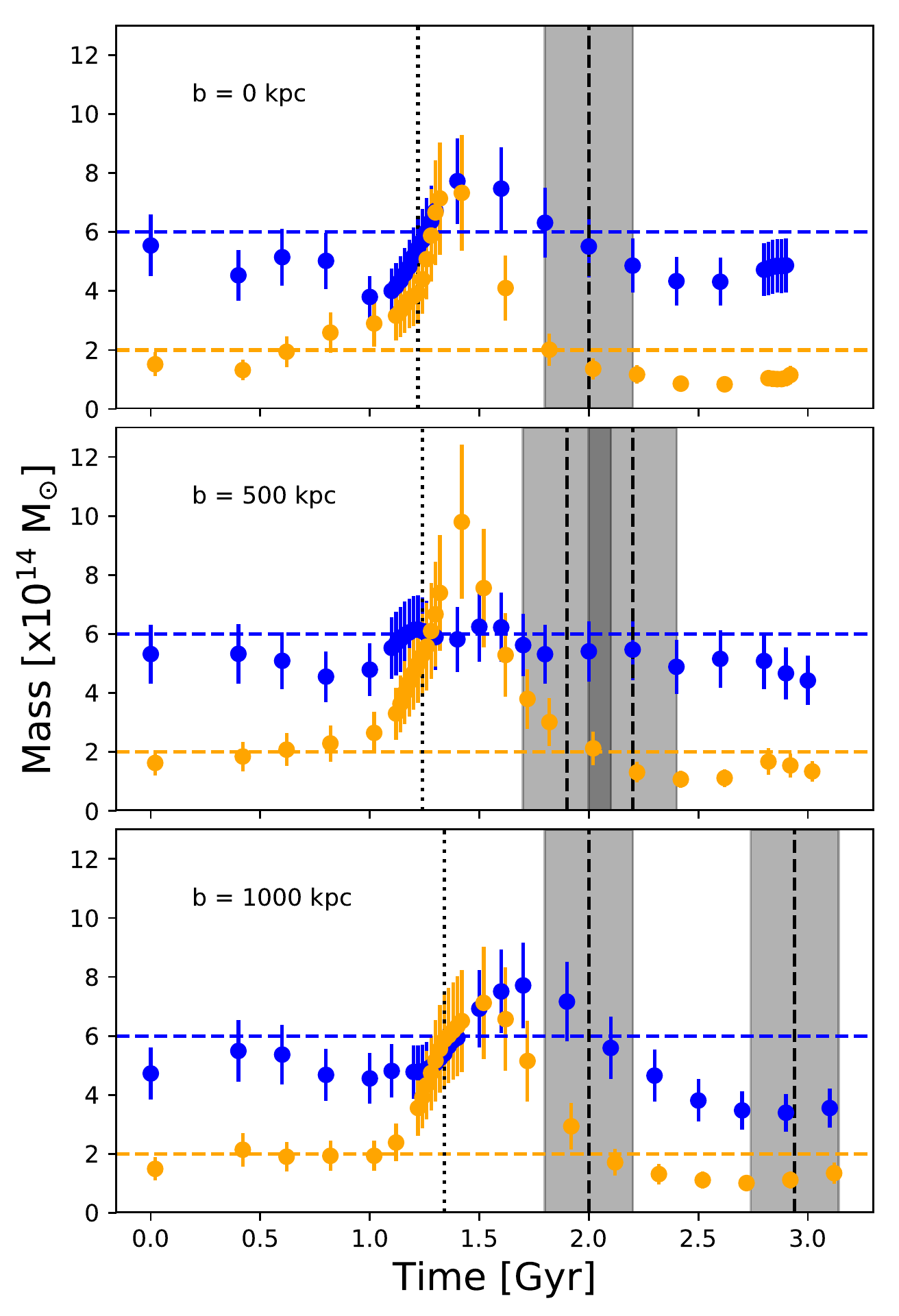}
    \vskip-0.13in
    \caption{Velocity dispersion derived masses for the 1:3 mass ratio simulations used in this work, with different $b$. The x-axis is the time since the simulation started running, with the blue and orange dots corresponding to the main cluster and the secondary cluster, respectively. The blue and orange dashed lines represent the masses of $6\times10^{14}$ and $2\times10^{14}$ M$_\odot$, respectively. Black dotted lines mark the collision times estimated following $\S$\ref{sec:history}. Vertical black dashed lines mark the estimated TSP0 and TSP1 shown in Table \ref{tab:simulation_prop}, with the gray area being the errors on this estimation.}
    \label{fig:masses_sims}
\end{figure}

To check how masses derived from the velocity dispersion of merging galaxy clusters could be overestimated, we estimate the masses, following the equations from \cite{Munari2013}, of the simulated clusters from the 1:3 merging simulation (from $\S$\ref{sec:history}) at all times (and $b$) using their velocity dispersion. It is worth noting that we cannot separate RCS members to estimate the velocity dispersions, since the simulation does not give information regarding the galaxy population. Fig.~\ref{fig:masses_sims} shows the $\sigma_v$ derived masses at different times for the 1:3 mass ratio simulation for different values of $b$. The black dotted lines represent the collision time and the dashed lines with the gray shaded areas represent the TSPs and their errors from Table \ref{tab:simulation_prop}, respectively. Before the collision and some Gyr after it, the  masses are overestimated, especially for the case of the smaller mass cluster. However, near the TSP0 times, the derived masses are in agreement, within the errors, with respect to the real masses. This is true also for the TSP1 with $b=500$ kpc, but for the same time with $b=1000$ kpc, the main cluster's mass is actually underestimated. Although we cannot further constrain the masses from the simulation using only RCS members, this information does suggest that our derived masses are not very affected by the merging itself given the possible times since collision.

\citet{bleem15b} estimated a total Sunyaev-Zeldovich based mass of M$_{500,\rm SZ}= 5.06\pm0.90\times 10^{14}$ $h_{70}^{-1}$ M$_{\odot}$, corresponding to M$_{200,\rm SZ}=  7.63\pm1.37\times 10^{14}$ $h_{70}^{-1}$ M$_{\odot}$ \citep{zenteno2020}, which is in agreement to our estimation of the total dynamical mass from scaling relations M$_{200,\rm dyn}$ = M$_{\rm S}$ + M$_{\rm N}$ = $5.55 \pm 2.33 \times 10^{14}$ M$_{\odot}$, at the 1$\sigma$ level.

\subsubsection{Recovery of the merger orbit}\label{sec:dis_recoverymh}

\begin{figure*}
    \centering
    \includegraphics[width=\linewidth]{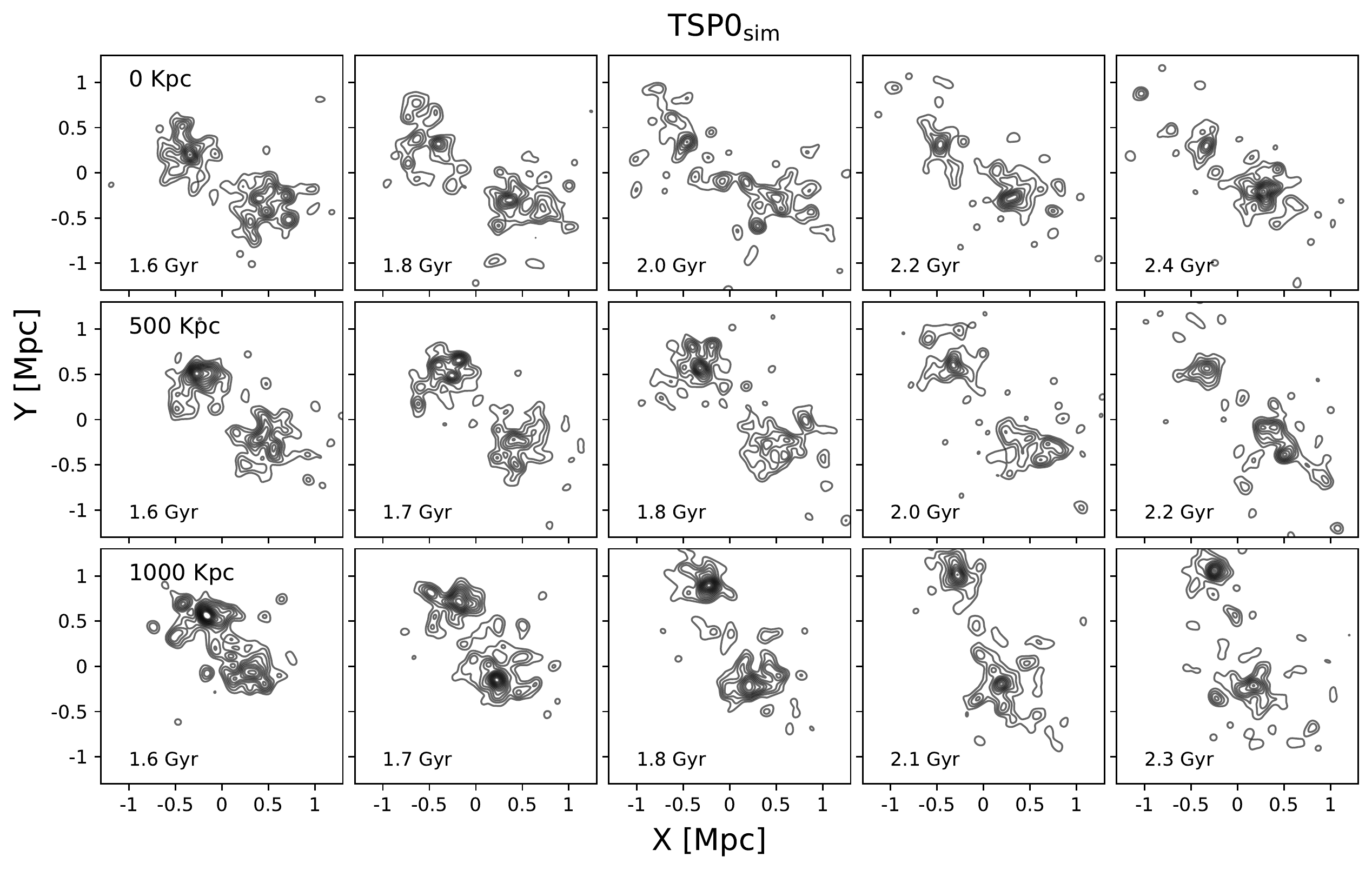}
    \vskip-0.13in
    \caption{Density maps for the simulated 1:3 mass ratio cluster merger. Each row represents the time evolution around the TSP0 for the different impact parameters $b=0,500,1000$ kpc shown at the top, middle and bottom rows, respectively. For each panel, the simulation time is written on the bottom left.}
    \label{fig:density_sims_all}
\end{figure*}

\texttt{MCMAC} gives as a result two different time since collision, TSP0=\tscin{} Gyr and TSP1=\tscout{} Gyr, for an outgoing and an incoming merger, respectively, after the first pericentric passage. A more detailed analysis of the X--ray could further constrain both the \texttt{MCMAC} output, e.g. by constraining the merging angle \citep{Monteiro2017, Monteiro2018} and the TSP \citep{Dawson2013, Ng2015, Monteiro2017} from shocks (if any), and also the merging scenario from hydrodynamical simulations, e.g. by comparing the temperature maps or by running a simulation which recovers the features (both of the galaxies and of the ICM) of this particular merger. This is particularly interesting given that the simulations that we use to compare have a merger axis angle of $\alpha = 0.0$ deg. \cite{Dawson2013} runs \texttt{MCMAC} on the Bullet Cluster data and finds $\alpha = 50^{+23}_{-23}$ deg, however, by adding a prior using the X--ray shock information, he is able to constrain the angle to $\alpha = 24^{+14}_{-8}$ deg, which is closer to the plane of the sky and also decreases significantly the error bars on the estimated collision times.

For instance, if we assume that the merger is nearly on the plane of the sky and constrain the merging angle, $\alpha$, from \texttt{MCMAC} to be between 0$^\circ$ and $45^\circ$, then the resulting values are $\alpha = 25^{+6}_{-6}$ deg, TSP0=$0.73^{+0.09}_{-0.09}$ and TSP1=$2.10^{+0.51}_{-0.30}$, which are still within the previous estimated values (within the errors) and have smaller error bars. However, the estimated TSP1 is still higher than any of the ones estimated from the simulations (see Table~\ref{tab:simulation_prop}). 

A similar system is the one studied by \citet{Dawson2012}; DLSCL J0916.2+2951, a major merging at $z=0.53$, with a projected distance of $1.0^{+0.11}_{-0.14}$ Mpc. Their dynamical analysis gives masses similar to that of our structures (when using $\sigma_v - M$ scaling relations), with the mass ratio between their northern and southern structures of $M_{\rm S}/ M_{\rm N} = 1.11 \pm 0.81$. Using an analytical model, they were able to recover a merging angle $\alpha = 34^{+20}_{-14}$ degrees and a physical separation of $d3D = 1.3^{+0.97}_ {-0.18}$, both values in agreement with what we found. Furthermore, their time since collision is also similar to the one found for our outgoing system TSP$ = 0.7^{+0.2}_{-0.1}$, however they do not differentiate between an outgoing or incoming system.

Regarding 0307-6225C, the estimated velocity dispersion is very high ($\sigma_v = 1415$ km s$^{-1}$) and the density map shows that this region is not as dense as the other two, with no dominant massive galaxy. To check whether it is common for a merging of two galaxy clusters, we take a look at how the density map varies in the 1:3 mass ratio simulations near the estimated TSP0. We show in Fig.~\ref{fig:density_sims_all}, on each row, the density maps of the simulations with the corresponding time shown at the bottom left, and the impact parameter of the row at the top left of the first figure of each row. At different times, the density maps for the same impact parameter show to be rather irregular, with the in-between region changing from snapshot to snapshot. In particular, both $b=0$ kpc and $b=1000$ kpc show an overdense in-between area near the TSP0. However, this is not the case in other snapshots, so we cannot state with confidence that this is common for a merging cluster to show such a pronounced in-between overdense region.  

\subsubsection{Constraining the TSP with simulations}\label{sec:constraining_tsc_with_sims}

We compare the results derived by \texttt{MCMAC} with those estimated from a hydrodynamical simulation of two merging structures with a mass ratio of 1:3 \citep{ZuHone2011, ZuHone2018}. We chose this ratio since the X--ray morphologies of both the simulation and the system are a better match than the 1:1 mass ratio, where the X--ray intensity from the simulation is similar for the two structures (see Fig.~\ref{fig:sim_TSP0_xray} and \ref{fig:sim_TSP1_xray}), unlike our system, which have two distinctly different structures (see the orange contour in Fig.~\ref{fig:rgb_image}). 

Using dark matter only simulations, \cite{Wittman2019} looked for halos with similar configurations to those of observed merging clusters (such as the Bullet and Musket Ball clusters) and compared the time since collisions to those derived by \texttt{MCMAC} and other hydrodynamical simulations, finding that with respect to the latter the derived merging angles and TSP are consistent. However, both the outgoing and incoming TSP and the angles are lower than those derived by \texttt{MCMAC}, attributing the differences to the \texttt{MCMAC} assumption of zero distance between the structures at the collision time.

\citet{Sarazin2002} discuss that most merging systems should have a small impact parameter, of the order of a few kpc. \citet{Dawson2012} argues that, given the displayed gas morphology, the dissociative merging galaxy cluster DLSCL J0916.2+2951, has a small impact parameter. The argument is that simulations show that the morphology for mergers with small impact parameters, is elongated transverse to the merger direction \citep{Schindler1993,Poole2006, Machado2013}. The X--ray morphology shown in this paper is similar to that from \citet{Dawson2012}. It is also similar to that of Abell 3376 \citep{Monteiro2017}, a merging galaxy cluster which was simulated by \citet{Machado2013} with different impact parameters ($b=0, 150, 350$ and $500$ kpc), with their results suggesting that a model with $b<150$ kpc is preferred. Given the similitude between SPT-CL J0307-6225 X--ray morphology and that of other systems such as Abell 3376 and DLSCL J0916.2+2951, then we suggest that the simulations with $b=0$ kpc or $b=500$ kpc are better representations of  our system. This implies that the preferred scenario for this merging cluster is that of an outgoing system or a system very close to turnaround. This can also be seen when comparing the X--ray morphology of SPT-CL J0307-6225 with that of the 1:3 mass ratio simulations at the estimated TSP0$_{\rm sim}$ and TSP1$_{\rm sim}$, shown in Fig.~\ref{fig:sim_TSP0_xray} and Fig.~\ref{fig:sim_TSP1_xray}, respectively, with the X--ray contours at TSP0$_{\rm sim}$ being more similar than the ones at TSP1$_{\rm sim}$ for $b=500,1000$ kpc.

\subsubsection{Proposed merger scenario}
We propose that the merger scenario that best describes the observations of 0307-6225 is that of a post-merger seen \tscin{} Gyr after collision. Combining the simulations with results from literature we constraint the impact parameter to be $b<500$ kpc. Simulations also support a mass ratio closer to 1:3 than 1:1, given the X--ray morphology.

\subsection{Galaxy population in a merging galaxy cluster}\label{sec:disc_galpop}

\begin{figure}
    \centering
    \includegraphics[width=\linewidth]{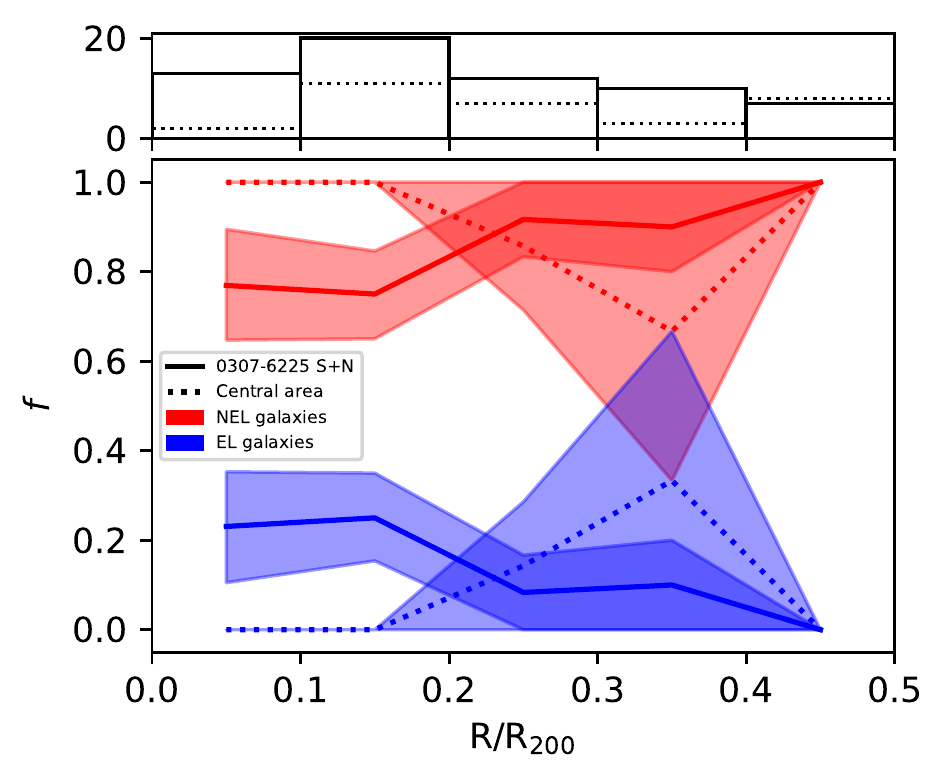}
    \vskip-0.13in
    \caption{Fraction of EL (blue) and NEL (red) galaxies with respect to the distance to the centers of 0307-6225S and 0307-6225N (continuous lines) and the SZ center (dotted lines). To compute the errors we do 10000 Poisson realizations around the true number of NEL and EL galaxies within a radial bin, and re-estimate the fractions for each area. We then compute the 16th and 84th percentiles as the 1$\sigma$ error regions, which are shown as shaded areas. The top panel shows the total number of galaxies per bin and per area (continuous lines for 0307-6225 S+N and dotted lines for the central area).}
    \label{fig:dressler_plot}
\end{figure}
An interesting feature of our system is that 90\% of the EL galaxies belong to any of the main substructures (Fig.~\ref{fig:phase_spectro}). \cite{stroe21} found that, for merging galaxy clusters, 40\% (80\%) of EL galaxies are located within 1.5 Mpc (3 Mpc) of the cluster center. To study this behavior further and analyse if our EL galaxies favour a spatial position within the substructures, we compare their galaxy radial distribution to that of the central region. We combine the galaxy distributions of 0307-6225S and 0307-6225N by normalizing the clustercentric distances R by the virial radius, R$_{200}$, of each substructure (1.16 Mpc for 0307-6225S and 1.06 Mpc for 0307-6225N) and then estimate the fraction of EL and NEL galaxies within bins of R/R$_{200}$. In the case of the central region we use the SZ position as the center and average the R$_{200}$ of the main substructures as the normalization radius (choosing only one of the radius does not affect the results). Fig.~\ref{fig:dressler_plot} shows the estimated fractions as a function of the clustercentric distance for $R < 0.5\times$R$_{200}$, with the total number of galaxies per bin shown in the upper panel. The fraction of EL galaxies towards the inner regions of the substructures (blue continuous line) is higher compared to that of the central area (blue dotted line), which is non-existent, at the 1$\sigma$ level. Overall, EL galaxies are preferentially located at distances of $R < 0.2\times$R$_{200}$ from the substructures centers.

\begin{figure}
    \centering
    \includegraphics[width=\linewidth]{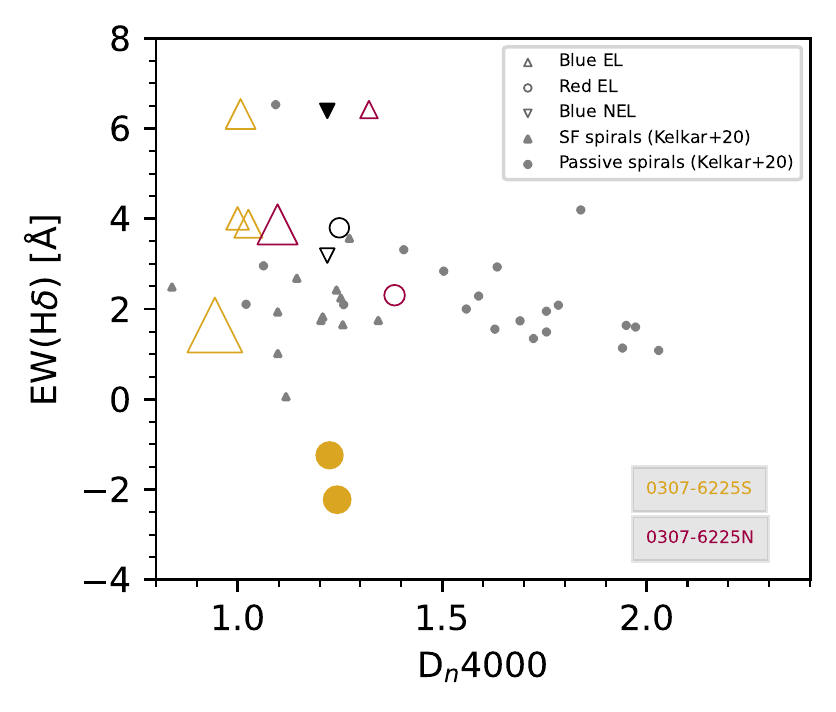}
    \vskip-0.13in
    \caption{EW(H$\delta$) vs D$_n$4000 index for our EL and blue NEL galaxies, continuing the symbols of Fig.~\ref{fig:phase_spectro}. Gray markers show SF (triangles) and passive (circles) spiral galaxies from \citet{Kelkar20}.}
    \label{fig:hdelta_d4000}
\end{figure}

We will divide the discussion of the galaxy population by studying the differences between the two clumps, analysing the red EL galaxy population and also the population in the area in-between 0307-6225S and 0307-6225N. Following the work of \cite{Kelkar20} we also study the EW(H$\delta$) vs D$_n$4000 plane in order to analyze the properties of the galaxy population. \cite{Kelkar20} studied the galaxy population in the merging cluster Abell 3376 (A3376), a young post merger ($\sim$0.6 Gyr) cluster at $z\sim0.046$ with clear merger shock features, analyzing the location of the galaxies, in particular of PSB galaxies. The D$_n 4000$ index corresponds to the ratio between the flux redward and blueward the 4000 \AA\ break, indicating the ages of the stellar population of the galaxies, which makes it an interesting measurement against the EW(H$\delta$) in absorption. Fig.~\ref{fig:hdelta_d4000} shows the EW(H$\delta$)-D$_n$4000 plane for our EL and blue NEL galaxies, where we estimate the D$_n$4000 index following \cite{balogh99}. We will further discuss the positions within the plane of the different galaxy types in the following subsections.

\subsubsection{Comparison between the northern and southern sub-clusters}

One interesting optical feature of 0307-6225S, is the two bright galaxies ($d_{\rm proj}=41$ kpc) at the center of its distribution (Fig.~\ref{fig:southernstr}). A similar, but rather extreme case is that of the galaxy cluster Abell 3827 at $z=0.099$, which shows evidence for a recent merger with four nearly equally bright galaxies within 10 kpc from the central region \citep{Carrasco2010, Massey2015}. Using GMOS data, \cite{Carrasco2010} found that the peculiar velocities of at least 3 of these galaxies are within $\sim$300 km s$^{-1}$ from the cluster redshift, with the remaining one having an offset of $\sim$1000 km s$^{-1}$.

BCGs have low peculiar velocities in relaxed clusters, whereas for disturbed clusters it is expected that their peculiar velocity is  20-30\% the velocity dispersion of the cluster \citep{Yoshikawa2003, Ye2017}. For 0307-6225S, one of the bright galaxies has a peculiar velocity of $\sim$666 km s$^{-1}$, which is $\sim$88\% the velocity dispersion of this subcluster.  This could be evidence  of a past merging between 0307-6225S and another cluster previous to the merger with 0307-6225N. The AD test gives a Gaussian distribution, where the results do not change by applying a 3-$\sigma$ iteration, which could indicate that the substructure is a post-merger. 

We apply the \cite{Raouf2019R} magnitude gap method to separate between relaxed and unrelaxed systems, to 0307-6225S and 0307-6225N independently. They use the magnitude difference between the first and second brigthest galaxy and select relaxed clusters as those with $\Delta M_{12} < 1.7$, whereas for unrelaxed clusters they use $\Delta M_{12} < 0.5$. We find that for 0307-6225S the magnitude difference is $\Delta M_{12} = 0.0152 < 0.5$, which supports the scenario that 0307-6225S suffered a previous merger prior to the one with 0307-6225N. Central galaxies take $\approx$1 Gyr to settle to the cluster centre during the post-merger phase \citep{White1976,Bird1994}, meaning that this previous merger must have taken place over 1 Gyr before the observed merger between 0307-6225S and 0307-6225N. On the other hand, for 0307-6225N the value is $\Delta M_{12} \approx 1.8 > 1.7$, meaning 0307-6225N was a relaxed system prior to this merger. 

Regarding the overall galaxy population, the fraction of EL galaxies in 0307-6225S (24\%) is nearly two times that of 0307-6225N ($\sim$13\%), although consistent within 1$\sigma$. All EL galaxies from 0307-6225N have small peculiar velocities (within 1$\sigma_v$), while for 0307-6225S 75\% (50\%) of the blue SF galaxies have peculiar velocities higher than 2$\sigma_v$ (3$\sigma_v$), as seen in Fig.~\ref{fig:phase_spectro}. These galaxies, which are bluer than the blue EL galaxies of 0307-6225N (Fig.~\ref{fig:gal_properties}), could be in the process of being accreted. 

Fig.~\ref{fig:hdelta_d4000} shows that blue EL galaxies located in 0307-6225N tend to have older stellar populations than their blue counterparts from 0307-6225S. Apart from the PSB galaxy (black filled triangle), there are 2 other blue galaxies with similar measured EW(H$\delta$). Both of this galaxies might be dusty star forming galaxies \citep[spectral type A+em, ][]{balogh99}, with the one from 0307-6225S having the smallest peculiar velocity of the blue galaxies from this subcluster ($\approx -1400 km s^{-1}$). The recent infall of this galaxy might be the reason behind the truncated star formation, whereas for the blue galaxy from 0307-6225N, with an older stellar population and a peculiar velocity within $1\sigma_v$ (within the errors), the merger itself might be the reason.

\cite{Stroe2015} found that the increase of H$\alpha$ emission of galaxies in the ``Sausage'' merging galaxy cluster, compared to galaxies in the ``Toothbrush'' merging galaxy cluster could be explained by their time since collision, with the ``Toothbrush'' cluster being more evolved \citep[TSP$\sim$2 Gyr,][]{Bruggen2012} than the ``Sausage'' \citep[TSP$\sim$1Gyr,][]{vanWeeren2011}. This timescales are similar to what we see from the merger of 0307-6225N and 0307-6225S (TSP=\tscin) and the possible previous merger of 0307-6225S, which happened at least $\approx$1 Gyr prior to the collision with 0307-6225N. This previous merger could have exhausted the star formation of the galaxies of 0307-6225S, which might be the reason that there are no blue star forming galaxies towards the central region (within 1$\sigma_v$) of 0307-6225S compared to 0307-6225N.

\subsubsection{Red EL galaxies}

Of particular interest are our EL galaxies located in the RCS. Out of the 4 red EL galaxies, 3 are located in the cores of the two main structures,  with 2 of them classified as SSB. Most of the blue SF galaxies are best matched by a high-redshift star forming or late-type emission galaxy template, whereas most of the red SF galaxies are best matched with an early-type absorption galaxy template. Our red EL galaxies have older stellar populations than our blue EL galaxies (except for 1, Fig.~\ref{fig:hdelta_d4000}), with the red EL galaxy from 0307-6225N having older stellar populations than those of 0307-6225S, which might be expected given that they are SSB.

\cite{Koyama2011} studied the region in and around the $z=0.41$ rich cluster CL0939+4713 (A851) using H$\alpha$ imaging to distinguish SF emission line galaxies. A851 is a dynamically young cluster with numerous groups at the outskirts. They found that the red H$\alpha$ emitters are preferentially located in low-density environments, such as the groups and the outskirts, whereas in the core of the cluster they did not find red H$\alpha$ emitters. Similar results were found by \cite{Einasto2018} for the galaxy cluster Abell 2142, with star forming galaxies (which includes red star-forming galaxies) located at 1.5-2.0 $h^{-1}$ Mpc from the cluster centre. \cite{ma10} studied the galaxy population of the merging galaxy cluster MACS J0025.4-1225 at $z=0.586$. In the areas around the cluster cores (with a radius of 150 kpc) they find emission line galaxies corresponding to two spiral galaxies (one for each subcluster), plus some spiral galaxies without spectroscopic information, accounting for 14\% of the total galaxies within the radius. Their Fig. 15 shows that they also have red EL galaxies, however they don't specify whether the 2 spiral galaxies within the cluster core are part of this population. Results from \cite{ma10}, \cite{Koyama2011} and \cite{Einasto2018} indicate that red EL galaxies are not likely to be found within the cores of dense regions.

\cite{Sobral2016} studied the population of H$\alpha$ emitters in the super-cluster Abell 851, finding that galaxies with higher dust extinctions to be preferentially located towards the densest environments. The results deviate from the expected extinctions given the masses of the galaxies. There is evidence for a population of RCS sequence galaxies with residual star formation in galaxy clusters as seen using ultra violet images. \cite{Crossett2014} found these galaxies to be red spirals located in low-density environments and towards the outskirts of massive clusters, concluding that they are either spirals with truncated star formation given their infall or high-mass spirals.  \cite{Sheen2016} found that for four rich Abell clusters at $z\leq0.1$, the fraction of red sequence galaxies with recent star formation that show signs of recent mergers is $\sim30$\%, implying internal processes playing a significant role for the supply of cold gas to this galaxy population.

75\% of our red EL galaxies do not have close neighbours which can supplement their gas reserves  (Fig.~\ref{fig:cutouts_spectra}). It is possible then that these objects accreted gas from the ICM, with the merger triggering then the SF. Given the peculiar velocity of the two SSB galaxy from our sample (which is classified as red), at least one of them was most likely part of the merging event. If, for example, merger shocks travelling through the ICM can trigger a starburst episode on galaxies with gas reservoirs for a few 100 Myr \citep[][]{Caldwell1997,Owers2012,Stroe2014,Stroe2015}, then these galaxies would make the outgoing scenario a better candidate than the incoming one. Another mechanism that can trigger a starburst of the gas is the rapid change of the tidal gravitational field due to the merger, which can drive gas to the inner part of galaxies \citep[][]{Bekki1999, Ferrari2003,Yoon2019}.

Unfortunately, we do not see evidence of shocks in our X--ray data, likely due to it being shallow given the redshift. Shocks lasting 1-2 Gyr, are expected to generate in mergers of clumps with M$\geq 10^{13}$ M$_\odot$ with colliding velocities of $10^3$ km s$^{-1}$, generating kinetic energies of over $10^{62}$ erg \citep[][]{Markevitch2007}. \cite{Ha2018} found evidence for shocks using hydrodynamical simulations of merging galaxy clusters with mass ratio $\sim$2, average virial masses similar to that of 0307-6225S and low impact parameters $b\leq140$ kpc. They found that shocks are likely to be observed $\sim$1 Gyr after the shock generation, at distances of 1-2 Mpc from the merger center, with mean mach numbers $M_S=2-3$. Thus, we expect shocks to be have taken part in our system given the similar mass properties and the collision velocity we estimate with MCMAC (2300$^{122}_{-96}$ km s$^{-1}$, Table~\ref{tab:mcmac}).

\subsubsection{Area in-between the main substructures}

The central area, meaning 0307-6225C and other galaxies not associated to any substructure, is comprised of $\sim$86\% red passive galaxies, with the only EL galaxy belonging to the RCS. Moreover, the 2 blue galaxies are classified as a passive and a PSB. \citet{ma10} found a fraction of post-starburst galaxies in the major cluster merger MACS J0025.4-1225, on the region in-between the collision between the two merging components, where, given the timescales, the starburst episode of them occurred during first passage. Similarly to our blue galaxies in this region, they found that their colors are located between those of blue EL galaxies and red passive galaxies (Fig.~\ref{fig:gal_properties}). 

\cite{Kelkar20} divided the PSB population in three subsamples: bright, faint and blue. Although they don't find a trend for the first two, they find that blue PSB tend to be concentrated between the two BCGs, along the merger axis, although showing a wide variety of line-of-sight velocities. Fig.~\ref{fig:sky_pos_xray} shows a similar trend for the PSB (red filled triangle) and passive (red unfilled triangle) blue galaxies in the central region. However, the velocity of the PSB
(Fig.~\ref{fig:phase_spectro}) indicates that this might be the result of the infall in the cluster rather than an outcome of the merger. This does not seem to be the case for the blue passive galaxy, with a velocity of $\sim$310 km s$^{-1}$.

\cite{Pranger2013} found a high fraction of NEL spiral galaxies towards the cluster core ($< 1.2$ Mpc) of the merging galaxy cluster Abell 3921 ($z=0.093$). Their results are in agreement with the idea of passive spirals being preferentially located in high density environments in relaxed clusters \citep[e.g.]{Bosch2013}, being an intermediate stage before developing to S0 galaxies \citep[e.g.][]{Vogt2004, Moran2007}. Passive spirals are believed to be the results of ram pressure stripping during their infall onto galaxy clusters \citep[e.g.][]{Vogt2004}, which correlates with the small velocity and EW(H$\delta$) of our blue passive galaxy. It is worth noting that our photometric data does not have the resolution to morphologically classify our galaxy population, meaning that some of our red passive galaxies might be passive spirals with colors similar to those of elliptical galaxies \citep[e.g.][]{Goto2003}.

\section{Summary and Conclusions}
\label{sec:conclusions}

In this paper we use deep optical imaging and new MUSE spectroscopic data along with archival GMOS data to study the photometric and spectral properties of the merging cluster candidate SPT-CL J0307-6225, estimating redshifts for 69 new galaxy cluster members. We used the data to characterize (a) its merging history by means of a dynamical analysis and (b) its galaxy population by means of their spectroscopic and photometric properties. 

With respect to the merging history, we were able to confirm the merging state of the cluster and conclude that:

\begin{itemize}
    \item Using the galaxy surface density map of the RCS galaxies we can see a bi-modality in the galaxy distribution. However, the cluster does not show signs of substructures along the line-of-sight.
    \item We assign galaxy members to each substructure by means of the \textsc{DBSCAN} algorithm. We name the two main substructures as 0307-6225N and 0307-6225S, referring to the northern and southern overdensities, respectively.
    \item For each substructure we measured the redshift, velocity dispersion and velocity-derived masses from scaling relations. We find a mass ratio of $M_{\rm S}/M_{\rm N} \approx$ \massSNnE{} and a velocity difference of $v_{\rm N}-v_{\rm S}=342$ km s$^{-1}$ between the northern and southern structures.
    \item To estimate the time since collision we use the \textsc{MCMAC} algorithm, which gave us the times for an outgoing and incoming system. By means of hydrodynamical simulations we constrained the most likely time to that of an outgoing system with TSP=\tscin{} Gyr. 
    \item The outgoing configuration is also supported by the comparison between the observed and simulated X--ray morphologies. This comparison between the X--ray morphologies also provide a constraint on the masses, where a merger with a mass ratio of 1:3 seems more likely than that of a 1:1 mass merger.
\end{itemize}

With respect to the galaxy population, we find that:

\begin{itemize}
    \item EL galaxies are located preferentially near the cluster cores (projected separations), where the average low peculiar velocities of red SF galaxies indicates that they were most likely accreted before the merger between 0307-6225N and 0307-6225S occurred.
    \item EL galaxies on 0307-6225N have smaller peculiar velocities and older stellar populations than those of 0307-6225S, where in the latter it appears that blue SF galaxies were either recently accreted or are in the process of being accreted.
    \item 0307-6225S shows two possible BCGs, which are very close in projected space. The magnitude and velocity differences between them are $\sim0$ mag and $\sim$674 km s$^{-1}$, respectively, with one of them having a peculiar velocity close to 0 km s$^{-1}$ with respect to 0307-6225S, while the other is close to the estimated $1\sigma_v$. However, the velocity distribution of the cluster shows no signs of being perturbed. This suggests that 0307-6225S could be the result of a previous merger which was at its last stage when the observed merger occurred.
    \item With respect to the in-between region, the galaxy population is comprised mostly of red galaxies, with the population of blue galaxies classified as passive or PSB, with colors close to the RCS.
\end{itemize}

In summary, our work supports a nearly face-on, in the plane of the sky, major merger scenario for SPT-CL J0307-6225. This interaction  accelerates the quenching of galaxies as a result of a rapid enhancement of their star formation activity and the subsequent gas depletion. This is in line with literature findings indicating that the dynamical state of a cluster merger has a strong  impact on galaxy population.  Of particular importance is to differentiate dynamically young and old mergers. Comparisons between such systems will further increase our understanding on the connection between mergers and the quenching of star formation in galaxies. In future studies, we will replicate  the analysis performed on  SPT-CL J0307-6225, to a larger cluster sample, including the most disturbed cluster candidates on the SPT sample. These studies will be the basis for a comprehensive analysis of star formation in mergers with a wide dynamical range.

\section*{Acknowledgements}
We thank the anonymous referee for the helpful comments and suggestions that greatly improved the article. DHL acknowledges financial support from the MPG Faculty Fellowship program, the new ORIGINS cluster funded by the Deutsche Forschungsgemeinschaft (DFG, German Research Foundation) under Germany's Excellence Strategy - EXC-2094 - 390783311, and the Ludwig-Maximilians-Universit\"at Munich. FA was supported by the doctoral thesis scholarship of ANID-Chile, grant 21211648. FAG acknowledges financial support from FONDECYT Regular 1211370. FAG and FA acknowledge funding from the Max Planck Society through a Partner Group grant. J. L. N. C. is grateful for the financial support received from the Southern Office of Aerospace Research and Development of the Air Force Office of the Scientific Research International Office of the United States  (SOARD/AFOSR) through grants FA9550-18-1-0018 and  FA9550-22-1-0037. AS is supported by the ERC-StG `ClustersXCosmo' grant agreement 716762, by the FARE-MIUR grant `ClustersXEuclid' R165SBKTMA, and by INFN InDark Grant.

\section*{Data Availability}
\textit{Chandra} and Megacam/Magellan are available upon request from McDonalds, M. and Stalder, B., respectively. GMOS data are available in \cite{bayliss16}. The raw MUSE data are available from the ESO Science Archive (\url{https://archive.eso.org/}, with programs IDs: 097.A-0922(A) and 100.A-0645(A)). Additional data on derived physical parameters are available in this paper.

\section*{Affiliations}
\noindent
{\it
$^{1}$Faculty of Physics, Ludwig-Maximilians-Universit\"{a}t, Scheinerstr.\ 1, 81679 Munich, Germany \\
$^{2}$Cerro Tololo Inter-American Observatory, NSF's National Optical-Infrared Astronomy Research Laboratory, Casilla 603, La Serena, Chile\\
$^{3}$Departamento de Astronom\'ia, Universidad de La Serena, Avenida Juan Cisternas 1200, La Serena, Chile\\
$^{4}$School of Physics, University of Melbourne, Parkville, VIC 3010, Australia \\
$^{5}$Instituto de F\'isica y Astronom\'ia, Universidad de Valpara\'iso, Avda. Gran Bretaña 1111, Valpara\'iso, Chile\\
$^{6}$Academia Sinica, Institute of Astronomy and Astrophysics, 11F of AS/NTU Astronomy-Mathematics Building, No.1, Sec. 4, Roosevelt Rd, Taipei 10617, Taiwan, R.O.C.\\
$^{7}$Universidade Estadual de Santa Cruz, Laborat\'orio de Astrof\'isica Te\'orica e Observacional - 45650-000, Ilh\'eus-BA, Brazil\\ 
$^{8}$Instituto de Investigaci\'on Multidisciplinar en Ciencia y Tecnolog\'ia, Universidad de La Serena, Ra\'ul Bitr\'an 1305, La Serena, Chile\\
$^{9}$ Gemini Observatory, NSF’s National Optical-Infrared Astronomy Research Laboratory, Casilla 603, La Serena, Chile \\
$^{10}$European Southern Observatory, Alonso de Cordova 3107, Vitacura, Casilla 19001, Santiago de Chile, Chile \\
$^{11}$Center for Astrophysics — Harvard \& Smithsonian, 60 Garden Street, Cambridge, MA 02138, USA\\
$^{12}$Vera C. Rubin Observatory Project Office, 950 N. Cherry Ave, Tucson, AZ 85719, USA\\
$^{13}$Kavli Institute for Astrophysics and Space Research, Massachusetts Institute of Technology, 77 Massachusetts Avenue, Cambridge, MA 02139 \\
$^{14}$ Department of Physics, University of Cincinnati, Cincinnati, OH 45221, USA\\
$^{15}$ Department of Physics and Astronomy, University of Missouri--Kansas City, 5110 Rockhill Road, Kansas City, MO 64110, USA\\
$^{16}$ Huntingdon Institute for X-ray Astronomy, LLC, 10677 Franks Road, Huntingdon, PA 16652, USA\\
$^{17}$Department of Geography, Ludwig-Maximilians-Universit\"at, Luisenstr 37, D-80333 Munich, Germany \\
$^{18}$Potsdam Institute for Climate Impact Research, Telegrafenberg, 14473 Potsdam, Germany\\
$^{19}$Department of Physics and Astronomy, University of Potsdam, Karl-Liebknecht-Str. 24/25, 14476 Potsdam-Golm, Germany\\
$^{20}$Department of Astronomy, University of Michigan, 1085 South University Ave, Ann Arbor, MI 48109, USA\\
$^{21}$Direcci\'on de Investigaci\'on y Desarrollo, Universidad de La Serena, Av. Ra\'ul Bitr\'an Nachary Nº 1305, La Serena, Chile.\\
$^{22}$Dipartimento di Fisica, Sezione di Astronomia, Universit\'a di Trieste, Via Tiepolo 11, I-34143 Trieste, Italy
\\
$^{23}$INAF – Osservatorio Astronomico di Trieste, via Tiepolo 11, I-34131 Trieste, Italy \\
$^{22}$IFPU – Institute for Fundamental Physics of the Universe, via Beirut 2, 34151, Trieste, Italy \\
$^{24}$INFN – Sezione di Trieste, I-34100 Trieste, Italy
}
\bibliographystyle{mnras}
\bibliography{newreference} 

\begin{thebibliography}{}
\makeatletter
\relax
\def\mn@urlcharsother{\let\do\@makeother \do\$\do\&\do\#\do\^\do\_\do\%\do\~}
\def\mn@doi{\begingroup\mn@urlcharsother \@ifnextchar [ {\mn@doi@}
  {\mn@doi@[]}}
\def\mn@doi@[#1]#2{\def\@tempa{#1}\ifx\@tempa\@empty \href
  {http://dx.doi.org/#2} {doi:#2}\else \href {http://dx.doi.org/#2} {#1}\fi
  \endgroup}
\def\mn@eprint#1#2{\mn@eprint@#1:#2::\@nil}
\def\mn@eprint@arXiv#1{\href {http://arxiv.org/abs/#1} {{\tt arXiv:#1}}}
\def\mn@eprint@dblp#1{\href {http://dblp.uni-trier.de/rec/bibtex/#1.xml}
  {dblp:#1}}
\def\mn@eprint@#1:#2:#3:#4\@nil{\def\@tempa {#1}\def\@tempb {#2}\def\@tempc
  {#3}\ifx \@tempc \@empty \let \@tempc \@tempb \let \@tempb \@tempa \fi \ifx
  \@tempb \@empty \def\@tempb {arXiv}\fi \@ifundefined
  {mn@eprint@\@tempb}{\@tempb:\@tempc}{\expandafter \expandafter \csname
  mn@eprint@\@tempb\endcsname \expandafter{\@tempc}}}

\bibitem[\protect\citeauthoryear{{Abbott} et~al.,}{{Abbott}
  et~al.}{2018}]{abbott18}
{Abbott} T.~M.~C.,  et~al., 2018, \mn@doi [\apjs] {10.3847/1538-4365/aae9f0},
  \href {https://ui.adsabs.harvard.edu/abs/2018ApJS..239...18A} {239, 18}

\bibitem[\protect\citeauthoryear{{Bacon} et~al.,}{{Bacon}
  et~al.}{2012}]{bacon12}
{Bacon} R.,  et~al., 2012, The Messenger, \href
  {http://adsabs.harvard.edu/abs/2012Msngr.147....4B} {147, 4}

\bibitem[\protect\citeauthoryear{{Bacon}, {Piqueras}, {Conseil}, {Richard}  \&
  {Shepherd}}{{Bacon} et~al.}{2016}]{bacon16}
{Bacon} R.,  {Piqueras} L.,  {Conseil} S.,  {Richard} J.,   {Shepherd} M.,
  2016, {MPDAF: MUSE Python Data Analysis Framework}, Astrophysics Source Code
  Library, record ascl:1611.003 (\mn@eprint {ascl} {1611.003})

\bibitem[\protect\citeauthoryear{{Balogh}, {Morris}, {Yee}, {Carlberg}  \&
  {Ellingson}}{{Balogh} et~al.}{1999}]{balogh99}
{Balogh} M.~L.,  {Morris} S.~L.,  {Yee} H.~K.~C.,  {Carlberg} R.~G.,
  {Ellingson} E.,  1999, \mn@doi [\apj] {10.1086/308056}, \href
  {https://ui.adsabs.harvard.edu/abs/1999ApJ...527...54B} {527, 54}

\bibitem[\protect\citeauthoryear{{Bayliss} et~al.,}{{Bayliss}
  et~al.}{2016}]{bayliss16}
{Bayliss} M.~B.,  et~al., 2016, \mn@doi [\apjs] {10.3847/0067-0049/227/1/3},
  \href {http://adsabs.harvard.edu/abs/2016ApJS..227....3B} {227, 3}

\bibitem[\protect\citeauthoryear{{Bayliss} et~al.,}{{Bayliss}
  et~al.}{2017}]{Bayliss2017}
{Bayliss} M.~B.,  et~al., 2017, \mn@doi [\apj] {10.3847/1538-4357/aa607c},
  \href {https://ui.adsabs.harvard.edu/abs/2017ApJ...837...88B} {837, 88}

\bibitem[\protect\citeauthoryear{{Beers}, {Flynn}  \& {Gebhardt}}{{Beers}
  et~al.}{1990}]{Beers1990}
{Beers} T.~C.,  {Flynn} K.,   {Gebhardt} K.,  1990, \mn@doi [\aj]
  {10.1086/115487}, \href {http://adsabs.harvard.edu/abs/1990AJ....100...32B}
  {100, 32}

\bibitem[\protect\citeauthoryear{{Bekki}}{{Bekki}}{1999}]{Bekki1999}
{Bekki} K.,  1999, \mn@doi [\apjl] {10.1086/311796}, \href
  {https://ui.adsabs.harvard.edu/abs/1999ApJ...510L..15B} {510, L15}

\bibitem[\protect\citeauthoryear{{Bertin}}{{Bertin}}{2011}]{Bertin2011}
{Bertin} E.,  2011, in {Evans} I.~N.,  {Accomazzi} A.,  {Mink} D.~J.,   {Rots}
  A.~H.,  eds,  Astronomical Society of the Pacific Conference Series Vol. 442,
  Astronomical Data Analysis Software and Systems XX. p.~435

\bibitem[\protect\citeauthoryear{{Bertin} \& {Arnouts}}{{Bertin} \&
  {Arnouts}}{1996}]{bertin96}
{Bertin} E.,  {Arnouts} S.,  1996, \aaps, \href
  {http://adsabs.harvard.edu/abs/1996A%26AS..117..393B} {117, 393}

\bibitem[\protect\citeauthoryear{{Bird}}{{Bird}}{1994}]{Bird1994}
{Bird} C.~M.,  1994, \mn@doi [\aj] {10.1086/116973}, \href
  {https://ui.adsabs.harvard.edu/abs/1994AJ....107.1637B} {107, 1637}

\bibitem[\protect\citeauthoryear{{Bleem} et~al.,}{{Bleem}
  et~al.}{2015}]{bleem15b}
{Bleem} L.~E.,  et~al., 2015, \mn@doi [\apjs] {10.1088/0067-0049/216/2/27},
  \href {http://adsabs.harvard.edu/abs/2015ApJS..216...27B} {216, 27}

\bibitem[\protect\citeauthoryear{{B{\"o}sch} et~al.,}{{B{\"o}sch}
  et~al.}{2013}]{Bosch2013}
{B{\"o}sch} B.,  et~al., 2013, \mn@doi [\aap] {10.1051/0004-6361/201219244},
  \href {https://ui.adsabs.harvard.edu/abs/2013A&A...549A.142B} {549, A142}

\bibitem[\protect\citeauthoryear{{Br{\"u}ggen}, {van Weeren}  \&
  {R{\"o}ttgering}}{{Br{\"u}ggen} et~al.}{2012}]{Bruggen2012}
{Br{\"u}ggen} M.,  {van Weeren} R.~J.,   {R{\"o}ttgering} H.~J.~A.,  2012,
  \mn@doi [\mnras] {10.1111/j.1745-3933.2012.01304.x}, \href
  {https://ui.adsabs.harvard.edu/abs/2012MNRAS.425L..76B} {425, L76}

\bibitem[\protect\citeauthoryear{{Caldwell} \& {Rose}}{{Caldwell} \&
  {Rose}}{1997}]{Caldwell1997}
{Caldwell} N.,  {Rose} J.~A.,  1997, \mn@doi [\aj] {10.1086/118271}, \href
  {https://ui.adsabs.harvard.edu/abs/1997AJ....113..492C} {113, 492}

\bibitem[\protect\citeauthoryear{{Carlstrom} et~al.,}{{Carlstrom}
  et~al.}{2011}]{carlstrom11}
{Carlstrom} J.~E.,  et~al., 2011, \mn@doi [\pasp] {10.1086/659879}, \href
  {http://adsabs.harvard.edu/abs/2011PASP..123..568C} {123, 568}

\bibitem[\protect\citeauthoryear{{Carrasco} et~al.,}{{Carrasco}
  et~al.}{2010}]{Carrasco2010}
{Carrasco} E.~R.,  et~al., 2010, \mn@doi [\apjl]
  {10.1088/2041-8205/715/2/L160}, \href
  {https://ui.adsabs.harvard.edu/abs/2010ApJ...715L.160C} {715, L160}

\bibitem[\protect\citeauthoryear{{Chiu} et~al.,}{{Chiu} et~al.}{2016}]{chiu16b}
{Chiu} I.,  et~al., 2016, \mn@doi [\mnras] {10.1093/mnras/stw190}, \href
  {http://adsabs.harvard.edu/abs/2016MNRAS.457.3050C} {457, 3050}

\bibitem[\protect\citeauthoryear{{Clowe}, {Brada{\v c}}, {Gonzalez},
  {Markevitch}, {Randall}, {Jones}  \& {Zaritsky}}{{Clowe}
  et~al.}{2006}]{clowe06}
{Clowe} D.,  {Brada{\v c}} M.,  {Gonzalez} A.~H.,  {Markevitch} M.,  {Randall}
  S.~W.,  {Jones} C.,   {Zaritsky} D.,  2006, \mn@doi [\apjl] {10.1086/508162},
  \href {http://adsabs.harvard.edu/abs/2006ApJ...648L.109C} {648, L109}

\bibitem[\protect\citeauthoryear{Collaboration: et~al.,}{Collaboration:
  et~al.}{2016}]{des16}
Collaboration: D. E.~S.,  et~al., 2016, \mn@doi [\mnras]
  {10.1093/mnras/stw641}, 460, 1270

\bibitem[\protect\citeauthoryear{{Cortese}, {Gavazzi}, {Boselli},
  {Iglesias-Paramo}  \& {Carrasco}}{{Cortese} et~al.}{2004}]{Cortese2004}
{Cortese} L.,  {Gavazzi} G.,  {Boselli} A.,  {Iglesias-Paramo} J.,   {Carrasco}
  L.,  2004, \mn@doi [\aap] {10.1051/0004-6361:20040381}, \href
  {https://ui.adsabs.harvard.edu/abs/2004A&A...425..429C} {425, 429}

\bibitem[\protect\citeauthoryear{{Crossett}, {Pimbblet}, {Stott}  \&
  {Jones}}{{Crossett} et~al.}{2014}]{Crossett2014}
{Crossett} J.~P.,  {Pimbblet} K.~A.,  {Stott} J.~P.,   {Jones} D.~H.,  2014,
  \mn@doi [\mnras] {10.1093/mnras/stt2065}, \href
  {https://ui.adsabs.harvard.edu/abs/2014MNRAS.437.2521C} {437, 2521}

\bibitem[\protect\citeauthoryear{{Dawson}}{{Dawson}}{2013}]{Dawson2013}
{Dawson} W.~A.,  2013, \mn@doi [\apj] {10.1088/0004-637X/772/2/131}, \href
  {http://adsabs.harvard.edu/abs/2013ApJ...772..131D} {772, 131}

\bibitem[\protect\citeauthoryear{{Dawson} et~al.,}{{Dawson}
  et~al.}{2012}]{Dawson2012}
{Dawson} W.~A.,  et~al., 2012, \mn@doi [\apjl] {10.1088/2041-8205/747/2/L42},
  \href {https://ui.adsabs.harvard.edu/abs/2012ApJ...747L..42D} {747, L42}

\bibitem[\protect\citeauthoryear{{Dawson} et~al.,}{{Dawson}
  et~al.}{2015}]{Dawson2015}
{Dawson} W.~A.,  et~al., 2015, \mn@doi [\apj] {10.1088/0004-637X/805/2/143},
  \href {https://ui.adsabs.harvard.edu/abs/2015ApJ...805..143D} {805, 143}

\bibitem[\protect\citeauthoryear{{Desai} et~al.,}{{Desai}
  et~al.}{2012}]{Desai2012}
{Desai} S.,  et~al., 2012, \mn@doi [\apj] {10.1088/0004-637X/757/1/83}, \href
  {https://ui.adsabs.harvard.edu/abs/2012ApJ...757...83D} {757, 83}

\bibitem[\protect\citeauthoryear{{Dietrich} et~al.,}{{Dietrich}
  et~al.}{2019}]{Dietrich2019}
{Dietrich} J.~P.,  et~al., 2019, \mn@doi [\mnras] {10.1093/mnras/sty3088},
  \href {https://ui.adsabs.harvard.edu/abs/2019MNRAS.483.2871D} {483, 2871}

\bibitem[\protect\citeauthoryear{{Doubrawa}, {Machado}, {Lagan{\'a}}, {Lima
  Neto}, {Monteiro-Oliveira}  \& {Cypriano}}{{Doubrawa}
  et~al.}{2020}]{Doubrawa2020}
{Doubrawa} L.,  {Machado} R.~E.~G.,  {Lagan{\'a}} T.~F.,  {Lima Neto} G.~B.,
  {Monteiro-Oliveira} R.,   {Cypriano} E.~S.,  2020, \mn@doi [\mnras]
  {10.1093/mnras/staa1051}, \href
  {https://ui.adsabs.harvard.edu/abs/2020MNRAS.495.2022D} {495, 2022}

\bibitem[\protect\citeauthoryear{{Dressler} \& {Gunn}}{{Dressler} \&
  {Gunn}}{1983}]{Dressler1983}
{Dressler} A.,  {Gunn} J.~E.,  1983, \mn@doi [\apj] {10.1086/161093}, \href
  {https://ui.adsabs.harvard.edu/abs/1983ApJ...270....7D} {270, 7}

\bibitem[\protect\citeauthoryear{{Dressler} \& {Shectman}}{{Dressler} \&
  {Shectman}}{1988}]{dressler88}
{Dressler} A.,  {Shectman} S.~A.,  1988, \aj, \href
  {http://adsabs.harvard.edu/cgi-bin/nph-bib_query?bibcode=1988AJ.....95..985D&db_key=AST}
  {95, 985}

\bibitem[\protect\citeauthoryear{{Drlica-Wagner} et~al.,}{{Drlica-Wagner}
  et~al.}{2018}]{Drlica-Wagner2018}
{Drlica-Wagner} A.,  et~al., 2018, \mn@doi [\apjs] {10.3847/1538-4365/aab4f5},
  \href {https://ui.adsabs.harvard.edu/abs/2018ApJS..235...33D} {235, 33}

\bibitem[\protect\citeauthoryear{{Einasto} et~al.,}{{Einasto}
  et~al.}{2018}]{Einasto2018}
{Einasto} M.,  et~al., 2018, \mn@doi [\aap] {10.1051/0004-6361/201731600},
  \href {https://ui.adsabs.harvard.edu/abs/2018A&A...610A..82E} {610, A82}

\bibitem[\protect\citeauthoryear{Ester, Kriegel, Sander  \& Xu}{Ester
  et~al.}{1996}]{Ester1996}
Ester M.,  Kriegel H.-P.,  Sander J.,   Xu X.,  1996, in Proceedings of the
  Second International Conference on Knowledge Discovery and Data Mining.
  KDD'96.
AAAI Press, pp 226--231, \url
  {http://dl.acm.org/citation.cfm?id=3001460.3001507}

\bibitem[\protect\citeauthoryear{{Evrard} et~al.,}{{Evrard}
  et~al.}{2008}]{Evrard2008}
{Evrard} A.~E.,  et~al., 2008, \mn@doi [\apj] {10.1086/521616}, \href
  {https://ui.adsabs.harvard.edu/abs/2008ApJ...672..122E} {672, 122}

\bibitem[\protect\citeauthoryear{Fabricant, Cheimets, Caldwell  \&
  Geary}{Fabricant et~al.}{1998}]{Fabricant_1998}
Fabricant D.,  Cheimets P.,  Caldwell N.,   Geary J.,  1998, \mn@doi [\pasp]
  {10.1086/316111}, 110, 79

\bibitem[\protect\citeauthoryear{{Ferragamo}, {Rubi{\~n}o-Mart{\'\i}n},
  {Betancort-Rijo}, {Munari}, {Sartoris}  \& {Barrena}}{{Ferragamo}
  et~al.}{2020}]{Ferragamo2020}
{Ferragamo} A.,  {Rubi{\~n}o-Mart{\'\i}n} J.~A.,  {Betancort-Rijo} J.,
  {Munari} E.,  {Sartoris} B.,   {Barrena} R.,  2020, \mn@doi [\aap]
  {10.1051/0004-6361/201834837}, \href
  {https://ui.adsabs.harvard.edu/abs/2020A&A...641A..41F} {641, A41}

\bibitem[\protect\citeauthoryear{{Ferrari}, {Maurogordato}, {Cappi}  \&
  {Benoist}}{{Ferrari} et~al.}{2003}]{Ferrari2003}
{Ferrari} C.,  {Maurogordato} S.,  {Cappi} A.,   {Benoist} C.,  2003, \mn@doi
  [\aap] {10.1051/0004-6361:20021741}, \href
  {https://ui.adsabs.harvard.edu/abs/2003A&A...399..813F} {399, 813}

\bibitem[\protect\citeauthoryear{{Gladders} \& {Yee}}{{Gladders} \&
  {Yee}}{2000}]{Gladders2000}
{Gladders} M.~D.,  {Yee} H.~K.~C.,  2000, \mn@doi [\aj] {10.1086/301557}, \href
  {https://ui.adsabs.harvard.edu/abs/2000AJ....120.2148G} {120, 2148}

\bibitem[\protect\citeauthoryear{{Gonzalez} et~al.,}{{Gonzalez}
  et~al.}{2018}]{Gonzalez2018}
{Gonzalez} E.~J.,  et~al., 2018, \mn@doi [\aap] {10.1051/0004-6361/201732003},
  \href {https://ui.adsabs.harvard.edu/abs/2018A&A...611A..78G} {611, A78}

\bibitem[\protect\citeauthoryear{{Goto} et~al.,}{{Goto}
  et~al.}{2003}]{Goto2003}
{Goto} T.,  et~al., 2003, \mn@doi [\pasj] {10.1093/pasj/55.4.757}, \href
  {https://ui.adsabs.harvard.edu/abs/2003PASJ...55..757G} {55, 757}

\bibitem[\protect\citeauthoryear{{Gunn} \& {Gott}}{{Gunn} \&
  {Gott}}{1972}]{GunnGott72}
{Gunn} J.~E.,  {Gott} III J.~R.,  1972, \mn@doi [\apj] {10.1086/151605}, \href
  {http://adsabs.harvard.edu/abs/1972ApJ...176....1G} {176, 1}

\bibitem[\protect\citeauthoryear{{Ha}, {Ryu}  \& {Kang}}{{Ha}
  et~al.}{2018}]{Ha2018}
{Ha} J.-H.,  {Ryu} D.,   {Kang} H.,  2018, \mn@doi [\apj]
  {10.3847/1538-4357/aab4a2}, \href
  {https://ui.adsabs.harvard.edu/abs/2018ApJ...857...26H} {857, 26}

\bibitem[\protect\citeauthoryear{{Harvey}, {Massey}, {Kitching}, {Taylor}  \&
  {Tittley}}{{Harvey} et~al.}{2015}]{harvey15}
{Harvey} D.,  {Massey} R.,  {Kitching} T.,  {Taylor} A.,   {Tittley} E.,  2015,
  \mn@doi [Science] {10.1126/science.1261381}, \href
  {http://adsabs.harvard.edu/abs/2015Sci...347.1462H} {347, 1462}

\bibitem[\protect\citeauthoryear{{Hennig} et~al.,}{{Hennig}
  et~al.}{2017}]{hennig17}
{Hennig} C.,  et~al., 2017, \mn@doi [\mnras] {10.1093/mnras/stx175}, \href
  {https://ui.adsabs.harvard.edu/abs/2017MNRAS.467.4015H} {467, 4015}

\bibitem[\protect\citeauthoryear{{High}, {Stubbs}, {Rest}, {Stalder}  \&
  {Challis}}{{High} et~al.}{2009}]{High2009}
{High} F.~W.,  {Stubbs} C.~W.,  {Rest} A.,  {Stalder} B.,   {Challis} P.,
  2009, \mn@doi [\aj] {10.1088/0004-6256/138/1/110}, \href
  {https://ui.adsabs.harvard.edu/abs/2009AJ....138..110H} {138, 110}

\bibitem[\protect\citeauthoryear{{High} et~al.,}{{High}
  et~al.}{2012}]{High2012}
{High} F.~W.,  et~al., 2012, \mn@doi [\apj] {10.1088/0004-637X/758/1/68}, \href
  {https://ui.adsabs.harvard.edu/abs/2012ApJ...758...68H} {758, 68}

\bibitem[\protect\citeauthoryear{{Hinton}, {Davis}, {Lidman}, {Glazebrook}  \&
  {Lewis}}{{Hinton} et~al.}{2016}]{Hinton2016}
{Hinton} S.~R.,  {Davis} T.~M.,  {Lidman} C.,  {Glazebrook} K.,   {Lewis}
  G.~F.,  2016, \mn@doi [Astronomy and Computing]
  {10.1016/j.ascom.2016.03.001}, \href
  {http://adsabs.harvard.edu/abs/2016A%26C....15...61H} {15, 61}

\bibitem[\protect\citeauthoryear{{Hou}, {Parker}, {Harris}  \& {Wilman}}{{Hou}
  et~al.}{2009}]{Hou2009}
{Hou} A.,  {Parker} L.~C.,  {Harris} W.~E.,   {Wilman} D.~J.,  2009, \mn@doi
  [\apj] {10.1088/0004-637X/702/2/1199}, \href
  {http://adsabs.harvard.edu/abs/2009ApJ...702.1199H} {702, 1199}

\bibitem[\protect\citeauthoryear{{Hou} et~al.,}{{Hou} et~al.}{2012}]{Hou2012}
{Hou} A.,  et~al., 2012, \mn@doi [\mnras] {10.1111/j.1365-2966.2012.20586.x},
  \href {https://ui.adsabs.harvard.edu/abs/2012MNRAS.421.3594H} {421, 3594}

\bibitem[\protect\citeauthoryear{{Kalita} \& {Ebeling}}{{Kalita} \&
  {Ebeling}}{2019}]{Kalita2019}
{Kalita} B.~S.,  {Ebeling} H.,  2019, \mn@doi [\apj]
  {10.3847/1538-4357/ab5184}, \href
  {https://ui.adsabs.harvard.edu/abs/2019ApJ...887..158K} {887, 158}

\bibitem[\protect\citeauthoryear{{Kapferer}, {Sluka}, {Schindler}, {Ferrari}
  \& {Ziegler}}{{Kapferer} et~al.}{2009}]{Kapferer2009}
{Kapferer} W.,  {Sluka} C.,  {Schindler} S.,  {Ferrari} C.,   {Ziegler} B.,
  2009, \mn@doi [\aap] {10.1051/0004-6361/200811551}, \href
  {https://ui.adsabs.harvard.edu/abs/2009A&A...499...87K} {499, 87}

\bibitem[\protect\citeauthoryear{{Kelkar} et~al.,}{{Kelkar}
  et~al.}{2020}]{Kelkar20}
{Kelkar} K.,  et~al., 2020, \mn@doi [\mnras] {10.1093/mnras/staa1547}, \href
  {https://ui.adsabs.harvard.edu/abs/2020MNRAS.496..442K} {496, 442}

\bibitem[\protect\citeauthoryear{{Kim}, {Peter}  \& {Wittman}}{{Kim}
  et~al.}{2017}]{kim2017}
{Kim} S.~Y.,  {Peter} A. H.~G.,   {Wittman} D.,  2017, \mn@doi [\mnras]
  {10.1093/mnras/stx896}, \href
  {https://ui.adsabs.harvard.edu/abs/2017MNRAS.469.1414K} {469, 1414}

\bibitem[\protect\citeauthoryear{{Komatsu} et~al.,}{{Komatsu}
  et~al.}{2011}]{Komatsu2011}
{Komatsu} E.,  et~al., 2011, \mn@doi [\apjs] {10.1088/0067-0049/192/2/18},
  \href {https://ui.adsabs.harvard.edu/abs/2011ApJS..192...18K} {192, 18}

\bibitem[\protect\citeauthoryear{{Koyama}, {Kodama}, {Nakata}, {Shimasaku}  \&
  {Okamura}}{{Koyama} et~al.}{2011}]{Koyama2011}
{Koyama} Y.,  {Kodama} T.,  {Nakata} F.,  {Shimasaku} K.,   {Okamura} S.,
  2011, \mn@doi [\apj] {10.1088/0004-637X/734/1/66}, \href
  {https://ui.adsabs.harvard.edu/abs/2011ApJ...734...66K} {734, 66}

\bibitem[\protect\citeauthoryear{{Lin}, {Mohr}  \& {Stanford}}{{Lin}
  et~al.}{2004}]{lin04a}
{Lin} Y.,  {Mohr} J.~J.,   {Stanford} S.~A.,  2004, \apj, \href
  {http://adsabs.harvard.edu/cgi-bin/nph-bib_query?bibcode=2004ApJ...610..745L&amp;db_key=AST}
  {610, 745}

\bibitem[\protect\citeauthoryear{{L{\'o}pez-Cruz}, {Barkhouse}  \&
  {Yee}}{{L{\'o}pez-Cruz} et~al.}{2004}]{lopez04}
{L{\'o}pez-Cruz} O.,  {Barkhouse} W.~A.,   {Yee} H.~K.~C.,  2004, \mn@doi
  [\apj] {10.1086/423664}, \href
  {http://adsabs.harvard.edu/abs/2004ApJ...614..679L} {614, 679}

\bibitem[\protect\citeauthoryear{{Ma}, {Ebeling}, {Marshall}  \&
  {Schrabback}}{{Ma} et~al.}{2010}]{ma10}
{Ma} C.-J.,  {Ebeling} H.,  {Marshall} P.,   {Schrabback} T.,  2010, \mn@doi
  [\mnras] {10.1111/j.1365-2966.2010.16673.x}, \href
  {http://adsabs.harvard.edu/abs/2010MNRAS.406..121M} {406, 121}

\bibitem[\protect\citeauthoryear{{Machado} \& {Lima Neto}}{{Machado} \& {Lima
  Neto}}{2013}]{Machado2013}
{Machado} R. E.~G.,  {Lima Neto} G.~B.,  2013, \mn@doi [\mnras]
  {10.1093/mnras/stt127}, \href
  {https://ui.adsabs.harvard.edu/abs/2013MNRAS.430.3249M} {430, 3249}

\bibitem[\protect\citeauthoryear{{Machado}, {Monteiro-Oliveira}, {Lima Neto}
  \& {Cypriano}}{{Machado} et~al.}{2015}]{Machado+2015}
{Machado} R.~E.~G.,  {Monteiro-Oliveira} R.,  {Lima Neto} G.~B.,   {Cypriano}
  E.~S.,  2015, \mn@doi [\mnras] {10.1093/mnras/stv1162}, \href
  {http://adsabs.harvard.edu/abs/2015MNRAS.451.3309M} {451, 3309}

\bibitem[\protect\citeauthoryear{{Mahler} et~al.,}{{Mahler}
  et~al.}{2020}]{Mahler2020}
{Mahler} G.,  et~al., 2020, \mn@doi [\apj] {10.3847/1538-4357/ab886b}, \href
  {https://ui.adsabs.harvard.edu/abs/2020ApJ...894..150M} {894, 150}

\bibitem[\protect\citeauthoryear{{Marinacci} et~al.,}{{Marinacci}
  et~al.}{2018}]{marinacci18}
{Marinacci} F.,  et~al., 2018, \mn@doi [\mnras] {10.1093/mnras/sty2206}, \href
  {https://ui.adsabs.harvard.edu/abs/2018MNRAS.480.5113M} {480, 5113}

\bibitem[\protect\citeauthoryear{{Markevitch} \& {Vikhlinin}}{{Markevitch} \&
  {Vikhlinin}}{2007}]{Markevitch2007}
{Markevitch} M.,  {Vikhlinin} A.,  2007, \mn@doi [\physrep]
  {10.1016/j.physrep.2007.01.001}, \href
  {https://ui.adsabs.harvard.edu/abs/2007PhR...443....1M} {443, 1}

\bibitem[\protect\citeauthoryear{{Markevitch}, {Gonzalez}, {Clowe},
  {Vikhlinin}, {Forman}, {Jones}, {Murray}  \& {Tucker}}{{Markevitch}
  et~al.}{2004}]{markevitch04}
{Markevitch} M.,  {Gonzalez} A.~H.,  {Clowe} D.,  {Vikhlinin} A.,  {Forman} W.,
   {Jones} C.,  {Murray} S.,   {Tucker} W.,  2004, \mn@doi [\apj]
  {10.1086/383178}, \href {http://adsabs.harvard.edu/abs/2004ApJ...606..819M}
  {606, 819}

\bibitem[\protect\citeauthoryear{{Massey} et~al.,}{{Massey}
  et~al.}{2015}]{Massey2015}
{Massey} R.,  et~al., 2015, \mn@doi [\mnras] {10.1093/mnras/stv467}, \href
  {https://ui.adsabs.harvard.edu/abs/2015MNRAS.449.3393M} {449, 3393}

\bibitem[\protect\citeauthoryear{{Mastropietro} \& {Burkert}}{{Mastropietro} \&
  {Burkert}}{2008}]{Mastropietro2008}
{Mastropietro} C.,  {Burkert} A.,  2008, \mn@doi [\mnras]
  {10.1111/j.1365-2966.2008.13626.x}, \href
  {https://ui.adsabs.harvard.edu/abs/2008MNRAS.389..967M} {389, 967}

\bibitem[\protect\citeauthoryear{{McDonald} et~al.,}{{McDonald}
  et~al.}{2013}]{mcdonald13}
{McDonald} M.,  et~al., 2013, \mn@doi [\apj] {10.1088/0004-637X/774/1/23},
  \href {http://adsabs.harvard.edu/abs/2013ApJ...774...23M} {774, 23}

\bibitem[\protect\citeauthoryear{{McDonald} et~al.,}{{McDonald}
  et~al.}{2017}]{McDonald2017}
{McDonald} M.,  et~al., 2017, \mn@doi [\apj] {10.3847/1538-4357/aa7740}, \href
  {https://ui.adsabs.harvard.edu/abs/2017ApJ...843...28M} {843, 28}

\bibitem[\protect\citeauthoryear{{McPartland}, {Ebeling}, {Roediger}  \&
  {Blumenthal}}{{McPartland} et~al.}{2016}]{mcpartland16}
{McPartland} C.,  {Ebeling} H.,  {Roediger} E.,   {Blumenthal} K.,  2016,
  \mn@doi [\mnras] {10.1093/mnras/stv2508}, \href
  {http://adsabs.harvard.edu/abs/2016MNRAS.455.2994M} {455, 2994}

\bibitem[\protect\citeauthoryear{{Menanteau} et~al.,}{{Menanteau}
  et~al.}{2012}]{Menanteau2012}
{Menanteau} F.,  et~al., 2012, \mn@doi [\apj] {10.1088/0004-637X/748/1/7},
  \href {https://ui.adsabs.harvard.edu/abs/2012ApJ...748....7M} {748, 7}

\bibitem[\protect\citeauthoryear{{Menci} \& {Fusco-Femiano}}{{Menci} \&
  {Fusco-Femiano}}{1996}]{Menci1996}
{Menci} N.,  {Fusco-Femiano} R.,  1996, \mn@doi [\apj] {10.1086/178040}, \href
  {https://ui.adsabs.harvard.edu/abs/1996ApJ...472...46M} {472, 46}

\bibitem[\protect\citeauthoryear{{Monteiro-Oliveira}, {Lima Neto}, {Cypriano},
  {Machado}, {Capelato}, {Lagan{\'a}}, {Durret}  \&
  {Bagchi}}{{Monteiro-Oliveira} et~al.}{2017}]{Monteiro2017}
{Monteiro-Oliveira} R.,  {Lima Neto} G.~B.,  {Cypriano} E.~S.,  {Machado}
  R.~E.~G.,  {Capelato} H.~V.,  {Lagan{\'a}} T.~F.,  {Durret} F.,   {Bagchi}
  J.,  2017, \mn@doi [\mnras] {10.1093/mnras/stx791}, \href
  {https://ui.adsabs.harvard.edu/abs/2017MNRAS.468.4566M} {468, 4566}

\bibitem[\protect\citeauthoryear{{Monteiro-Oliveira}, {Cypriano}, {Vitorelli},
  {Ribeiro}, {Sodr{\'e}}, {Dupke}  \& {Mendes de Oliveira}}{{Monteiro-Oliveira}
  et~al.}{2018}]{Monteiro2018}
{Monteiro-Oliveira} R.,  {Cypriano} E.~S.,  {Vitorelli} A.~Z.,  {Ribeiro}
  A.~L.~B.,  {Sodr{\'e}} L.,  {Dupke} R.,   {Mendes de Oliveira} C.,  2018,
  \mn@doi [\mnras] {10.1093/mnras/sty2349}, \href
  {http://adsabs.harvard.edu/abs/2018MNRAS.481.1097M} {481, 1097}

\bibitem[\protect\citeauthoryear{{Monteiro-Oliveira}, {Doubrawa}, {Machado},
  {Lima Neto}, {Castejon}  \& {Cypriano}}{{Monteiro-Oliveira}
  et~al.}{2020}]{Monteiro2020}
{Monteiro-Oliveira} R.,  {Doubrawa} L.,  {Machado} R.~E.~G.,  {Lima Neto}
  G.~B.,  {Castejon} M.,   {Cypriano} E.~S.,  2020, \mn@doi [\mnras]
  {10.1093/mnras/staa1218}, \href
  {https://ui.adsabs.harvard.edu/abs/2020MNRAS.495.2007M} {495, 2007}

\bibitem[\protect\citeauthoryear{{Monteiro-Oliveira}, {Soja}, {Ribeiro},
  {Bagchi}, {Sankhyayan}, {Candido}  \& {Flores}}{{Monteiro-Oliveira}
  et~al.}{2021}]{Monteiro-Oliveira21}
{Monteiro-Oliveira} R.,  {Soja} A.~C.,  {Ribeiro} A.~L.~B.,  {Bagchi} J.,
  {Sankhyayan} S.,  {Candido} T.~O.,   {Flores} R.~R.,  2021, \mn@doi [\mnras]
  {10.1093/mnras/staa3575}, \href
  {https://ui.adsabs.harvard.edu/abs/2021MNRAS.501..756M} {501, 756}

\bibitem[\protect\citeauthoryear{{Moran}, {Ellis}, {Treu}, {Smith}, {Rich}  \&
  {Smail}}{{Moran} et~al.}{2007}]{Moran2007}
{Moran} S.~M.,  {Ellis} R.~S.,  {Treu} T.,  {Smith} G.~P.,  {Rich} R.~M.,
  {Smail} I.,  2007, \mn@doi [\apj] {10.1086/522303}, \href
  {https://ui.adsabs.harvard.edu/abs/2007ApJ...671.1503M} {671, 1503}

\bibitem[\protect\citeauthoryear{Morganson et~al.,}{Morganson
  et~al.}{2018}]{morganson18}
Morganson E.,  et~al., 2018, \mn@doi [\pasp] {10.1088/1538-3873/aab4ef}, 130,
  074501

\bibitem[\protect\citeauthoryear{{Moura}, {Machado}  \&
  {Monteiro-Oliveira}}{{Moura} et~al.}{2021}]{Moura21}
{Moura} M.~T.,  {Machado} R. E.~G.,   {Monteiro-Oliveira} R.,  2021, \mn@doi
  [\mnras] {10.1093/mnras/staa3399}, \href
  {https://ui.adsabs.harvard.edu/abs/2021MNRAS.500.1858M} {500, 1858}

\bibitem[\protect\citeauthoryear{{Munari}, {Biviano}, {Borgani}, {Murante}  \&
  {Fabjan}}{{Munari} et~al.}{2013}]{Munari2013}
{Munari} E.,  {Biviano} A.,  {Borgani} S.,  {Murante} G.,   {Fabjan} D.,  2013,
  \mn@doi [\mnras] {10.1093/mnras/stt049}, \href
  {https://ui.adsabs.harvard.edu/abs/2013MNRAS.430.2638M} {430, 2638}

\bibitem[\protect\citeauthoryear{Naiman et~al.,}{Naiman
  et~al.}{2018}]{naiman18}
Naiman J.~P.,  et~al., 2018, \mn@doi [\mnras] {10.1093/mnras/sty618}, 477,
  1206–1224

\bibitem[\protect\citeauthoryear{{Navarro}, {Frenk}  \& {White}}{{Navarro}
  et~al.}{1996}]{NFW0}
{Navarro} J.~F.,  {Frenk} C.~S.,   {White} S. D.~M.,  1996, \mn@doi [\apj]
  {10.1086/177173}, \href
  {https://ui.adsabs.harvard.edu/abs/1996ApJ...462..563N} {462, 563}

\bibitem[\protect\citeauthoryear{{Navarro}, {Frenk}  \& {White}}{{Navarro}
  et~al.}{1997}]{NFW1}
{Navarro} J.~F.,  {Frenk} C.~S.,   {White} S. D.~M.,  1997, \mn@doi [\apj]
  {10.1086/304888}, \href
  {https://ui.adsabs.harvard.edu/abs/1997ApJ...490..493N} {490, 493}

\bibitem[\protect\citeauthoryear{{Nelson}, {Rudd}, {Shaw}  \& {Nagai}}{{Nelson}
  et~al.}{2012}]{nelson12}
{Nelson} K.,  {Rudd} D.~H.,  {Shaw} L.,   {Nagai} D.,  2012, \mn@doi [\apj]
  {10.1088/0004-637X/751/2/121}, \href
  {http://adsabs.harvard.edu/abs/2012ApJ...751..121N} {751, 121}

\bibitem[\protect\citeauthoryear{{Nelson}, {Lau}, {Nagai}, {Rudd}  \&
  {Yu}}{{Nelson} et~al.}{2014}]{nelson14}
{Nelson} K.,  {Lau} E.~T.,  {Nagai} D.,  {Rudd} D.~H.,   {Yu} L.,  2014,
  \mn@doi [\apj] {10.1088/0004-637X/782/2/107}, \href
  {http://adsabs.harvard.edu/abs/2014ApJ...782..107N} {782, 107}

\bibitem[\protect\citeauthoryear{Nelson et~al.,}{Nelson
  et~al.}{2017}]{nelson17}
Nelson D.,  et~al., 2017, \mn@doi [\mnras] {10.1093/mnras/stx3040}, 475,
  624–647

\bibitem[\protect\citeauthoryear{{Ng}, {Dawson}, {Wittman}, {Jee}, {Hughes},
  {Menanteau}  \& {Sif{\'o}n}}{{Ng} et~al.}{2015}]{Ng2015}
{Ng} K.~Y.,  {Dawson} W.~A.,  {Wittman} D.,  {Jee} M.~J.,  {Hughes} J.~P.,
  {Menanteau} F.,   {Sif{\'o}n} C.,  2015, \mn@doi [\mnras]
  {10.1093/mnras/stv1713}, \href
  {https://ui.adsabs.harvard.edu/abs/2015MNRAS.453.1531N} {453, 1531}

\bibitem[\protect\citeauthoryear{{Nurgaliev} et~al.,}{{Nurgaliev}
  et~al.}{2017}]{Nurgaliev2017}
{Nurgaliev} D.,  et~al., 2017, \mn@doi [\apj] {10.3847/1538-4357/aa6db4}, \href
  {https://ui.adsabs.harvard.edu/abs/2017ApJ...841....5N} {841, 5}

\bibitem[\protect\citeauthoryear{{Ochsenbein}, {Bauer}  \&
  {Marcout}}{{Ochsenbein} et~al.}{2000}]{Vizier}
{Ochsenbein} F.,  {Bauer} P.,   {Marcout} J.,  2000, \mn@doi [\aaps]
  {10.1051/aas:2000169}, \href
  {https://ui.adsabs.harvard.edu/abs/2000A&AS..143...23O} {143, 23}

\bibitem[\protect\citeauthoryear{{Olave-Rojas}, {Cerulo}, {Demarco},
  {Jaff{\'e}}, {Mercurio}, {Rosati}, {Balestra}  \& {Nonino}}{{Olave-Rojas}
  et~al.}{2018}]{Olave2018}
{Olave-Rojas} D.,  {Cerulo} P.,  {Demarco} R.,  {Jaff{\'e}} Y.~L.,  {Mercurio}
  A.,  {Rosati} P.,  {Balestra} I.,   {Nonino} M.,  2018, \mn@doi [\mnras]
  {10.1093/mnras/sty1669}, \href
  {https://ui.adsabs.harvard.edu/abs/2018MNRAS.479.2328O} {479, 2328}

\bibitem[\protect\citeauthoryear{{Owers}, {Couch}, {Nulsen}  \&
  {Randall}}{{Owers} et~al.}{2012}]{Owers2012}
{Owers} M.~S.,  {Couch} W.~J.,  {Nulsen} P. E.~J.,   {Randall} S.~W.,  2012,
  \mn@doi [\apjl] {10.1088/2041-8205/750/1/L23}, \href
  {https://ui.adsabs.harvard.edu/abs/2012ApJ...750L..23O} {750, L23}

\bibitem[\protect\citeauthoryear{{Paccagnella}, {Vulcani}, {Poggianti},
  {Moretti}, {Fritz}, {Gullieuszik}  \& {Fasano}}{{Paccagnella}
  et~al.}{2019}]{Paccagnella2019}
{Paccagnella} A.,  {Vulcani} B.,  {Poggianti} B.~M.,  {Moretti} A.,  {Fritz}
  J.,  {Gullieuszik} M.,   {Fasano} G.,  2019, \mn@doi [\mnras]
  {10.1093/mnras/sty2728}, \href
  {https://ui.adsabs.harvard.edu/abs/2019MNRAS.482..881P} {482, 881}

\bibitem[\protect\citeauthoryear{{Pallero}, {G{\'o}mez}, {Padilla}, {Bah{\'e}},
  {Vega-Mart{\'\i}nez}  \& {Torres-Flores}}{{Pallero}
  et~al.}{2022}]{Pallero2021}
{Pallero} D.,  {G{\'o}mez} F.~A.,  {Padilla} N.~D.,  {Bah{\'e}} Y.~M.,
  {Vega-Mart{\'\i}nez} C.~A.,   {Torres-Flores} S.,  2022, \mn@doi [\mnras]
  {10.1093/mnras/stab3318}, \href
  {https://ui.adsabs.harvard.edu/abs/2022MNRAS.511.3210P} {511, 3210}

\bibitem[\protect\citeauthoryear{{Paulino-Afonso}, {Sobral}, {Darvish},
  {Ribeiro}, {Smail}, {Best}, {Stroe}  \& {Cairns}}{{Paulino-Afonso}
  et~al.}{2020}]{Paulino2019}
{Paulino-Afonso} A.,  {Sobral} D.,  {Darvish} B.,  {Ribeiro} B.,  {Smail} I.,
  {Best} P.,  {Stroe} A.,   {Cairns} J.,  2020, \mn@doi [\aap]
  {10.1051/0004-6361/201834244}, \href
  {https://ui.adsabs.harvard.edu/abs/2020A&A...633A..70P} {633, A70}

\bibitem[\protect\citeauthoryear{Pillepich et~al.,}{Pillepich
  et~al.}{2017}]{pillepich17}
Pillepich A.,  et~al., 2017, \mn@doi [\mnras] {10.1093/mnras/stx3112}, 475,
  648–675

\bibitem[\protect\citeauthoryear{{Pinkney}, {Roettiger}, {Burns}  \&
  {Bird}}{{Pinkney} et~al.}{1996}]{Pinkney1996}
{Pinkney} J.,  {Roettiger} K.,  {Burns} J.~O.,   {Bird} C.~M.,  1996, \mn@doi
  [\apjs] {10.1086/192290}, \href
  {http://adsabs.harvard.edu/abs/1996ApJS..104....1P} {104, 1}

\bibitem[\protect\citeauthoryear{{Poggianti}, {Bridges}, {Komiyama}, {Yagi},
  {Carter}, {Mobasher}, {Okamura}  \& {Kashikawa}}{{Poggianti}
  et~al.}{2004}]{Poggianti2004}
{Poggianti} B.~M.,  {Bridges} T.~J.,  {Komiyama} Y.,  {Yagi} M.,  {Carter} D.,
  {Mobasher} B.,  {Okamura} S.,   {Kashikawa} N.,  2004, \mn@doi [\apj]
  {10.1086/380195}, \href
  {https://ui.adsabs.harvard.edu/abs/2004ApJ...601..197P} {601, 197}

\bibitem[\protect\citeauthoryear{{Poggianti} et~al.,}{{Poggianti}
  et~al.}{2016}]{poggianti16}
{Poggianti} B.~M.,  et~al., 2016, \mn@doi [\aj] {10.3847/0004-6256/151/3/78},
  \href {http://adsabs.harvard.edu/abs/2016AJ....151...78P} {151, 78}

\bibitem[\protect\citeauthoryear{{Poole}, {Fardal}, {Babul}, {McCarthy},
  {Quinn}  \& {Wadsley}}{{Poole} et~al.}{2006}]{Poole2006}
{Poole} G.~B.,  {Fardal} M.~A.,  {Babul} A.,  {McCarthy} I.~G.,  {Quinn} T.,
  {Wadsley} J.,  2006, \mn@doi [\mnras] {10.1111/j.1365-2966.2006.10916.x},
  \href {https://ui.adsabs.harvard.edu/abs/2006MNRAS.373..881P} {373, 881}

\bibitem[\protect\citeauthoryear{{Pranger} et~al.,}{{Pranger}
  et~al.}{2013}]{Pranger2013}
{Pranger} F.,  et~al., 2013, \mn@doi [\aap] {10.1051/0004-6361/201321929},
  \href {https://ui.adsabs.harvard.edu/abs/2013A&A...557A..62P} {557, A62}

\bibitem[\protect\citeauthoryear{{Pranger}, {B{\"o}hm}, {Ferrari},
  {Maurogordato}, {Benoist}, {H{\"o}ller}  \& {Schindler}}{{Pranger}
  et~al.}{2014}]{pranger14}
{Pranger} F.,  {B{\"o}hm} A.,  {Ferrari} C.,  {Maurogordato} S.,  {Benoist} C.,
   {H{\"o}ller} H.,   {Schindler} S.,  2014, \mn@doi [\aap]
  {10.1051/0004-6361/201424727}, \href
  {http://adsabs.harvard.edu/abs/2014A%26A...570A..40P} {570, A40}

\bibitem[\protect\citeauthoryear{Quintana, Carrasco  \& Reisenegger}{Quintana
  et~al.}{2000}]{Quintana_2000}
Quintana H.,  Carrasco E.~R.,   Reisenegger A.,  2000, \mn@doi [\aj]
  {10.1086/301476}, 120, 511

\bibitem[\protect\citeauthoryear{{Raouf}, {Smith}, {Khosroshahi}, {Dariush},
  {Driver}, {Ko}  \& {Hwang}}{{Raouf} et~al.}{2019}]{Raouf2019R}
{Raouf} M.,  {Smith} R.,  {Khosroshahi} H.~G.,  {Dariush} A.~A.,  {Driver} S.,
  {Ko} J.,   {Hwang} H.~S.,  2019, \mn@doi [\apj] {10.3847/1538-4357/ab5581},
  \href {https://ui.adsabs.harvard.edu/abs/2019ApJ...887..264R} {887, 264}

\bibitem[\protect\citeauthoryear{{Ribeiro}, {Lopes}  \& {Rembold}}{{Ribeiro}
  et~al.}{2013}]{ribeiro13}
{Ribeiro} A.~L.~B.,  {Lopes} P.~A.~A.,   {Rembold} S.~B.,  2013, \mn@doi [\aap]
  {10.1051/0004-6361/201220801}, \href
  {http://adsabs.harvard.edu/abs/2013A%26A...556A..74R} {556, A74}

\bibitem[\protect\citeauthoryear{{Rines}, {Geller}, {Diaferio}  \&
  {Kurtz}}{{Rines} et~al.}{2013}]{Rines2013}
{Rines} K.,  {Geller} M.~J.,  {Diaferio} A.,   {Kurtz} M.~J.,  2013, \mn@doi
  [\apj] {10.1088/0004-637X/767/1/15}, \href
  {https://ui.adsabs.harvard.edu/abs/2013ApJ...767...15R} {767, 15}

\bibitem[\protect\citeauthoryear{{Ruel} et~al.,}{{Ruel}
  et~al.}{2014}]{Ruel2014}
{Ruel} J.,  et~al., 2014, \mn@doi [\apj] {10.1088/0004-637X/792/1/45}, \href
  {https://ui.adsabs.harvard.edu/abs/2014ApJ...792...45R} {792, 45}

\bibitem[\protect\citeauthoryear{{Sarazin}}{{Sarazin}}{2002}]{Sarazin2002}
{Sarazin} C.~L.,  2002, in {Feretti} L.,  {Gioia} I.~M.,   {Giovannini} G.,
  eds,  Astrophysics and Space Science Library Vol. 272, Merging Processes in
  Galaxy Clusters. pp 1--38 (\mn@eprint {arXiv} {astro-ph/0105418}),
  \mn@doi{10.1007/0-306-48096-4_1}

\bibitem[\protect\citeauthoryear{{Sarazin}}{{Sarazin}}{2004}]{Sarazin2004}
{Sarazin} C.~L.,  2004, \mn@doi [Journal of Korean Astronomical Society]
  {10.5303/JKAS.2004.37.5.433}, \href
  {https://ui.adsabs.harvard.edu/abs/2004JKAS...37..433S} {37, 433}

\bibitem[\protect\citeauthoryear{{Saro}, {Mohr}, {Bazin}  \& {Dolag}}{{Saro}
  et~al.}{2013}]{Saro2013}
{Saro} A.,  {Mohr} J.~J.,  {Bazin} G.,   {Dolag} K.,  2013, \mn@doi [\apj]
  {10.1088/0004-637X/772/1/47}, \href
  {https://ui.adsabs.harvard.edu/abs/2013ApJ...772...47S} {772, 47}

\bibitem[\protect\citeauthoryear{{Schindler} \& {Muller}}{{Schindler} \&
  {Muller}}{1993}]{Schindler1993}
{Schindler} S.,  {Muller} E.,  1993, \aap, \href
  {https://ui.adsabs.harvard.edu/abs/1993A&A...272..137S} {272, 137}

\bibitem[\protect\citeauthoryear{{Sheen}, {Yi}, {Ree}, {Jaff{\'e}}, {Demarco}
  \& {Treister}}{{Sheen} et~al.}{2016}]{Sheen2016}
{Sheen} Y.-K.,  {Yi} S.~K.,  {Ree} C.~H.,  {Jaff{\'e}} Y.,  {Demarco} R.,
  {Treister} E.,  2016, \mn@doi [\apj] {10.3847/0004-637X/827/1/32}, \href
  {https://ui.adsabs.harvard.edu/abs/2016ApJ...827...32S} {827, 32}

\bibitem[\protect\citeauthoryear{{Skrutskie} et~al.,}{{Skrutskie}
  et~al.}{2006}]{Skrutskie2006}
{Skrutskie} M.~F.,  et~al., 2006, \mn@doi [\aj] {10.1086/498708}, \href
  {https://ui.adsabs.harvard.edu/abs/2006AJ....131.1163S} {131, 1163}

\bibitem[\protect\citeauthoryear{{Sobral}, {Stroe}, {Koyama}, {Darvish},
  {Calhau}, {Afonso}, {Kodama}  \& {Nakata}}{{Sobral}
  et~al.}{2016}]{Sobral2016}
{Sobral} D.,  {Stroe} A.,  {Koyama} Y.,  {Darvish} B.,  {Calhau} J.,  {Afonso}
  A.,  {Kodama} T.,   {Nakata} F.,  2016, \mn@doi [\mnras]
  {10.1093/mnras/stw534}, \href
  {https://ui.adsabs.harvard.edu/abs/2016MNRAS.458.3443S} {458, 3443}

\bibitem[\protect\citeauthoryear{{Song} et~al.,}{{Song} et~al.}{2012}]{song12b}
{Song} J.,  et~al., 2012, \mn@doi [\apj] {10.1088/0004-637X/761/1/22}, \href
  {http://adsabs.harvard.edu/abs/2012ApJ...761...22S} {761, 22}

\bibitem[\protect\citeauthoryear{Springel et~al.,}{Springel
  et~al.}{2017}]{springel17}
Springel V.,  et~al., 2017, \mn@doi [\mnras] {10.1093/mnras/stx3304}, 475,
  676–698

\bibitem[\protect\citeauthoryear{{Stoehr} et~al.,}{{Stoehr}
  et~al.}{2007}]{Stoehr2007}
{Stoehr} F.,  et~al., 2007, Space Telescope European Coordinating Facility
  Newsletter, \href {https://ui.adsabs.harvard.edu/abs/2007STECF..42....4S}
  {42, 4}

\bibitem[\protect\citeauthoryear{{Stroe} \& {Sobral}}{{Stroe} \&
  {Sobral}}{2021}]{stroe21}
{Stroe} A.,  {Sobral} D.,  2021, \mn@doi [\apj] {10.3847/1538-4357/abe7f8},
  \href {https://ui.adsabs.harvard.edu/abs/2021ApJ...912...55S} {912, 55}

\bibitem[\protect\citeauthoryear{{Stroe}, {Sobral}, {R{\"o}ttgering}  \& {van
  Weeren}}{{Stroe} et~al.}{2014}]{Stroe2014}
{Stroe} A.,  {Sobral} D.,  {R{\"o}ttgering} H. J.~A.,   {van Weeren} R.~J.,
  2014, \mn@doi [\mnras] {10.1093/mnras/stt2286}, \href
  {https://ui.adsabs.harvard.edu/abs/2014MNRAS.438.1377S} {438, 1377}

\bibitem[\protect\citeauthoryear{{Stroe} et~al.,}{{Stroe}
  et~al.}{2015}]{Stroe2015}
{Stroe} A.,  et~al., 2015, \mn@doi [\mnras] {10.1093/mnras/stu2519}, \href
  {https://ui.adsabs.harvard.edu/abs/2015MNRAS.450..646S} {450, 646}

\bibitem[\protect\citeauthoryear{{Stroe}, {Sobral}, {Paulino-Afonso}, {Alegre},
  {Calhau}, {Santos}  \& {van Weeren}}{{Stroe} et~al.}{2017}]{stroe17}
{Stroe} A.,  {Sobral} D.,  {Paulino-Afonso} A.,  {Alegre} L.,  {Calhau} J.,
  {Santos} S.,   {van Weeren} R.,  2017, \mn@doi [\mnras]
  {10.1093/mnras/stw2939}, \href
  {http://adsabs.harvard.edu/abs/2017MNRAS.465.2916S} {465, 2916}

\bibitem[\protect\citeauthoryear{{Takizawa}, {Nagino}  \&
  {Matsushita}}{{Takizawa} et~al.}{2010}]{takizawa10}
{Takizawa} M.,  {Nagino} R.,   {Matsushita} K.,  2010, \mn@doi [\pasj]
  {10.1093/pasj/62.4.951}, \href
  {http://adsabs.harvard.edu/abs/2010PASJ...62..951T} {62, 951}

\bibitem[\protect\citeauthoryear{{Thompson}, {Dav{\'e}}  \&
  {Nagamine}}{{Thompson} et~al.}{2015}]{thompson15}
{Thompson} R.,  {Dav{\'e}} R.,   {Nagamine} K.,  2015, \mn@doi [\mnras]
  {10.1093/mnras/stv1433}, \href
  {http://adsabs.harvard.edu/abs/2015MNRAS.452.3030T} {452, 3030}

\bibitem[\protect\citeauthoryear{{Tonry} \& {Davis}}{{Tonry} \&
  {Davis}}{1979}]{Tonry1979}
{Tonry} J.,  {Davis} M.,  1979, \mn@doi [\aj] {10.1086/112569}, \href
  {https://ui.adsabs.harvard.edu/abs/1979AJ.....84.1511T} {84, 1511}

\bibitem[\protect\citeauthoryear{{Toomre} \& {Toomre}}{{Toomre} \&
  {Toomre}}{1972}]{ToomreToomre72}
{Toomre} A.,  {Toomre} J.,  1972, \mn@doi [\apj] {10.1086/151823}, \href
  {http://adsabs.harvard.edu/abs/1972ApJ...178..623T} {178, 623}

\bibitem[\protect\citeauthoryear{{Treu}, {Ellis}, {Kneib}, {Dressler}, {Smail},
  {Czoske}, {Oemler}  \& {Natarajan}}{{Treu} et~al.}{2003}]{treu03}
{Treu} T.,  {Ellis} R.~S.,  {Kneib} J.-P.,  {Dressler} A.,  {Smail} I.,
  {Czoske} O.,  {Oemler} A.,   {Natarajan} P.,  2003, \mn@doi [\apj]
  {10.1086/375314}, \href {http://adsabs.harvard.edu/abs/2003ApJ...591...53T}
  {591, 53}

\bibitem[\protect\citeauthoryear{{Vogt}, {Haynes}, {Giovanelli}  \&
  {Herter}}{{Vogt} et~al.}{2004}]{Vogt2004}
{Vogt} N.~P.,  {Haynes} M.~P.,  {Giovanelli} R.,   {Herter} T.,  2004, \mn@doi
  [\aj] {10.1086/420702}, \href
  {https://ui.adsabs.harvard.edu/abs/2004AJ....127.3300V} {127, 3300}

\bibitem[\protect\citeauthoryear{{Weilbacher}, {Streicher}, {Urrutia},
  {P{\'e}contal-Rousset}, {Jarno}  \& {Bacon}}{{Weilbacher}
  et~al.}{2014}]{weilbacher14}
{Weilbacher} P.~M.,  {Streicher} O.,  {Urrutia} T.,  {P{\'e}contal-Rousset} A.,
   {Jarno} A.,   {Bacon} R.,  2014, in {Manset} N.,  {Forshay} P.,  eds,
  Astronomical Society of the Pacific Conference Series Vol. 485, Astronomical
  Data Analysis Software and Systems XXIII. p.~451 (\mn@eprint {arXiv}
  {1507.00034})

\bibitem[\protect\citeauthoryear{{Weilbacher}, {Streicher}  \&
  {Palsa}}{{Weilbacher} et~al.}{2016}]{Weilbacher2016}
{Weilbacher} P.~M.,  {Streicher} O.,   {Palsa} R.,  2016, {MUSE-DRP: MUSE Data
  Reduction Pipeline}, Astrophysics Source Code Library, record ascl:1610.004
  (\mn@eprint {ascl} {1610.004})

\bibitem[\protect\citeauthoryear{{White}}{{White}}{1976}]{White1976}
{White} S.~D.~M.,  1976, \mn@doi [\mnras] {10.1093/mnras/174.1.19}, \href
  {https://ui.adsabs.harvard.edu/abs/1976MNRAS.174...19W} {174, 19}

\bibitem[\protect\citeauthoryear{{White} et~al.,}{{White}
  et~al.}{2015}]{White2015}
{White} J.~A.,  et~al., 2015, \mn@doi [\mnras] {10.1093/mnras/stv1831}, \href
  {http://adsabs.harvard.edu/abs/2015MNRAS.453.2718W} {453, 2718}

\bibitem[\protect\citeauthoryear{{Williamson} et~al.,}{{Williamson}
  et~al.}{2011}]{williamson11}
{Williamson} R.,  et~al., 2011, \mn@doi [\apj] {10.1088/0004-637X/738/2/139},
  \href {http://adsabs.harvard.edu/abs/2011ApJ...738..139W} {738, 139}

\bibitem[\protect\citeauthoryear{{Wittman}}{{Wittman}}{2019}]{Wittman2019}
{Wittman} D.,  2019, \mn@doi [\apj] {10.3847/1538-4357/ab3052}, \href
  {https://ui.adsabs.harvard.edu/abs/2019ApJ...881..121W} {881, 121}

\bibitem[\protect\citeauthoryear{{Wittman}, {Cornell}  \& {Nguyen}}{{Wittman}
  et~al.}{2018}]{Wittman2018}
{Wittman} D.,  {Cornell} B.~H.,   {Nguyen} J.,  2018, \mn@doi [\apj]
  {10.3847/1538-4357/aacf3e}, \href
  {https://ui.adsabs.harvard.edu/abs/2018ApJ...862..160W} {862, 160}

\bibitem[\protect\citeauthoryear{{Ye}, {Guo}, {Zheng}  \& {Zehavi}}{{Ye}
  et~al.}{2017}]{Ye2017}
{Ye} J.-N.,  {Guo} H.,  {Zheng} Z.,   {Zehavi} I.,  2017, \mn@doi [\apj]
  {10.3847/1538-4357/aa70e7}, \href
  {https://ui.adsabs.harvard.edu/abs/2017ApJ...841...45Y} {841, 45}

\bibitem[\protect\citeauthoryear{{Yoon} \& {Im}}{{Yoon} \&
  {Im}}{2020}]{Yoon2020}
{Yoon} Y.,  {Im} M.,  2020, \mn@doi [\apj] {10.3847/1538-4357/ab8008}, \href
  {https://ui.adsabs.harvard.edu/abs/2020ApJ...893..117Y} {893, 117}

\bibitem[\protect\citeauthoryear{{Yoon}, {Im}, {Lee}, {Lee}  \& {Lim}}{{Yoon}
  et~al.}{2019}]{Yoon2019}
{Yoon} Y.,  {Im} M.,  {Lee} G.-H.,  {Lee} S.-K.,   {Lim} G.,  2019, \mn@doi
  [Nature Astronomy] {10.1038/s41550-019-0799-7}, \href
  {https://ui.adsabs.harvard.edu/abs/2019NatAs...3..844Y} {3, 844}

\bibitem[\protect\citeauthoryear{{Yoshikawa}, {Jing}  \&
  {B{\"o}rner}}{{Yoshikawa} et~al.}{2003}]{Yoshikawa2003}
{Yoshikawa} K.,  {Jing} Y.~P.,   {B{\"o}rner} G.,  2003, \mn@doi [\apj]
  {10.1086/375148}, \href
  {https://ui.adsabs.harvard.edu/abs/2003ApJ...590..654Y} {590, 654}

\bibitem[\protect\citeauthoryear{{Zenteno} et~al.,}{{Zenteno}
  et~al.}{2011}]{zenteno11}
{Zenteno} A.,  et~al., 2011, \mn@doi [\apj] {10.1088/0004-637X/734/1/3}, \href
  {http://adsabs.harvard.edu/abs/2011ApJ...734....3Z} {734, 3}

\bibitem[\protect\citeauthoryear{{Zenteno} et~al.,}{{Zenteno}
  et~al.}{2016}]{zenteno16}
{Zenteno} A.,  et~al., 2016, \mn@doi [\mnras] {10.1093/mnras/stw1649}, \href
  {http://adsabs.harvard.edu/abs/2016MNRAS.462..830Z} {462, 830}

\bibitem[\protect\citeauthoryear{{Zenteno} et~al.,}{{Zenteno}
  et~al.}{2020}]{zenteno2020}
{Zenteno} A.,  et~al., 2020, \mn@doi [\mnras] {10.1093/mnras/staa1157}, \href
  {https://ui.adsabs.harvard.edu/abs/2020MNRAS.495..705Z} {495, 705}

\bibitem[\protect\citeauthoryear{{Zhang}, {Verdugo}, {Klein}  \&
  {Schneider}}{{Zhang} et~al.}{2012}]{Zhang2012}
{Zhang} Y.-Y.,  {Verdugo} M.,  {Klein} M.,   {Schneider} P.,  2012, \mn@doi
  [\aap] {10.1051/0004-6361/201218979}, \href
  {http://adsabs.harvard.edu/abs/2012A%26A...542A.106Z} {542, A106}

\bibitem[\protect\citeauthoryear{{ZuHone}}{{ZuHone}}{2011}]{ZuHone2011}
{ZuHone} J.~A.,  2011, \mn@doi [\apj] {10.1088/0004-637X/728/1/54}, \href
  {https://ui.adsabs.harvard.edu/abs/2011ApJ...728...54Z} {728, 54}

\bibitem[\protect\citeauthoryear{{ZuHone}, {Kowalik}, {{\"O}hman}, {Lau}  \&
  {Nagai}}{{ZuHone} et~al.}{2018}]{ZuHone2018}
{ZuHone} J.~A.,  {Kowalik} K.,  {{\"O}hman} E.,  {Lau} E.,   {Nagai} D.,  2018,
  \mn@doi [\apjs] {10.3847/1538-4365/aa99db}, \href
  {https://ui.adsabs.harvard.edu/abs/2018ApJS..234....4Z} {234, 4}

\bibitem[\protect\citeauthoryear{{van Weeren}, {Br{\"u}ggen}, {R{\"o}ttgering}
  \& {Hoeft}}{{van Weeren} et~al.}{2011}]{vanWeeren2011}
{van Weeren} R.~J.,  {Br{\"u}ggen} M.,  {R{\"o}ttgering} H.~J.~A.,   {Hoeft}
  M.,  2011, \mn@doi [\mnras] {10.1111/j.1365-2966.2011.19478.x}, \href
  {https://ui.adsabs.harvard.edu/abs/2011MNRAS.418..230V} {418, 230}

\makeatother
\end{thebibliography}


\appendix

\section{Completeness of MUSE catalog}
\label{sec:completeness}

Since our aim is to look at the properties of the galaxy population, we need to first characterise a limiting magnitude to define that population. Fig.~\ref{fig:cmd} shows that the population of spectroscopic RS galaxies stops at $i_{\rm auto} \approx 22.8$, with blue galaxies going as deep as $i_{\rm auto} \approx 23.3$. In order to find out the limiting magnitude we want to use, we compare our photometric catalog inside the cubes footprints within magnitude bins, checking the fraction of spectroscopically confirmed galaxies within each bin. This check allows us to (1) validate our method for selecting RCS members, which will become important when looking for substructures (see $\S$\ref{sec:subcluster}), and (2) to look for potential cluster members not found by \textsc{MARZ}.

\begin{figure}
    \centering
    \includegraphics[width=\linewidth]{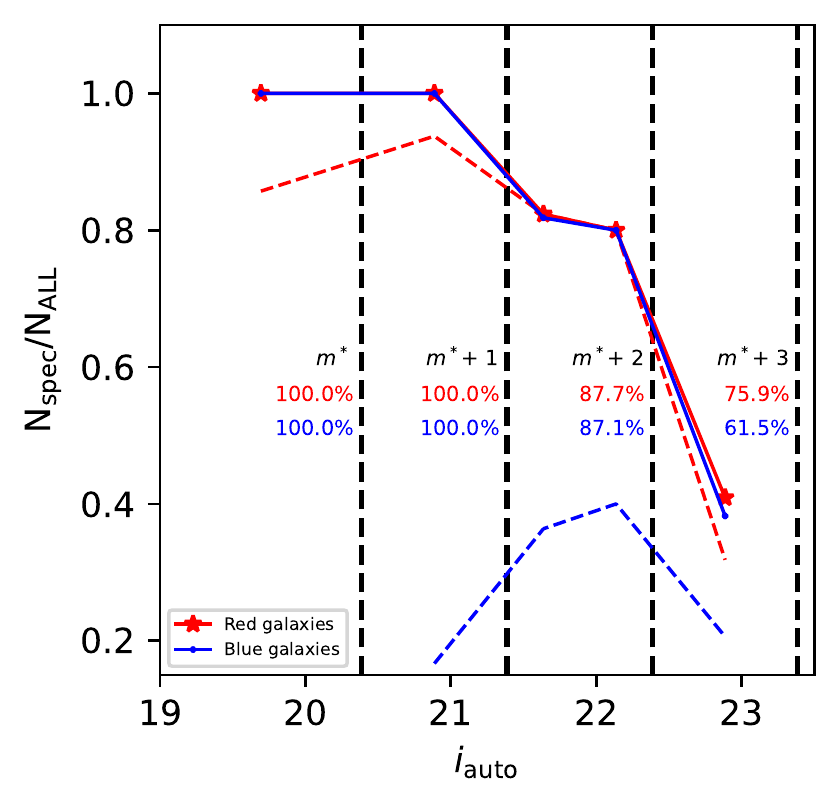}
    \vskip-0.13in
    \caption{Ratio of the spectroscopically confirmed members with respect to the galaxies from our catalog (photometrically and spectroscopically selected) at different bins of magnitudes. Continuous lines show the completeness of all the MUSE catalog with measured redshifts, while dashed lines are only those galaxies identified as cluster members. Lines are color coded according to the galaxy population. Black dashed lines denote the limits for $m^*$, $m^*+1$, $m^*+2$ and $m^*+3$, with the percentages being the accumulated completeness for a given limit of the MUSE catalog.}
    \label{fig:completeness}
\end{figure}

In Fig.~\ref{fig:completeness} we show the estimated completeness within different magnitude bins, where the lines are color coded according to the galaxy population. Continuous lines represent all the galaxies with spectroscopic information with MUSE, while dashed lines are only cluster members.

For the red galaxies, we have a completeness of 100\% up to $m^* +1$, with one galaxy at $i_{\rm auto}<m^*$ and z = 0.611 ($\Delta v=5,940$ km s$^{-1}$), while at $m^* \leq i_{\rm auto}<m^* +1$ we have two galaxies at $z=0.612$ and $z=0.716$ ($\Delta v=6,130$ km s$^{-1}$ and $\Delta v=25,867$ km s$^{-1}$, respectively). The latter one showed similar properties to the galaxies that belong to the cluster; size, visual color and spatially close to the BCG. Fig.~\ref{fig:manual_redshift} shows the spectra of this galaxy in cyan. Its $r-i$ color index was also part of, towards the higher end, the rather generous width used for our RCS catalog. At $i_{\rm auto}\geq m^* +2$, galaxies look like they belong to the cluster, but do not show strong spectral features with which we can estimate the redshift accurately. Blue galaxies show a similar trend as for red galaxies, with completeness of 100\% up to $m^* +1$, and over 80\% at $i_{\rm auto}<m^* +2$. However most of the blue galaxies, unlike red galaxies, do not belong to the cluster.

\section{Comparison to GMOS data}
\label{sec:gmos_comparison}

To estimate the redshifts of the \NgalallGMOS\ from the GMOS spectroscopic archival data we use the \textsc{IRAF} task \textsc{fxcor}. For these estimations we use 4 template spectra from the \textsc{IRAF} package \textsc{rvsao}; \textit{eltemp} and \textit{sptemp} that are composites of elliptical and spiral galaxies, respectively, produced with the \textsc{FAST} spectrograph for the Tillinghast Telescope \citep{Fabricant_1998}; \textit{habtemp0} produced with the \textsc{hectospec} spectrograph for the MMT as a composite of absorption line galaxies \citep{Fabricant_1998}; and a synthetic galaxy template \textit{syn4} from stellar spectra libraries constructed using stellar light ratios \citep{Quintana_2000}. The redshifts are solved in the spectrum mode of \textsc{fxcor} taking the $r$-value \citep{Tonry1979} as the main reliability factor of the correlation following \cite{Quintana_2000}. They consider $r > 4$ as the limit for a reliable result, here we use the resulting velocity only if it follows that (a) at least 3 out of the 4 estimated redshifts from the templates agree with the heliocentric velocity within $\pm$100 km s$^{-1}$ from the median and (b) at least 2 of those have $r > 5$. Finally, the radial heliocentric velocity of the galaxy and its error is calculated as the mean of the values from the ``on-redshift'' correlations.

\begin{figure}
    \centering
    \includegraphics[width=\linewidth]{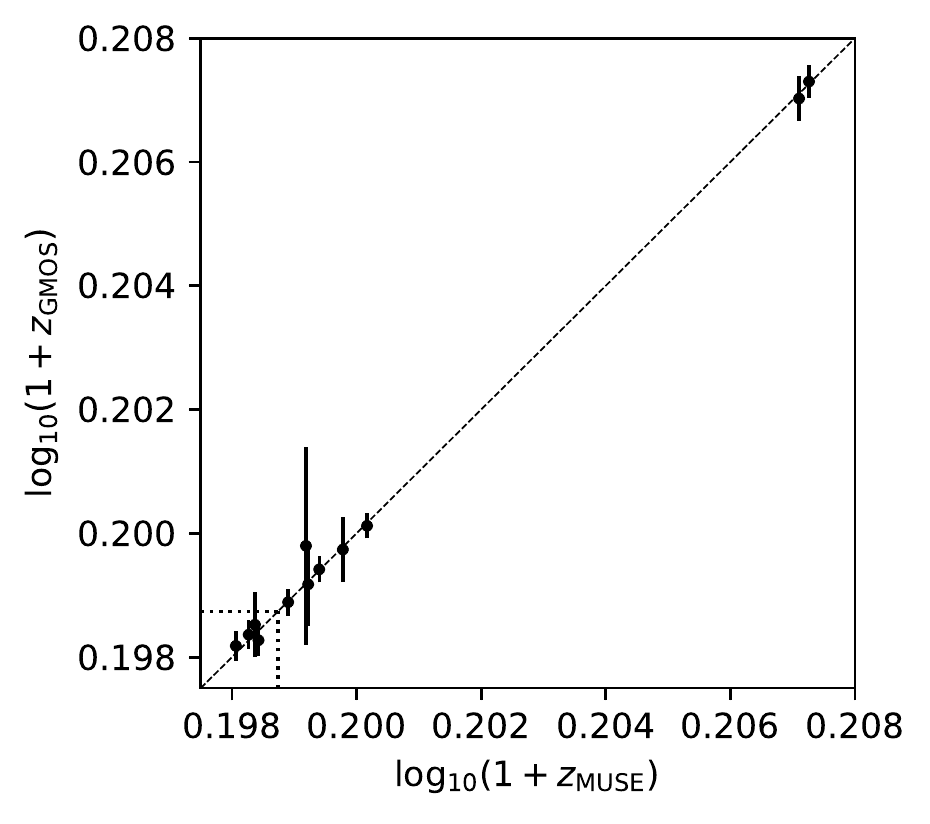}
    \vskip-0.13in
    \caption{Redshift comparison between the GMOS spectra (\textsc{fxcor}) and the MUSE spectra (\textsc{MARZ}) for the \NgalGMOSandMUSE\  galaxies in common. The dashed line shows the 1-to-1 relation while the dotted lines mark the redshift of the cluster at $z_{\rm cl}=0.5803$.}
    \label{fig:gmos_vs_muse}
\end{figure}

Out of the \NgalallGMOS\ GMOS spectra, we have \NgalGMOSandMUSE\  galaxies with a common MUSE measurement, \NgalallGMOScommon\ belonging to the cluster. We use these \NgalGMOSandMUSE\ galaxies in common to compare the results given by \textsc{fxcor} and \textsc{MARZ}, obtaining a mean difference of $60\pm205$ km s$^{-1}$ on the heliocentric reference frame. Fig.~\ref{fig:gmos_vs_muse} shows the estimated redshifts of these sources with the two different methods. Only one galaxy shows a velocity difference higher than 3$\sigma$. Excluding this galaxy from the analysis gives a mean velocity difference of $4\pm96$ km s$^{-1}$.

With respect to the redshift measurements presented in \cite{bayliss16}, we find that the velocity difference within $\pm$5000 km s$^{-1}$ from their redshift estimation of the cluster ($z_{\rm cl} = 0.5801$) is of $|\Delta cz| \approx 300$ km s$^{-1}$ with a big dispersion. Regarding potential cluster members, we select only galaxies where the redshifts reported by \cite{bayliss16} and the ones estimated using \textsc{fxcor} have a difference smaller than 500 km s$^{-1}$, which at $z_{\rm cl} = 0.5801$ corresponds to a difference of $\sim$0.1\%. This eliminates 2 potential cluster members, one from each method. Meaning that we add \NgalallGMOSnewClus\ cluster members from the GMOS data in the final sample

\section{Catalog of spectroscopically confirmed objects}\label{sec:appendix_catalog}

\begin{figure*}
    \centering
    \includegraphics[width=\linewidth]{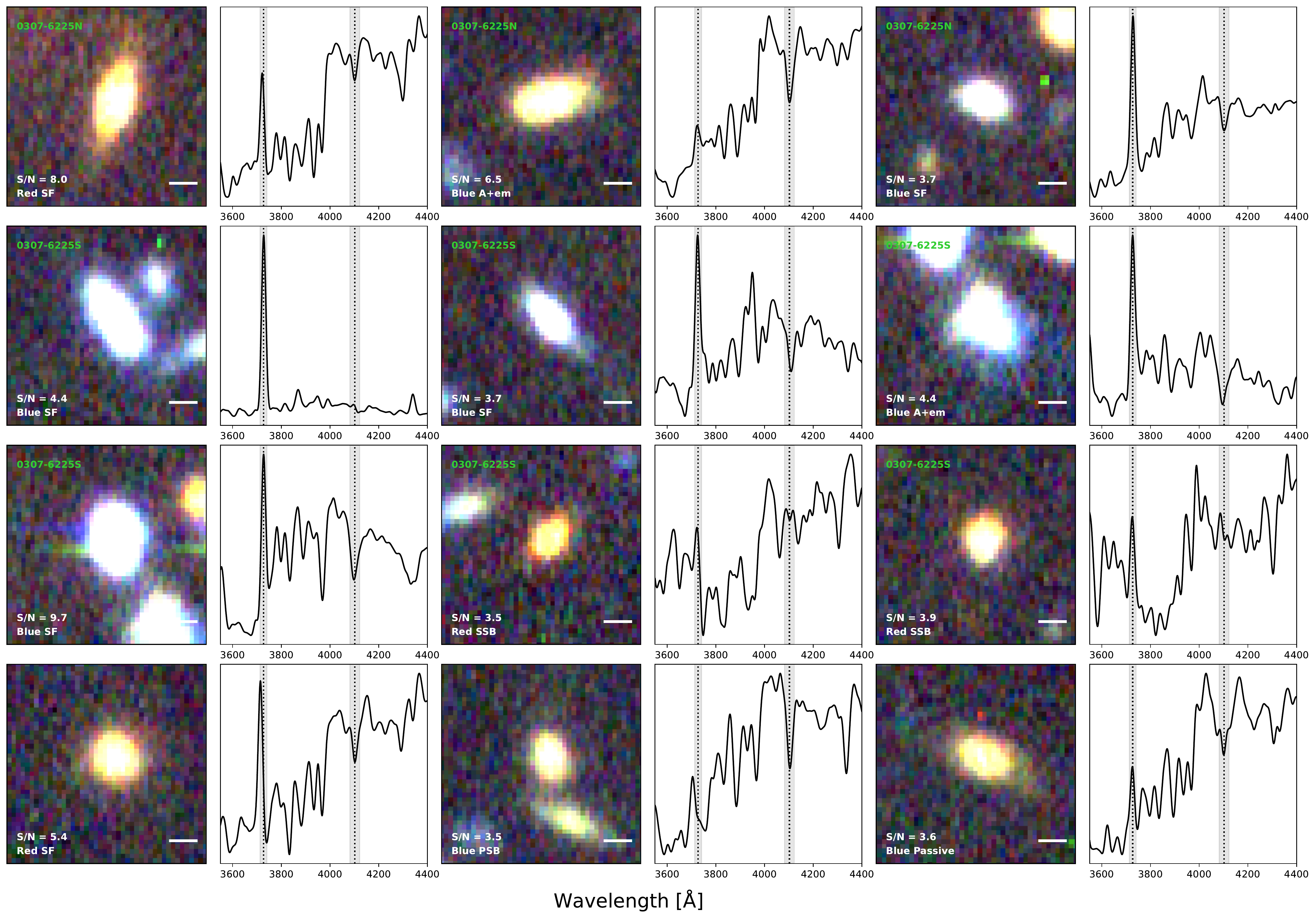}
    \vskip-0.13in
    \caption{Pseudo-color crop images (box size of 7$\times$7 arcsec$^2$) of the SF, A+em, SSB and PSB galaxies from our sample (plus one blue passive galaxy). On the bottom left of each image the spectral type of the galaxy is shown, with a white bar on the bottom right representing the scale size of 1 arcsec. Galaxies on the top and middle row belong to 0307-6225S and 0307-6225N, respectively, while galaxies on the bottom row are those that do not belong to any of the aforementioned. The doppler corrected spectra of each galaxy it's shown to the right, with the dotted lines showing the H$\delta$ and [OII] $\lambda 3727$ \AA\ lines and the gray area marking the width of each line that we use to classify the galaxy \citep{balogh99}.}
    \label{fig:cutouts_spectra}
\end{figure*}

Table~\ref{tab:all_objs_properties} shows the properties of the \NgalALLspec\ objects with spectroscopic information from MUSE (\NgalMUSEspec) or GMOS (22) within the field. The ``Field'' column is a combination of the instrument plus the number of the observed field. In the case of MUSE data this corresponds to the data cubes shown in Fig.~\ref{fig:rgb_image} and Table~\ref{tab:muse_data}, whereas in the case of GMOS, this corresponds to the first or second observed mask \citep[see][]{bayliss16}. The ID column are the object's unique ID within the observed field. Redshifts for MUSE objects correspond to the ones derived using \textsc{MARZ}, while for GMOS they correspond to the ones derived using \textsc{fxcor}. Magnitudes are the derived using \textsc{SExtractor}'s  \textsc{mag\_auto} parameter, while color indexes are derived using \textsc{SExtractor}'s \texttt{mag\_aper} parameter, with a fixed aperture of $\sim40$ kpc at the cluster's redshift. The last column, Q, corresponds to the cluster membership, with 1 for galaxies within the $\pm$3000 km s$^{-1}$ cut from the cluster's redshift, and 0 otherwise.

\begin{table*}
    \caption{Properties of the spectroscopically confirmed objects. The first and second columns are the sky coordinates of the objects. Columns (3) and (4) are the instrument (along with the corresponding field) and the object ID within the field. The heliocentric redshifts are listed in column (5). Columns (6) through (10) are the derived magnitudes and the $g-r$, $r-i$ color indexes (from aperture magnitudes). The last column corresponds to the cluster membership, where 1 means galaxies within the $\pm$3000 km s$^{-1}$ cut from the cluster's redshift $z_{\rm cl} = 0.5803$.}
    \label{tab:all_objs_properties}
    \centering
    \begin{threeparttable}
    \resizebox{0.498\textwidth}{0.41\textheight}{
    \begin{tabular}{|l|l|l|l|r|r|r|r|r|r|r|r}
    \hline
    \hline
      \multicolumn{1}{c}{R.A.} &
      \multicolumn{1}{c}{Dec.} &
      \multicolumn{1}{c}{Field} &
      \multicolumn{1}{c}{ID} &
      \multicolumn{1}{c}{$z_{\rm Helio}$} &
      \multicolumn{1}{c}{$g_{\rm auto}$} &
      \multicolumn{1}{c}{$r_{\rm auto}$} &
      \multicolumn{1}{c}{$i_{\rm auto}$} &
      \multicolumn{1}{c}{$g-r$} &
      \multicolumn{1}{c}{$r-i$} &
      \multicolumn{1}{c}{Q} \\
      \multicolumn{1}{c}{(J2000)} & \multicolumn{1}{c}{(J2000)} & & & & \multicolumn{1}{c}{mag} & \multicolumn{1}{c}{mag} & \multicolumn{1}{c}{mag} & \multicolumn{1}{c}{mag} & \multicolumn{1}{c}{mag} & \\
    \hline
      03:07:17.96 & -62:27:12.19 & MUSE-1 & 01 & 0.6116 & 22.992 & 21.859 & 21.039 & 1.092 & 0.812 & 0\\
  03:07:13.92 & -62:27:28.50 & MUSE-1 & 02 & 0.3711 & 19.972 & 20.888 & 20.336 & -4.737 & 0.567 & 0\\
  03:07:16.80 & -62:26:23.12 & MUSE-1 & 04 & 0.5761 & 25.292 & 23.188 & 22.13 & 2.198 & 1.100 & 1\\
  03:07:16.12 & -62:26:25.34 & MUSE-1 & 06 & 0.5855 & 22.298 & 20.752 & 19.94 & 1.572 & 0.808 & 1\\
  03:07:16.02 & -62:26:28.51 & MUSE-1 & 07 & 0.5716 & 23.813 & 22.289 & 21.487 & 1.412 & 0.816 & 1\\
  03:07:14.51 & -62:26:26.89 & MUSE-1 & 09 & 0.5761 & 24.473 & 22.600 & 21.651 & 1.884 & 0.961 & 1\\
  03:07:14.27 & -62:26:30.88 & MUSE-1 & 13 & 0.5780 & 23.503 & 21.574 & 20.718 & 1.915 & 0.868 & 1\\
  03:07:13.04 & -62:26:33.31 & MUSE-1 & 17 & 0.2153 & 21.541 & 20.511 & 20.163 & 1.012 & 0.359 & 0\\
  03:07:14.69 & -62:26:38.69 & MUSE-1 & 28 & 0.6128 & 24.248 & 23.121 & 22.272 & 0.836 & 0.626 & 0\\
  03:07:14.50 & -62:26:39.97 & MUSE-1 & 29 & 0.2405 & 22.284 & 21.324 & 20.911 & 0.935 & 0.414 & 0\\
  03:07:15.11 & -62:26:40.05 & MUSE-1 & 30 & 0.5694 & 21.449 & 21.037 & 20.809 & 0.419 & 0.234 & 1\\
  03:07:16.57 & -62:26:41.39 & MUSE-1 & 35$^{S_2}$ & 0.5827 & 22.270 & 20.261 & 19.329 & 1.878 & 0.950 & 1\\
  03:07:14.87 & -62:26:43.18 & MUSE-1 & 38 & 0.5729 & 22.122 & 21.722 & 21.598 & 0.393 & 0.104 & 1\\
  03:07:13.68 & -62:26:43.71 & MUSE-1 & 40 & 0.5806 & 24.252 & 22.775 & 21.914 & 1.474 & 0.897 & 1\\
  03:07:18.09 & -62:26:45.12 & MUSE-1 & 42 & 0.5961 & 23.000 & 22.413 & 22.129 & 0.532 & 0.327 & 1\\
  03:07:16.23 & -62:26:47.06 & MUSE-1 & 45 & 0.5813 & 24.609 & 22.317 & 21.333 & 2.235 & 0.979 & 1\\
  03:07:16.75 & -62:26:47.43 & MUSE-1 & 46$^{S_1}$ & 0.5792 & 22.222 & 20.303 & 19.344 & 1.912 & 0.984 & 1\\
  03:07:16.61 & -62:26:49.61 & MUSE-1 & 47 & 0.5740 & 24.553 & 22.549 & 21.600 & 2.008 & 0.950 & 1\\
  03:07:19.06 & -62:26:48.68 & MUSE-1 & 50 & 0.5792 & 25.137 & 22.992 & 22.082 & 2.830 & 0.986 & 1\\
  03:07:16.22 & -62:26:50.30 & MUSE-1 & 51 & 0.5786 & 24.482 & 22.556 & 21.596 & 2.091 & 0.967 & 1\\
  03:07:20.30 & -62:26:51.05 & MUSE-1 & 54 & 0.3284 & 23.391 & 22.886 & 22.846 & 0.319 & 0.014 & 0\\
  03:07:17.33 & -62:26:52.89 & MUSE-1 & 59 & 0.1599 & 23.182 & 22.511 & 22.081 & 0.752 & 0.473 & 0\\
  03:07:16.57 & -62:26:54.58 & MUSE-1 & 60 & 0.0000 & 23.038 & 21.449 & 20.528 & 1.635 & 0.925 & 0\\
  03:07:17.62 & -62:26:55.52 & MUSE-1 & 61 & 0.5830 & 25.192 & 23.469 & 22.601 & 2.023 & 0.855 & 1\\
  03:07:14.52 & -62:26:55.41 & MUSE-1 & 63 & 0.5920 & 23.568 & 23.349 & 23.142 & 0.133 & 0.179 & 1\\
  03:07:20.82 & -62:26:56.46 & MUSE-1 & 65 & 0.5778 & 24.663 & 22.861 & 21.961 & 1.445 & 0.882 & 1\\
  03:07:14.26 & -62:26:59.07 & MUSE-1 & 66 & 0.3713 & 22.570 & 21.772 & 21.617 & 0.790 & 0.157 & 0\\
  03:07:20.45 & -62:26:58.98 & MUSE-1 & 68 & 0.6110 & 23.229 & 21.268 & 20.307 & 1.965 & 0.962 & 0\\
  03:07:18.42 & -62:26:59.81 & MUSE-1 & 69 & 0.5749 & 24.896 & 23.217 & 22.356 & 1.530 & 0.948 & 1\\
  03:07:14.37 & -62:27:03.47 & MUSE-1 & 74 & 0.5879 & 24.141 & 23.651 & 23.327 & 0.493 & 0.225 & 1\\
  03:07:15.77 & -62:27:09.28 & MUSE-1 & 80 & 0.5949 & 22.051 & 21.692 & 21.540 & 0.355 & 0.154 & 1\\
  03:07:12.74 & -62:27:10.74 & MUSE-1 & 82 & 0.5766 & 23.951 & 21.885 & 20.898 & 2.124 & 0.982 & 1\\
  03:07:14.87 & -62:27:11.27 & MUSE-1 & 84 & 0.5855 & 25.151 & 23.060 & 22.152 & 1.918 & 0.960 & 1\\
  03:07:15.89 & -62:27:22.14 & MUSE-1 & 92 & 0.5807 & 24.375 & 22.546 & 21.520 & 1.649 & 1.044 & 1\\
  03:07:15.65 & -62:27:22.78 & MUSE-1 & 93 & 0.5834 & 23.585 & 21.895 & 20.951 & 1.686 & 0.947 & 1\\
  03:07:20.02 & -62:25:55.68 & MUSE-2 & 01 & 0.2151 & 19.335 & 18.233 & 17.874 & 1.115 & 0.366 & 0\\
  03:07:17.20 & -62:25:02.68 & MUSE-2 & 04 & 0.5787 & 24.702 & 22.803 & 22.021 & 1.968 & 0.811 & 1\\
  03:07:16.48 & -62:25:03.86 & MUSE-2 & 05 & 0.4989 & 22.090 & 20.726 & 20.156 & 1.367 & 0.568 & 0\\
  03:07:17.02 & -62:25:05.38 & MUSE-2 & 08 & 0.5724 & 24.943 & 23.234 & 22.432 & 1.539 & 0.736 & 1\\
  03:07:21.08 & -62:25:13.45 & MUSE-2 & 17 & 0.5850 & 24.040 & 23.281 & 22.917 & 0.740 & 0.186 & 1\\
  03:07:16.80 & -62:25:18.55 & MUSE-2 & 21 & 0.5867 & 24.156 & 23.286 & 23.073 & 0.965 & 0.052 & 1\\
  03:07:22.93 & -62:25:18.19 & MUSE-2 & 23 & 0.5856 & 25.186 & 23.491 & 22.706 & 1.767 & 0.733 & 1\\
  03:07:21.73 & -62:25:19.71 & MUSE-2 & 25 & 0.5829 & 24.928 & 23.157 & 22.273 & 1.763 & 0.933 & 1\\
  03:07:19.30 & -62:25:26.50 & MUSE-2 & 29 & 0.0001 & 20.574 & 18.993 & 17.758 & 1.570 & 1.243 & 0\\
  03:07:15.16 & -62:25:26.79 & MUSE-2 & 37 & 0.2146 & 22.552 & 21.950 & 21.804 & 0.606 & 0.160 & 0\\
  03:07:21.16 & -62:25:31.01 & MUSE-2 & 38 & 0.5894 & 22.802 & 21.098 & 20.166 & 1.695 & 0.937 & 1\\
  03:07:21.88 & -62:25:36.13 & MUSE-2 & 48 & 0.5829 & 24.595 & 22.628 & 21.681 & 1.640 & 0.918 & 1\\
  03:07:22.26 & -62:25:37.21 & MUSE-2 & 49 & 0.5749 & 23.628 & 21.943 & 20.991 & 1.698 & 0.956 & 1\\
  03:07:17.23 & -62:25:40.65 & MUSE-2 & 51 & 0.5705 & 24.285 & 22.422 & 21.494 & 1.881 & 0.952 & 1\\
  03:07:21.00 & -62:25:36.43 & MUSE-2 & 55 & 0.5922 & 24.272 & 22.430 & 21.593 & 1.927 & 0.846 & 1\\
  03:07:17.81 & -62:25:46.31 & MUSE-2 & 61 & 0.5780 & 23.132 & 21.614 & 20.797 & 1.528 & 0.816 & 1\\
  03:07:19.81 & -62:25:45.89 & MUSE-2 & 63 & 0.5910 & 24.260 & 23.520 & 23.285 & 0.959 & -0.010 & 1\\
  03:07:16.65 & -62:25:48.66 & MUSE-2 & 66 & 0.5899 & 25.097 & 23.280 & 22.409 & 2.206 & 0.842 & 1\\
  03:07:19.27 & -62:25:48.79 & MUSE-2 & 67 & 0.5797 & 24.414 & 22.930 & 22.072 & 1.427 & 0.774 & 1\\
  03:07:20.70 & -62:25:50.35 & MUSE-2 & 68 & 0.5786 & 21.944 & 20.670 & 19.892 & 1.313 & 0.837 & 1\\
  03:07:17.89 & -62:25:51.37 & MUSE-2 & 70 & 0.2156 & 22.441 & 21.887 & 21.752 & 0.532 & 0.137 & 0\\
  03:07:14.00 & -62:25:53.24 & MUSE-2 & 71 & 0.0002 & 20.385 & 19.029 & 18.514 & 1.343 & 0.526 & 0\\
  03:07:16.65 & -62:25:54.11 & MUSE-2 & 78 & 0.5690 & 25.203 & 23.675 & 22.783 & 1.251 & 0.948 & 1\\
  03:07:20.69 & -62:25:53.83 & MUSE-2 & 79 & 0.5797 & 23.168 & 22.083 & 21.708 & 1.079 & 0.406 & 1\\
  03:07:17.81 & -62:25:56.62 & MUSE-2 & 85 & 0.5859 & 23.620 & 21.788 & 20.898 & 1.767 & 0.890 & 1\\
  03:07:21.40 & -62:25:58.09 & MUSE-2 & 87 & 0.5795 & 24.844 & 22.919 & 22.016 & 2.217 & 0.872 & 1\\
  03:07:22.17 & -62:26:00.71 & MUSE-2 & 90 & 0.5736 & 25.099 & 23.136 & 22.234 & 2.199 & 1.002 & 1\\
  03:07:17.12 & -62:26:01.83 & MUSE-2 & 93 & 0.3696 & 22.866 & 22.028 & 21.840 & 0.855 & 0.201 & 0\\
  03:07:18.83 & -62:26:03.72 & MUSE-2 & 97 & 0.2754 & 23.377 & 22.672 & 22.485 & 0.724 & 0.170 & 0\\
  03:07:19.20 & -62:24:20.28 & MUSE-3 & 07 & -0.0001 & 21.096 & 19.504 & 18.474 & 1.566 & 1.037 & 0\\
  03:07:22.83 & -62:24:18.61 & MUSE-3 & 08 & 0.5830 & 23.713 & 21.908 & 21.033 & 1.815 & 0.894 & 1\\
  03:07:26.09 & -62:24:23.49 & MUSE-3 & 16 & 0.3977 & 23.064 & 21.461 & 20.977 & 1.488 & 0.478 & 0\\
  03:07:23.40 & -62:24:27.73 & MUSE-3 & 19 & 0.5790 & 23.336 & 21.591 & 20.792 & 1.780 & 0.829 & 1\\
  03:07:23.14 & -62:24:29.86 & MUSE-3 & 23 & 0.5783 & 23.821 & 22.027 & 21.164 & 1.729 & 0.855 & 1\\
  03:07:20.55 & -62:24:32.88 & MUSE-3 & 24 & 0.5720 & 24.798 & 23.100 & 22.272 & 1.597 & 0.825 & 1\\
    \hline
    \hline
    \end{tabular}}
    \resizebox{0.498\textwidth}{0.41\textheight}{
    \begin{tabular}{|l|l|l|l|r|r|r|r|r|r|r}
    \hline
    \hline
      \multicolumn{1}{c}{R.A.} &
      \multicolumn{1}{c}{Dec.} &
      \multicolumn{1}{c}{Field} &
      \multicolumn{1}{c}{ID} &
      \multicolumn{1}{c}{$z_{\rm Helio}$} &
      \multicolumn{1}{c}{$g_{\rm auto}$} &
      \multicolumn{1}{c}{$r_{\rm auto}$} &
      \multicolumn{1}{c}{$i_{\rm auto}$} &
      \multicolumn{1}{c}{$g-r$} &
      \multicolumn{1}{c}{$r-i$} &
      \multicolumn{1}{c}{Q} \\
      \multicolumn{1}{c}{(J2000)} & \multicolumn{1}{c}{(J2000)} & & & & \multicolumn{1}{c}{mag} & \multicolumn{1}{c}{mag} & \multicolumn{1}{c}{mag} & \multicolumn{1}{c}{mag} & \multicolumn{1}{c}{mag} &\\
    \hline
      03:07:26.25 & -62:24:37.72 & MUSE-3 & 30 & 0.8144 & 24.928 & 24.661 & 24.227 & 0.215 & -0.075 & 0\\
  03:07:21.46 & -62:24:42.92 & MUSE-3 & 34 & 0.5819 & 24.244 & 22.718 & 22.052 & 1.583 & 0.632 & 1\\
  03:07:23.08 & -62:24:45.50 & MUSE-3 & 39 & 0.2843 & 25.394 & 23.964 & 23.418 & 1.922 & 0.450 & 0\\
  03:07:23.09 & -62:24:47.93 & MUSE-3 & 42 & 0.5753 & 24.943 & 23.012 & 22.102 & 2.120 & 0.920 & 1\\
  03:07:21.52 & -62:24:50.91 & MUSE-3 & 48 & 0.8614 & 23.911 & 23.659 & 23.156 & 0.201 & 0.591 & 0\\
  03:07:20.51 & -62:24:50.94 & MUSE-3 & 49 & 0.5779 & 23.558 & 22.119 & 21.186 & 1.279 & 0.934 & 1\\
  03:07:22.81 & -62:24:54.49 & MUSE-3 & 51 & 0.1385 & 23.274 & 22.694 & 22.561 & 0.576 & 0.106 & 0\\
  03:07:22.94 & -62:24:59.24 & MUSE-3 & 58 & 0.5899 & 23.944 & 22.708 & 22.196 & 1.254 & 0.513 & 1\\
  03:07:22.83 & -62:25:01.47 & MUSE-3 & 62 & 0.5810 & 24.242 & 23.165 & 22.819 & 0.987 & 0.403 & 1\\
  03:07:25.25 & -62:25:03.18 & MUSE-3 & 64 & 0.5864 & 23.979 & 22.442 & 21.643 & 1.631 & 0.803 & 1\\
  03:07:23.71 & -62:25:06.47 & MUSE-3 & 67 & 0.3701 & 21.651 & 20.579 & 20.236 & 1.094 & 0.346 & 0\\
  03:07:24.01 & -62:25:07.89 & MUSE-3 & 68 & 0.7218 & 23.492 & 22.811 & 22.385 & 0.582 & 0.440 & 0\\
  03:07:23.33 & -62:25:06.18 & MUSE-3 & 73 & 0.6039 & 23.649 & 22.947 & 22.548 & 0.607 & 0.489 & 0\\
  03:07:22.16 & -62:25:09.17 & MUSE-3 & 76 & 0.5725 & 25.233 & 23.313 & 22.376 & 1.834 & 0.945 & 1\\
  03:07:26.14 & -62:23:18.43 & MUSE-4 & 03 & 0.0003 & 21.154 & 19.698 & 19.118 & 1.429 & 0.588 & 0\\
  03:07:26.83 & -62:23:21.89 & MUSE-4 & 05 & -0.0001 & 21.896 & 20.397 & 19.85 & 1.475 & 0.566 & 0\\
  03:07:24.84 & -62:23:22.06 & MUSE-4 & 06 & 0.8031 & 21.605 & 21.330 & 21.035 & 0.268 & 0.303 & 0\\
  03:07:25.41 & -62:23:25.29 & MUSE-4 & 12 & 0.8032 & 23.058 & 22.731 & 22.206 & 0.307 & 0.551 & 0\\
  03:07:21.89 & -62:23:28.04 & MUSE-4 & 13 & 0.5771 & 22.956 & 21.217 & 20.409 & 1.712 & 0.817 & 1\\
  03:07:26.90 & -62:23:28.12 & MUSE-4 & 14 & 0.5860 & 24.507 & 22.764 & 21.955 & 1.719 & 0.821 & 1\\
  03:07:30.15 & -62:23:31.80 & MUSE-4 & 15 & 0.5790 & 23.320 & 22.259 & 21.585 & 1.081 & 0.684 & 1\\
  03:07:28.25 & -62:23:36.05 & MUSE-4 & 16 & 0.5867 & 24.186 & 22.344 & 21.405 & 1.873 & 0.945 & 1\\
  03:07:26.92 & -62:23:36.95 & MUSE-4 & 18 & 0.5733 & 24.684 & 22.822 & 21.887 & 1.848 & 0.939 & 1\\
  03:07:28.56 & -62:23:37.56 & MUSE-4 & 19 & 0.5837 & 24.066 & 22.388 & 21.656 & 1.716 & 0.774 & 1\\
  03:07:23.54 & -62:23:39.02 & MUSE-4 & 21 & 0.5345 & 23.556 & 22.927 & 22.717 & 0.623 & 0.246 & 0\\
  03:07:29.30 & -62:23:41.94 & MUSE-4 & 23 & -0.0000 & 23.342 & 21.804 & 20.677 & 1.410 & 1.144 & 0\\
  03:07:22.44 & -62:23:43.38 & MUSE-4 & 24 & 0.5803 & 24.648 & 22.691 & 21.821 & 1.540 & 0.800 & 1\\
  03:07:26.13 & -62:23:44.20 & MUSE-4 & 25 & 0.5823 & 24.632 & 22.951 & 22.073 & 1.452 & 0.785 & 1\\
  03:07:28.48 & -62:23:44.96 & MUSE-4 & 26 & 0.5841 & 23.618 & 22.073 & 21.463 & 1.525 & 0.615 & 1\\
  03:07:22.96 & -62:23:48.47 & MUSE-4 & 28 & 0.1160 & 19.287 & 18.934 & 18.727 & 0.338 & 0.206 & 0\\
  03:07:22.87 & -62:23:57.32 & MUSE-4 & 32 & 0.1160 & 22.113 & 21.097 & 20.812 & 0.954 & 0.275 & 0\\
  03:07:26.77 & -62:23:51.79 & MUSE-4 & 33 & 0.5759 & 23.908 & 22.160 & 21.280 & 1.667 & 0.888 & 1\\
  03:07:26.15 & -62:23:52.93 & MUSE-4 & 34 & 0.7777 & 24.493 & 23.999 & 23.557 & 0.657 & 0.337 & 0\\
  03:07:25.74 & -62:23:54.13 & MUSE-4 & 35 & 0.5815 & 23.969 & 22.147 & 21.224 & 1.659 & 0.892 & 1\\
  03:07:27.18 & -62:23:54.43 & MUSE-4 & 37 & 0.5846 & 23.364 & 22.609 & 22.225 & 0.663 & 0.417 & 1\\
  03:07:22.48 & -62:24:04.09 & MUSE-4 & 40 & 0.3700 & 25.763 & 23.961 & 23.271 & 2.543 & 0.788 & 0\\
  03:07:27.50 & -62:23:59.31 & MUSE-4 & 44 & 0.5779 & 23.718 & 23.318 & 23.221 & 0.445 & -0.117 & 1\\
  03:07:24.51 & -62:24:00.96 & MUSE-4 & 46 & 0.5830 & 25.350 & 23.431 & 22.477 & 1.936 & 1.023 & 1\\
  03:07:23.86 & -62:24:02.18 & MUSE-4 & 47 & 0.5802 & 24.326 & 22.439 & 21.528 & 1.890 & 0.934 & 1\\
  03:07:24.76 & -62:24:02.30 & MUSE-4 & 48 & 0.5780 & 25.216 & 23.322 & 22.370 & 1.979 & 1.044 & 1\\
  03:07:26.41 & -62:24:03.99 & MUSE-4 & 49 & 0.3706 & 23.287 & 21.736 & 21.318 & 1.543 & 0.457 & 0\\
  03:07:28.06 & -62:24:03.86 & MUSE-4 & 50 & 0.5819 & 24.889 & 22.947 & 22.140 & 1.978 & 0.730 & 1\\
  03:07:28.16 & -62:24:04.95 & MUSE-4 & 51 & 0.5808 & 25.110 & 22.964 & 22.049 & 2.527 & 0.821 & 1\\
  03:07:24.55 & -62:24:06.62 & MUSE-4 & 54 & 0.5821 & 24.272 & 22.389 & 21.518 & 1.816 & 0.868 & 1\\
  03:07:24.18 & -62:24:07.57 & MUSE-4 & 56 & 0.5753 & 24.073 & 22.237 & 21.278 & 1.844 & 0.938 & 1\\
  03:07:23.85 & -62:24:10.12 & MUSE-4 & 57$^N$ & 0.5809 & 21.834 & 19.980 & 19.014 & 1.893 & 0.993 & 1\\
  03:07:25.10 & -62:24:11.12 & MUSE-4 & 60 & 0.5847 & 25.519 & 23.725 & 22.702 & 1.705 & 1.137 & 1\\
  03:07:28.71 & -62:24:00.09 & GMOS-1 & 07 & 0.6728 & 23.967 & 23.583 & 22.932 & 0.380 & 0.674 & 0\\
  03:07:19.96 & -62:23:53.24 & GMOS-1 & 09 & 0.6115 & 21.843 & 20.844 & 20.292 & 1.033 & 0.570 & 0\\
  03:07:07.53 & -62:24:35.80 & GMOS-1 & 11 & 0.5837 & 22.640 & 21.539 & 20.904 & 1.102 & 0.650 & 1\\
  03:07:10.56 & -62:24:29.11 & GMOS-1 & 12 & 0.5426 & 22.485 & 21.501 & 21.115 & 0.971 & 0.397 & 0\\
  03:07:30.38 & -62:25:23.19 & GMOS-1 & 13 & 0.4893 & 24.082 & 22.990 & 22.305 & 1.063 & 0.708 & 0\\
  03:07:29.24 & -62:25:04.10 & GMOS-1 & 14 & 0.5732 & 23.335 & 22.153 & 21.477 & 1.204 & 0.674 & 1\\
  03:07:26.39 & -62:25:37.16 & GMOS-1 & 15 & 0.5815 & 24.048 & 22.468 & 21.734 & 1.567 & 0.729 & 1\\
  03:07:27.26 & -62:25:13.13 & GMOS-1 & 17 & 0.5807 & 23.876 & 22.555 & 22.108 & 1.283 & 0.444 & 1\\
  03:07:19.96 & -62:24:50.10 & GMOS-1 & 18 & 0.4669 & 22.338 & 22.203 & 21.867 & 0.175 & 0.342 & 0\\
  03:07:30.26 & -62:26:01.74 & GMOS-1 & 23 & 0.4999 & 23.540 & 22.904 & 22.794 & 0.531 & -0.004 & 0\\
  03:07:22.09 & -62:28:03.40 & GMOS-1 & 30 & 0.5010 & 23.523 & 22.756 & 22.052 & 0.738 & 0.684 & 0\\
  03:07:21.13 & -62:27:49.79 & GMOS-1 & 31 & 0.5709 & 22.413 & 21.213 & 20.553 & 1.165 & 0.652 & 1\\
  03:06:59.11 & -62:27:39.75 & GMOS-1 & 33 & 0.5347 & 22.723 & 22.663 & 22.309 & 0.032 & 0.400 & 0\\
  03:07:13.96 & -62:28:30.21 & GMOS-1 & 34 & 0.6023 & 23.281 & 22.520 & 21.927 & 0.702 & 0.599 & 0\\
  03:07:06.69 & -62:24:36.80 & GMOS-2 & 09 & 0.4743 & 23.406 & 22.474 & 21.841 & 0.850 & 0.672 & 0\\
  03:07:26.50 & -62:25:18.88 & GMOS-2 & 10 & 0.5804 & 23.213 & 21.235 & 20.273 & 2.007 & 0.961 & 1\\
  03:07:08.71 & -62:24:51.92 & GMOS-2 & 12 & 0.5752 & 22.124 & 21.322 & 20.898 & 0.792 & 0.420 & 1\\
  03:07:25.63 & -62:25:43.17 & GMOS-2 & 14 & 0.6392 & 23.659 & 21.941 & 20.965 & 1.618 & 0.967 & 0\\
  03:07:03.68 & -62:25:35.52 & GMOS-2 & 15 & 0.8106 & 22.900 & 21.932 & 21.234 & 0.957 & 0.689 & 0\\
  03:07:36.22 & -62:25:54.68 & GMOS-2 & 20 & 0.5810 & 22.772 & 21.325 & 20.757 & 1.381 & 0.563 & 1\\
  03:07:09.78 & -62:27:02.67 & GMOS-2 & 23 & 0.6405 & 25.008 & 23.449 & 22.752 & 1.355 & 0.706 & 0\\
  03:07:17.94 & -62:27:54.13 & GMOS-2 & 27 & 0.4043 & 24.333 & 22.794 & 22.088 & 1.444 & 0.640 & 0\\\\
    \hline
    \hline
    \end{tabular}
    }
    \begin{tablenotes}
 \footnotesize
  \item[N]{BCG of 0307-6225N}
  \item[S$_1$]{First BCG of 0307-6225S}
  \item[S$_2$]{Second BCG of 0307-6225S}
  
 \end{tablenotes}
\end{threeparttable}
\end{table*}

\bsp	
\label{lastpage}
\end{document}